\documentclass[pdflatex,sn-nature]{sn-jnl}% Style for submissions to Nature Portfolio journals

\usepackage{graphicx}%
\usepackage{multirow}%
\usepackage{amsmath,amssymb,amsfonts}%
\usepackage{amsthm}%
\usepackage{mathrsfs}%
\usepackage{xcolor}%
\usepackage{textcomp}%
\usepackage{manyfoot}%
\usepackage{booktabs}%
\usepackage{algorithm}%
\usepackage{algorithmicx}%
\usepackage{algpseudocode}%
\usepackage{listings}%
\usepackage{makecell}%
\usepackage{bm}% bold math
\usepackage{float}% Provides [!htbp] placement specifier
\usepackage{tikz}%
\usetikzlibrary{positioning,arrows.meta,calc}%

\hypersetup{bookmarksdepth=4,hypertexnames=false}

\theoremstyle{thmstyleone}%
\theoremstyle{thmstyletwo}%
\theoremstyle{thmstylethree}%

\unnumbered

\makeatletter
\newcommand{\compactfirstpagefrontmatter}{%
  \def\Titlefont{\reset@font\fontsize{16bp}{19bp}\selectfont\titraggedcenter}%
  \def\Authorfont{\reset@font\fontsize{10.5bp}{12.5bp}\selectfont\boldmath\titraggedcenter}%
  \def\addressfont{\reset@font\fontsize{9.2bp}{10.5bp}\selectfont\titraggedcenter}%
  \def\abstractfont{\reset@font\fontsize{8.6bp}{9.8bp}\selectfont\leftskip=18pt\rightskip=18pt\parfillskip=0pt plus 1fil}%
  \def\keywordfont{\reset@font\fontsize{8bp}{9.2bp}\selectfont\leftskip=18pt\rightskip=18pt plus0.5fill}%
  \renewcommand{\@maketitle}{\newpage\null%
    \if@remarkboxon\vbox to 0pt{\vspace*{-78pt}\hspace*{-18pt}\FMremark}\else\vskip8pt\fi%
    \hsize\textwidth\parindent0pt%
    {\hbox to \textwidth{{\Artcatfont\ArtType\hfill}\par}}%
    \ifx\@title\empty\else%
        \removelastskip\vskip12pt\nointerlineskip%
        {\Titlefont\@title\par}%
    \fi%
    \ifx\@subtitle\empty\else%
        \vskip6pt%
        {{\SubTitlefont\@subtitle\par}}%
    \fi%
    \ifnum\aucount>0
        \global\punctcount\aucount%
        \vskip11pt%
        \artauthors\par%
        {\vskip5pt\addressfont\auaddress\par%
         \removelastskip\vskip8pt%
        \ifnum\emailcnt>0\relax%
           \ifx\corrauthemail\@empty\else{\ifnum\aucount>1*\fi}%
           Corresponding author(s). E-mail(s): \corrauthemail\par\fi%
           \ifx\authemail\@empty\else Contributing authors:\ \authemail\fi%
        \fi%
        \ifequalcont{\par$^{\dagger}$\@equalconttext\par}\fi%
         \removelastskip\vskip8pt%
        \ifpresentaddress{\par\@presentaddresstext\par}\fi%
        }%
     \fi%
     {\printabstract\par}%
     {\printkeywords\par}%
     \ifx\@pacs\empty\else%
       \loop\ifnum\PacsCount>0%
          \csname\romannumeral\PacsTmpCnt StorePacsTxt\endcsname\par%
          \StepDownCounter{\PacsCount}%
          \StepUpCounter{\PacsTmpCnt}%
       \repeat%
    \fi%
    \removelastskip\vskip14pt\vskip0pt}%
}
\makeatother

\newif\ifshowauthors
\showauthorstrue

\begin{document}

\title[Label-Free FVM-Residual Training of Attention Graph Neural Networks]{Label-Free Finite-Volume-Residual Training of Attention Graph Neural Networks for Coupled Thermo-Fluid Fields}

\ifshowauthors
\author[1]{\fnm{Tianyu} \sur{Li}}

\author[2]{\fnm{Zhiwei} \sur{Cao}}

\author[3]{\fnm{Qingang} \sur{Zhang}}

\author[4]{\fnm{Ruihang} \sur{Wang}}

\author[5]{\fnm{Binyang} \sur{Song}}

\author*[3]{\fnm{Yonggang} \sur{Wen}}\email{ygwen@ntu.edu.sg}

\affil[1]{\orgdiv{Interdisciplinary Graduate Programme}, \orgname{Nanyang Technological University}, \orgaddress{\city{Singapore}, \country{Singapore}}}

\affil[2]{\orgdiv{School of Future Technology}, \orgname{Fuzhou University}, \orgaddress{\city{Fuzhou}, \country{China}}}

\affil*[3]{\orgdiv{College of Computing and Data Science}, \orgname{Nanyang Technological University}, \orgaddress{\city{Singapore}, \country{Singapore}}}

\affil[4]{\orgdiv{School of Information Science and Engineering}, \orgname{Yunnan University}, \orgaddress{\city{Kunming}, \state{Yunnan}, \country{China}}}

\affil[5]{\orgdiv{School of Mechanical and Aerospace Engineering}, \orgname{Nanyang Technological University}, \orgaddress{\city{Singapore}, \country{Singapore}}}
\else
\author[1]{\fnm{Anonymous} \sur{Authors}}
\affil[1]{\orgname{Author and affiliation information withheld for anonymous circulation}}
\fi

\abstract{
Neural surrogates are widely used in scientific machine learning for fast prediction of three-dimensional (3D) thermo-fluid fields.
However, generating training data using conventional numerical solvers often incurs substantial computational and storage costs.
We propose to train an attention graph neural network by minimizing the finite-volume method (FVM) residuals of the governing equations.
These residuals are evaluated directly on the mesh,
requiring no labeled data.
We evaluate the trained surrogates against computational fluid dynamics (CFD) references and a data-supervised baseline across four scenarios.
On the two steady-state benchmarks, each solved on the same mesh as its CFD reference,
the FVM-loss model reproduces the reference flow and temperature fields with mean field-averaged normalized root-mean-square errors (nRMSEs) near 3\%, 
including the buoyancy--energy coupling.
On the two parametric transient cases,
the FVM-loss model attains lower rollout errors than the supervised baseline in almost every field,
while avoiding the data-generation cost entirely.
These results indicate that the FVM loss can provide a practical training signal for neural surrogates and reduce the model development cost.
}

\keywords{Graph Neural Network, Finite Volume Method, Physics-Constrained Neural Network, 3D Thermo-Fluid Simulation}

\compactfirstpagefrontmatter
\maketitle
\clearpage

%%==========================================================================%%
%%  INTRODUCTION -- Nature style: no subheadings, concise
%%==========================================================================%%
\section{Introduction}\label{sec:introduction}

Fast and accurate prediction of three-dimensional (3D) thermo-fluid fields underpins the design and operation of industrial-scale engineering systems.
For example, data centers already account for about 1\% of global electricity use despite decades of gains in computing efficiency~\cite{koomey2011computing_efficiency,masanet2020datacenter_energy},
and in data centers that run artificial intelligence workloads, compute racks dissipate heat at rising power densities~\cite{schmidt2005datacenter_cooling,latif2025gpu_power_demand}.
Inaccurate or slow thermal models thus risk costly failures and wasted energy.
Data-center operators require predictive models sufficiently fast to support repeated what-if studies over many operating conditions,
a regime in which repeated computational fluid dynamics (CFD) solves quickly become computationally expensive~\cite{wang2023kalibre_tomacs,sarkar2025fast_dc_surrogate,cao2022reducio}.
The problem couples momentum, continuity, and energy transport partial differential equations (PDEs) on complex 3D geometries under varying boundary conditions.
A useful model should therefore be physically accurate, computationally efficient, and inexpensive to develop.

Existing learning-based approaches partially meet these requirements.
Data-driven surrogates, including reduced-order models~\cite{cao2022reducio} and neural operators~\cite{li2021fourier_neural_operator,li2023gino,lu2021deeponet}, offer rapid inference.
However, their physical consistency is not ensured by construction,
and they rely largely on labeled training data.
Their data generation for 3D flow problems typically entails many numerical solver runs~\cite{guo2016cnn_steady_flow,leer2021fast_flow_field,duraisamy2019age_of_data,sanchez2020learning_to_simulate},
at the cost of long simulation times and substantial data storage.
In contrast, physics-informed neural networks (PINNs)~\cite{raissi2019physics_informed,raissi2020hidden_fluid_mechanics,mandl2024separable_deeponet,li2025continuous_dependence, karniadakis2021physics_informed_ml} and related strong-form methods, such as the deep Galerkin method~\cite{sirignano2018dgm} and parametric flow surrogates trained without simulation data~\cite{sun2020surrogate}, introduce physical laws into training,
reducing dependence on precomputed data by minimizing PDE residuals through automatic differentiation.
This approach nevertheless remains difficult to scale to complex coupled 3D cases.
The residuals of several coupled equations must be carefully balanced during training~\cite{wang2022ntk_pinn_fail},
which has motivated adaptive residual-weighting strategies~\cite{toscano2026residual_adaptivity}.
The training problem also tends to inherit the ill-conditioning of the discretized PDE system~\cite{cao2025pinn_illconditioning},
and the cost of differentiating the higher-order PDE terms through the network grows with the network size.

This differentiation cost, however, is tied to where the derivatives are evaluated rather than to the residual objective itself.
The finite-volume method (FVM)~\cite{moukalled2016fvm,versteeg2007fvm} evaluates derivatives through sparse, precomputable stencils on the mesh,
and the cost consequently scales with the mesh rather than the neural network size.
On such meshes, geometry-aware graph neural networks (GNNs)~\cite{pfaff2021meshgraphnets} operate naturally,
since message-passing operators are well-defined on the unstructured mesh topology;
yet, when trained on data alone, they carry no explicit mechanism that promotes cell-wise conservation of mass, momentum, or energy~\cite{pfaff2021meshgraphnets,brandstetter2022message_passing_pde,brunton2020ml_fluid}.
Message passing is moreover a local operation,
and standard GNNs are known to be limited in capturing long-range interactions~\cite{alon2021bottleneck},
a shortcoming that becomes consequential in 3D domains with global recirculation.
These requirements therefore call for a label-free training signal that is defined on complex 3D meshes and carries discrete conservation structure,
together with an architecture that also captures long-range coupling.
Here, \emph{label-free} means that no precomputed solution fields serve as regression targets, while the discretized conservation laws, together with the geometry and the boundary and initial conditions, still supervise the model.

Residual losses have also been evaluated on structured grids and unstructured meshes rather than pointwise.
Physics-constrained convolutional surrogates compute finite-difference residuals on regular~\cite{zhu2019physics_constrained} or mapped grids~\cite{gao2021phygeonet}.
A network trained on a staggered marker-and-cell grid learns incompressible flow without ground-truth data~\cite{wandel2021from_scratch},
whereas physics-informed neural operators combine training data with residual constraints and can also learn a solution operator from the residual alone on regular grids~\cite{li2024pino}.
On unstructured meshes, graph networks have been trained with weak-form residuals~\cite{gao2022graph_galerkin} and with physics-informed residuals complemented by a few labeled samples~\cite{peng2023pigcn}.
Finite-volume-informed networks embed the discretization into the network or its training objective,
demonstrated on subsurface contaminant transport~\cite{praditia2021finn}, steady convective heat transfer~\cite{zhang2025fvgpnet}, myocardial perfusion on 3D meshes~\cite{salle2024fvgnn}, and two-dimensional incompressible flow~\cite{li2024fvinn,li2025learning}.
Differentiable-solver approaches make the numerical solver itself differentiable~\cite{du2026difvm},
allowing networks trained through it to correct or accelerate the solve~\cite{um2020solver_in_loop,kochkov2021ml_accelerated_cfd}.
Discrete-loss optimization, in turn, solves each case by optimizing the field variables directly,
without an amortized network to reuse across conditions~\cite{karnakov2024odil}.
A graph network trained on a GPU-assembled finite-volume residual with weighted least-squares gradient reconstructions has solved two-dimensional steady incompressible flows for many boundary conditions in one training run, using a pool of evolving states with random resets~\cite{li2025learning},
and a finite-volume graph network trained with labels has predicted two-dimensional unsteady flow~\cite{li2024fvinn}.
Autoregressive surrogates trained without labels on discrete residuals of their own predictions were demonstrated on regular grids~\cite{geneva2020pde},
and a convolutional solver with an embedded finite-volume discretization solved steady three-dimensional incompressible flow and conjugate heat transfer without training data~\cite{ranade2021discretizationnet}.
Networks trained on a staggered grid learned incompressible flow without ground-truth data in two~\cite{wandel2021from_scratch} and three dimensions~\cite{wandel2021pof3d},
and graph networks trained on finite-element residuals of transient heat conduction transferred to unseen three-dimensional meshes~\cite{wurth2024pimgn}.
Beyond the two-dimensional steady setting of~\cite{li2025learning}, the present study addresses unsteady three-dimensional momentum--energy coupling and introduces the compound-cell graph with boundary-face entries, the face-control-volume momentum residual, slice attention and a disk-backed pool with first-in-first-out resets.

Relative to these lines of work, the present study trains one amortized surrogate for coupled, unsteady 3D thermo-fluid fields on unstructured meshes from the finite-volume residual alone,
is designed to suppress pressure--velocity decoupling in that residual through a complementary momentum balance on face control volumes, and reuses the trained model across operating conditions and unseen cross-sections.
The training signal comes from the conservation equations themselves.
They supply cell-wise residuals on unstructured meshes,
a label-free signal aligned with the reference solver.
Specifically, we develop a graphics processing unit (GPU)-accelerated finite-volume residual objective for three-dimensional unstructured meshes that supervises coupled velocity, pressure, and temperature fields without any labeled data;
because the discrete residual couples each cell to its neighbors,
it supervises each prediction jointly with its neighborhood.
The objective retains the coupled finite-volume operator chain, including the pressure--velocity coupling.
The network is trained on its own rollout states, which prepares it for long autoregressive horizons.
We couple this objective with an attention graph neural network, TransFVGN, which extends the finite-volume graph network (FVGN)~\cite{li2024fvinn} with a slice-attention layer adopted from Transolver~\cite{wu2024transolver}, so that long-range information propagates in one layer rather than over many hops of purely local message passing.
We then evaluate the resulting model on four benchmarks of increasing complexity,
culminating in a synthetic data center hall,
and demonstrate that the FVM residual alone can supply the training signal for coupled 3D thermo-fluid prediction.
The cylinder experiment crosses three architectures (TransFVGN, Transolver and an attention-free GNN) with two training objectives, the finite-volume residual (FVM loss) and supervised regression on CFD labels (data loss),
and the data-center experiment repeats the TransFVGN comparison on an application-oriented hall.
On held-out operating conditions of both parametric benchmarks, the FVM-loss model attains lower full-horizon rollout errors than the data-loss model in almost every field,
and on unseen elliptical cross-sections every FVM-loss model is more accurate than its data-loss counterpart.
The cost analysis separates data generation, training, network evaluation and rollout, and includes central processing unit (CPU)-parallel CFD references.

The remainder of the paper is organized as follows.
The Results section demonstrates the framework across the four benchmarks,
from the steady lid-driven cavity and natural-convection benchmarks to the unsteady heated cylinder wake, its transfer to unseen cross-section geometries, and the synthetic data center hall.
It also examines pressure--velocity decoupling under residual optimization and reports the accuracy and amortized-cost comparison against a data-loss baseline.
The Discussion then interprets these findings and states the present limitations.
The Methods section details the governing equations,
the graph representation, the network architecture, the differentiable finite-volume residual, and the training and evaluation protocols.

%%==========================================================================%%
%%  RESULTS -- Nature style: comes immediately after Introduction
%%==========================================================================%%
\section{Results}\label{sec:results}

\begin{figure}[!b]
\centering
\includegraphics[width=\textwidth,height=0.55\textheight,keepaspectratio]{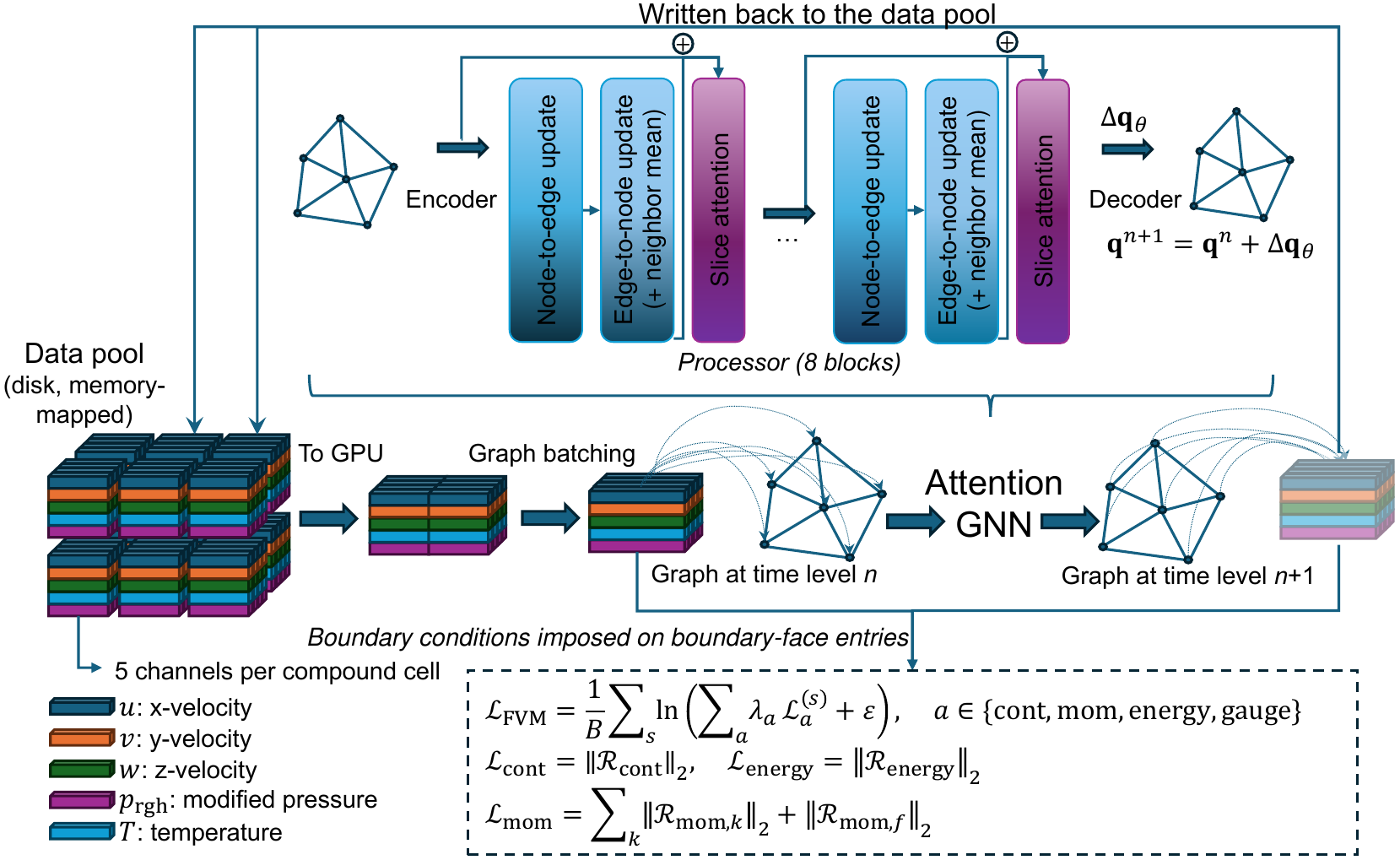}
\caption{Framework overview of the finite-volume-method (FVM)-loss model.
Computational fluid dynamics (CFD) meshes and boundary conditions are parsed into heterogeneous graphs carrying cell states, geometric attributes, and physical parameters.
Bottom row: the evolving cell states are streamed from the data pool to the graphics processing unit (GPU), batched into compound-cell graphs at time level $n$, advanced by the attention graph neural network (GNN) to time level $n+1$ and written back to the pool.
The colored layers of each state are the five channels listed in the key at the lower left, namely the velocity components $u$, $v$, $w$, the modified pressure $p_{\mathrm{rgh}}$ and the temperature $T$.
Top row: the encoder--processor--decoder of the attention GNN (TransFVGN, the finite-volume graph network extended with slice attention), whose processor layers pair local message passing with slice attention, with circled plus signs marking the residual additions of each block;
the decoder returns the state increment $\Delta\mathbf{q}_{\theta}$, and $\mathbf{q}^{n}$ is the state at time level $n$.
Dashed box: the FVM loss $\mathcal{L}_{\mathrm{FVM}}$ and the norms $\mathcal{L}_{\mathrm{cont}}$, $\mathcal{L}_{\mathrm{mom}}$ and $\mathcal{L}_{\mathrm{energy}}$ of the continuity, momentum, and energy residuals of the predicted state, with $\mathcal{L}_{\mathrm{mom}}$ also including the face momentum residual;
the index $a$ runs over the continuity, momentum, energy and pressure-gauge (gauge) terms with fixed weights $\lambda_a$, $s$ indexes the $B$ cases of a batch, $\varepsilon$ is a small floor, $\mathcal{R}$ denotes a discrete residual, with $\mathcal{R}_{\mathrm{mom},k}$ the momentum residual of component $k$ and $\mathcal{R}_{\mathrm{mom},f}$ the face momentum residual, and $\|\cdot\|_2$ is the $\ell_2$ norm.
Weighted least-squares gradient reconstructions, face interpolation, and conservative flux integration assemble these residuals after the boundary conditions are imposed on the boundary-face entries.
Eqs.~\eqref{eq:discrete_continuity}--\eqref{eq:discrete_loss} give the full discrete form.
The same network and residual assembly are used for all benchmarks reported in this study.}
\label{fig:framework_overview}
\end{figure}

We evaluate the FVM-loss model on four benchmarks of increasing complexity.
Two steady benchmarks (the lid-driven cavity and natural convection), each solved by a TransFVGN model trained on the finite-volume residual with the SOAP optimizer~\cite{vyas2024soap} (a second-order method that runs Adam in the eigenbasis of the Shampoo preconditioner), are followed by two parametric application benchmarks (the heated cylinder wake and the data center hall).
The 3D lid-driven cavity verifies incompressible velocity and pressure accuracy,
and the natural-convection cavity isolates buoyancy--energy coupling.
The heated cylinder wake introduces unsteady thermo-fluid dynamics on an unstructured mesh and tests accuracy on Reynolds numbers held out from training within the trained parameter range.
The synthetic data center hall serves as an application-oriented stress test for data-center thermal management.
The ``Transfer to unseen cross-section geometries'' section then applies the cylinder models to unseen cross-sections,
and the ``Pressure--velocity decoupling under residual optimization'' section examines the pressure--velocity decoupling that motivates the combined momentum residual.
The ``Computational cost and amortization'' section compares the cost of the surrogates with that of CFD.
Detailed problem specifications, mesh parameters, and solver configurations are given in Methods (``Experimental setup'').

\subsection{Problem definition and framework overview}\label{subsec:problem_definition}

The FVM-loss model learns a one-step update operator on the mesh
\begin{equation}
\mathbf{q}^{n+1} = \mathcal{G}_{\theta}(\mathcal{M},\mathcal{B},\boldsymbol{\mu},\mathbf{q}^{n}),
\qquad
\mathbf{q}=(u,v,w,p_{\mathrm{rgh}},T)^{\top},
\label{eq:problem_definition}
\end{equation}
where $\mathcal{M}$ denotes the 3D mesh (defined in the ``Heterogeneous graph representation'' section of Methods), $\mathcal{B}$ the boundary-condition set, $\boldsymbol{\mu}$ the physical parameters, $\mathbf{q}^{n}$ the cell-wise thermo-fluid state at time level $n$ (a pseudo-time level for the two steady benchmarks), and $\theta$ the trainable network parameters.
Unlike data-driven surrogates, $\mathcal{G}_{\theta}$ is not trained to regress labeled CFD snapshots.
It is trained toward the parametric solution operator of the finite-volume discretization rather than toward a given CFD dataset.
Its predictions accordingly inherit the discretization error of the finite-volume scheme, as those of a conventional solver would, in addition to the optimization and approximation errors of the network, which we refer to jointly as the learning error.
Comparisons with the CFD reference further contain the scheme differences between the residual and the reference solver (see ``Experimental setup'' in Methods).
Supplementary Tables~1 and~2 (Supplementary Note~2) collect the notation and the main formulas of the method in one place.

As shown in Figure~\ref{fig:framework_overview},
the neural network and FVM residual evaluation share the same mesh representation.
This is a key practical distinction from PINNs based on automatic differentiation (AD).
The optimization target is the discrete algebraic balance actually solved on the mesh, not a pointwise continuous residual evaluated through AD.
We report errors field by field, so that the large absolute temperature scale does not dominate a concatenated state norm.
For field $f$, case $s$ and time step $n$, the relative $\ell_2$ error, reported as the normalized root-mean-square error (nRMSE), and its cumulative geometric mean in time are
\begin{equation}
\begin{aligned}
e_{s,f}^{n}&=\frac{\|q_{s,f}^{n}-q_{s,f}^{\mathrm{ref},n}\|_2}{\|q_{s,f}^{\mathrm{ref},n}\|_2},\\
\bar e_{s,f}(N)&=\exp\!\left[\frac{1}{N}\sum_{n=1}^{N}\ln\!\bigl(\max(e_{s,f}^{n},\varepsilon)\bigr)\right],\\
E_f(N)&=\frac{1}{N_{\mathrm{test}}}\sum_{s=1}^{N_{\mathrm{test}}}\bar e_{s,f}(N),
\end{aligned}
\label{eq:nrmse}
\end{equation}
where the norms run over interior cells and $\varepsilon$ only prevents the logarithm of zero.
Here $N_{\mathrm{test}}$ is the number of held-out test cases.
The geometric mean in time follows the aggregated error of~\cite{koehler2024apebench}, which keeps the late steps of a growing error from dominating the aggregate;
whereas that benchmark averages over samples before aggregating in time, we aggregate each case in time and then average over cases.
The accuracy tables report these case means, and the field-averaged (``all fields'') score is the average of $E_f$ over $f\in\{U,p_{\mathrm{rgh}},T\}$.
The two steady benchmarks average the fields they report, namely velocity and pressure for the lid-driven cavity and velocity, pressure and temperature for natural convection.
The rollout figures instead show case medians, with interquartile or full-range bands as stated in each caption.
In the accuracy tables and the rollout figures, the velocity entry $|\mathbf U|$ is the relative $\ell_2$ error of the velocity vector, whose components are stated with each benchmark.
Data-center temperature is additionally reported as absolute error in kelvin because normalization by an approximately 300~K reference makes its nRMSE small.
In the two steady cases, the TransFVGN architecture is trained on the finite-volume residual separately for each condition, and the error of Eq.~\eqref{eq:nrmse} is evaluated once, on the final state ($N=1$).
This comparison against the CFD solution checks the trained discrete solve on the specified meshes and is separate from the parametric surrogate rollouts.

\subsection{3D lid-driven cavity}\label{subsec:cavity}

The lid-driven cavity (LDC) serves as a canonical benchmark for 3D incompressible flows.
The configuration consists of a unit cube ($L=1$~m) with a moving top lid (velocity $U_{\mathrm{lid}}$ along $+x$) and stationary no-slip walls elsewhere.
The Reynolds number $\mathrm{Re}=U_{\mathrm{lid}}L/\nu$ spans $[100,3200]$.
In this benchmark $y$ is the lid-normal vertical direction and $z$ the spanwise direction, as in the cubic natural-convection case;
the data center instead takes $z$ as vertical.
Here the FVM-loss model is trained separately at each of five representative Reynolds numbers ($\mathrm{Re}=100,400,1000,2000,3200$);
as the first benchmark,
it verifies that training on the finite-volume residual alone reproduces the CFD velocity and pressure fields computed on the same mesh (see ``Experimental setup'' in Methods for setup details).

\begin{figure}[!htbp]
\centering
\includegraphics[width=\textwidth,height=0.72\textheight,keepaspectratio]{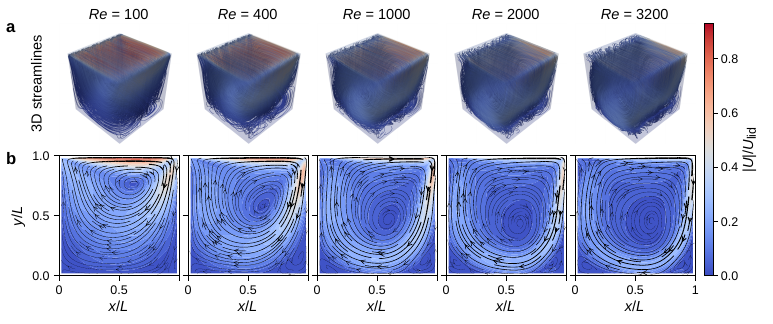}
\caption{Flow structure predicted by the FVM-loss model in the three-dimensional (3D) lid-driven cavity.
The FVM-loss model is trained on the finite-volume-method (FVM) residual without labeled data,
and the columns correspond to the Reynolds numbers $\mathrm{Re}=100$, $400$, $1000$, $2000$ and $3200$.
\textbf{a}, 3D streamlines inside the unit cube.
\textbf{b}, two-dimensional streamlines, drawn in black, on the mid-span plane ($z\approx0.49$), with $x$ and $y$ the lid-parallel and vertical coordinates, $z$ the spanwise coordinate and $L$ the cube side.
The streamlines of row a and the background of row b are colored by the velocity magnitude $|U|/U_{\mathrm{lid}}$ on a shared scale, where $U_{\mathrm{lid}}$ is the lid velocity.}
\label{fig:cavity_results}
\end{figure}

\begin{figure}[!htbp]
\centering
\includegraphics[width=\textwidth,height=0.72\textheight,keepaspectratio]{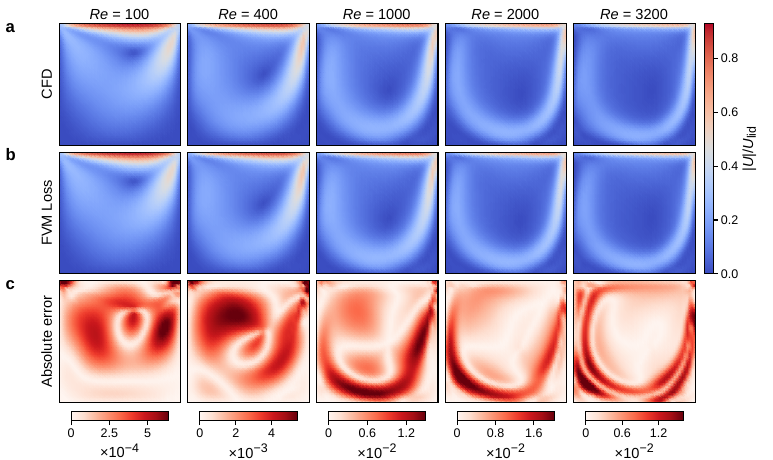}
\caption{Mid-span velocity-magnitude comparison for the three-dimensional lid-driven cavity.
Columns correspond to the five solved Reynolds numbers $\mathrm{Re}=100$, $400$, $1000$, $2000$ and $3200$.
\textbf{a}, computational fluid dynamics (CFD) reference.
\textbf{b}, prediction of the model trained on the finite-volume-method residual loss (FVM Loss).
\textbf{c}, absolute error $\big|\Delta|U|\big|$ in units of the lid velocity $U_{\mathrm{lid}}$.
Rows a and b share one color scale of the normalized velocity magnitude $|U|/U_{\mathrm{lid}}$;
each error panel has its own scale, capped at its 99th percentile.
At $\mathrm{Re}=3200$ the CFD reference is a transient time average over $t=50$--$150$~s.}
\label{fig:ldc_zslice_comparison}
\end{figure}

\begin{figure}[!htbp]
\centering
\includegraphics[width=\textwidth,height=0.72\textheight,keepaspectratio]{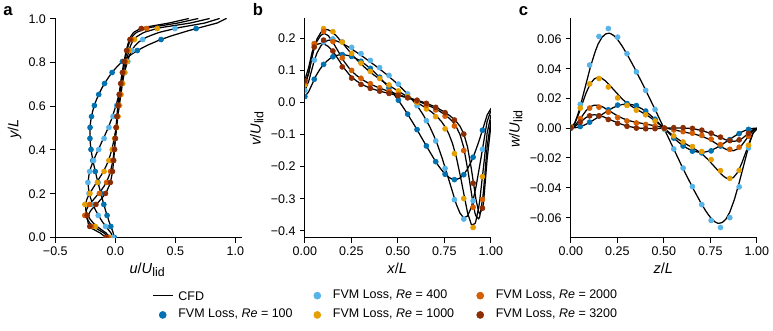}
\caption{Centerline velocity profiles of the three-dimensional lid-driven cavity.
The profiles pass through the cube center ($x=y=z=0.5$, coordinates normalized by the cube side $L$) at five Reynolds numbers $\mathrm{Re}$,
and $u$, $v$ and $w$ are the velocity components along the lid direction $x$, the vertical $y$ and the spanwise $z$, normalized by the lid velocity $U_{\mathrm{lid}}$.
\textbf{a}, $u(y)$ along the vertical centerline,
showing the lid-driven shear layer.
\textbf{b}, $v(x)$ along the horizontal centerline, showing recirculation strength.
\textbf{c}, $w(z)$ along the spanwise centerline, revealing three-dimensional effects.
Black lines are the computational fluid dynamics (CFD) reference and colored markers the model trained on the finite-volume-method residual loss (FVM Loss), with one color per Reynolds number as identified in the key below the panels.}
\label{fig:ldc_centerline}
\end{figure}

\begin{figure}[!htbp]
\centering
\includegraphics[width=\textwidth,height=0.72\textheight,keepaspectratio]{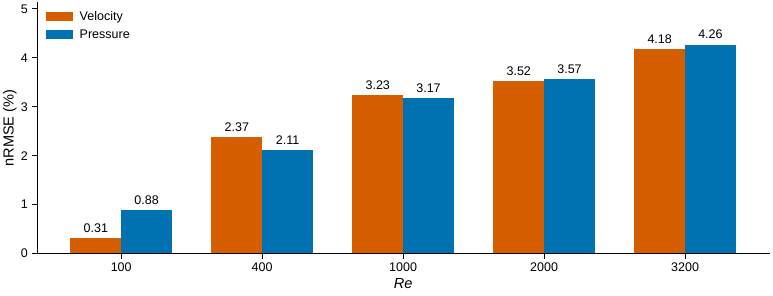}
\caption{Accuracy of the FVM-loss model across Reynolds numbers in the lid-driven cavity.
Bars give the velocity (vermillion) and pressure (blue) normalized root-mean-square error (nRMSE, in percent) of the model trained on the finite-volume-method (FVM) residual loss, over all interior cells of the $48^3$ mesh at each Reynolds number $\mathrm{Re}$ of the lid-driven cavity (LDC) benchmark,
with the value of each bar printed above it.
At $\mathrm{Re}=3200$ the reference is the transient computational fluid dynamics (CFD) time average described in Figure~\ref{fig:ldc_zslice_comparison}.}
\label{fig:ldc_error_heatmap}
\end{figure}

The FVM-loss model reproduces the flow topology across the full Reynolds-number range,
with velocity nRMSE of $0.31$--$4.18\%$ (Eq.~\eqref{eq:nrmse}).
Figure~\ref{fig:cavity_results} shows the FVM-loss predictions as 3D streamlines inside the unit cube and as streamlines on the mid-span plane, colored by velocity magnitude.
The mid-span streamlines show the primary recirculation zone,
its descent toward mid-height as $\mathrm{Re}$ increases,
and the growth of the secondary corner eddies at higher Reynolds numbers.
To quantify the spatial error distribution beyond this qualitative agreement, Figure~\ref{fig:ldc_zslice_comparison} compares the velocity magnitude on the mid-span plane ($z\approx0.49$) between the CFD reference and the FVM-loss model at the five solved Reynolds numbers;
all LDC figures (Figures~\ref{fig:cavity_results}, \ref{fig:ldc_centerline}, \ref{fig:ldc_error_heatmap}) use these conditions and therefore characterize accuracy at the conditions each solve was trained for.
The absolute error remains small, at most $0.04\,U_{\mathrm{lid}}$ on this plane.
At $\mathrm{Re}=100$ and $400$ it is spread over the recirculation region,
whereas from $\mathrm{Re}=1000$ upward it concentrates along the bottom and side walls, at the outer edge of the primary vortex.

Figure~\ref{fig:ldc_centerline} compares centerline velocity profiles at five Reynolds numbers.
The predictions follow the CFD profiles closely,
including the primary vortex position, the shear layer and the weak spanwise velocity.
Figure~\ref{fig:ldc_error_heatmap} shows the velocity and pressure nRMSE over all interior cells at these five solved Reynolds numbers.
Both errors are below $1\%$ at $\mathrm{Re}=100$ and rise to about $4\%$ at $\mathrm{Re}=3200$,
with a mean of $2.76\%$ across the two fields and five cases.
At $\mathrm{Re}=3200$, where the steady semi-implicit method for pressure-linked equations (SIMPLE) does not converge, the reference is a transient CFD time average over $t=50$--$150$~s.
The comparison at this Reynolds number therefore concerns the mean flow, which a steady solution need not reproduce exactly, because the time average also contains the contribution of the velocity fluctuations.
The growth of the error with the Reynolds number is consistent with the thinner shear layers and larger corner eddies at higher $\mathrm{Re}$, where the remaining scheme differences between the residual and the reference solver (see ``Experimental setup'' in Methods) matter more.

\subsection{3D natural convection}\label{subsec:natural_convection}

The natural-convection benchmark tests whether residual training resolves the two-way coupling between buoyancy-driven flow and heat transport.
It uses a three-dimensional, differentially heated cube of side $H=1$~m;
as for the lid-driven cavity, its fields are obtained by training the same architecture on the finite-volume residual.
The cold and hot walls are at $x=0$ and $x=H$;
gravity acts in $-y$, and $z$ is spanwise.
All walls are no-slip and the four non-isothermal walls are adiabatic.
We solve Rayleigh numbers $\mathrm{Ra}=10^3$, $10^4$, $10^5$ and $10^6$ at Prandtl number $\mathrm{Pr}=0.71$ on a wall-refined $48^3$ mesh.
The wall temperature difference $T_{\mathrm{hot}}-T_{\mathrm{cold}}$ is 1~K,
and temperatures are reported as the dimensionless $\theta=(T-T_{\mathrm{cold}})/(T_{\mathrm{hot}}-T_{\mathrm{cold}})$, which is 0 on the cold wall and 1 on the hot wall.
Velocity profiles and peak values are normalized by the free-fall velocity $U_f=\sqrt{g\beta(T_{\mathrm{hot}}-T_{\mathrm{cold}})H}\approx0.181$~m\,s$^{-1}$.

\begin{figure}[!htbp]
\centering
\includegraphics[width=\textwidth,height=0.72\textheight,keepaspectratio]{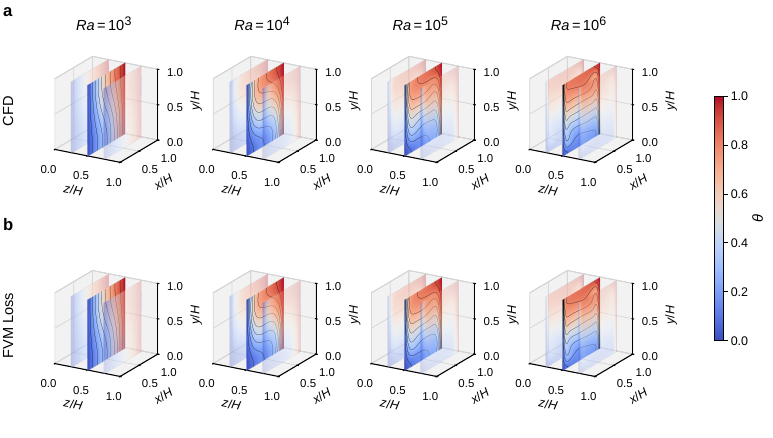}
\caption{Temperature fields of the three-dimensional natural-convection cavity on three spanwise planes.
The planes are $z/H=0.25$, 0.5 and 0.75, with $x$ the hot-wall-normal, $y$ the vertical and $z$ the spanwise coordinate and $H$ the cube side,
and the columns correspond to the Rayleigh numbers $\mathrm{Ra}=10^3$, $10^4$, $10^5$ and $10^6$.
\textbf{a}, computational fluid dynamics (CFD) solution.
\textbf{b}, solution of the model trained on the finite-volume-method residual loss (FVM Loss).
Both rows show the dimensionless temperature $\theta=(T-T_{\mathrm{cold}})/(T_{\mathrm{hot}}-T_{\mathrm{cold}})$, which is 0 on the cold wall and 1 on the hot wall, on the color scale at the right.
Black lines are isotherms on the mid-span plane at $\theta=0.1,0.2,\dots,0.9$.
The outer planes are drawn semi-transparent.}
\label{fig:nc_theta_planes}
\end{figure}

\begin{figure}[!htbp]
\centering
\includegraphics[width=\textwidth,height=0.72\textheight,keepaspectratio]{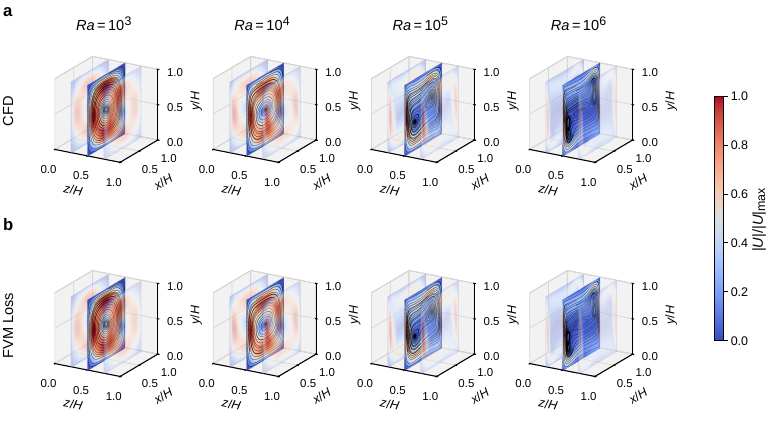}
\caption{Velocity-magnitude fields of the three-dimensional natural-convection cavity on three spanwise planes.
The planes are $z/H=0.25$, 0.5 and 0.75, with $x$ the hot-wall-normal, $y$ the vertical and $z$ the spanwise coordinate and $H$ the cube side.
The columns correspond to the Rayleigh numbers $\mathrm{Ra}=10^3$--$10^6$, and projected streamlines are drawn in black on the mid-span plane.
\textbf{a}, computational fluid dynamics (CFD) solution.
\textbf{b}, solution of the model trained on the finite-volume-method residual loss (FVM Loss).
Each speed field $|U|$ is normalized by its own maximum $|U|_{\max}$ over the cube, and the colors therefore compare the spatial organization of the flow.
The outer planes are drawn semi-transparent.
Absolute amplitudes are compared through the centerline profiles and the normalized root-mean-square errors over all interior cells (Figures~\ref{fig:nc_centerlines} and~\ref{fig:nc_error_heatmap}).}
\label{fig:nc_velocity_planes}
\end{figure}

\begin{figure}[!htbp]
\centering
\includegraphics[width=\textwidth,height=0.72\textheight,keepaspectratio]{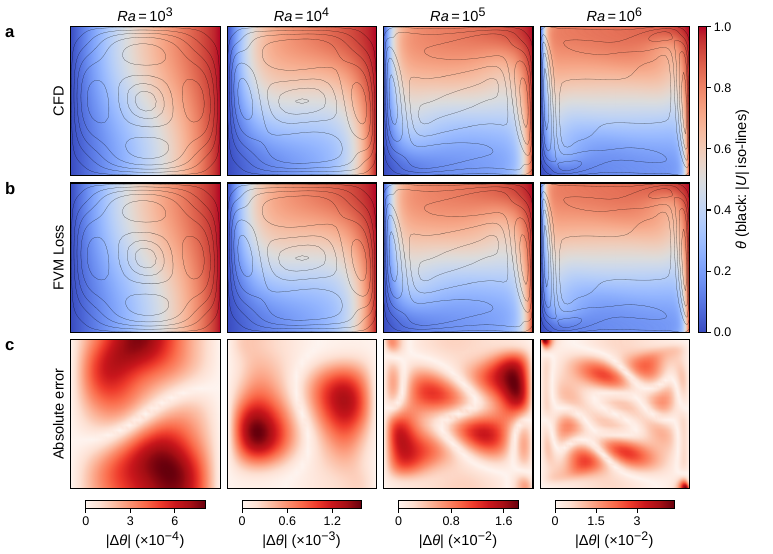}
\caption{Mid-span temperature fields of the three-dimensional natural-convection cavity.
The plane is $z=H/2$, with $z$ the spanwise coordinate and $H$ the cube side,
and the columns correspond to the Rayleigh numbers $\mathrm{Ra}=10^3$--$10^6$.
\textbf{a}, computational fluid dynamics (CFD) solution.
\textbf{b}, solution of the model trained on the finite-volume-method residual loss (FVM Loss).
\textbf{c}, absolute difference $|\Delta\theta|$, with a separate scale for each Rayleigh number.
Here $\theta$ is the dimensionless temperature, 0 on the cold wall and 1 on the hot wall.
Black lines in \textbf{a} and \textbf{b} are isolines of the velocity magnitude $|U|$ at 10, 30, 50, 70 and 90\% of its maximum on the plane.
The cold wall is on the left and the hot wall on the right.}
\label{fig:natural_convection_ra_comparison}
\end{figure}

\begin{figure}[!htbp]
\centering
\includegraphics[width=\textwidth,height=0.72\textheight,keepaspectratio]{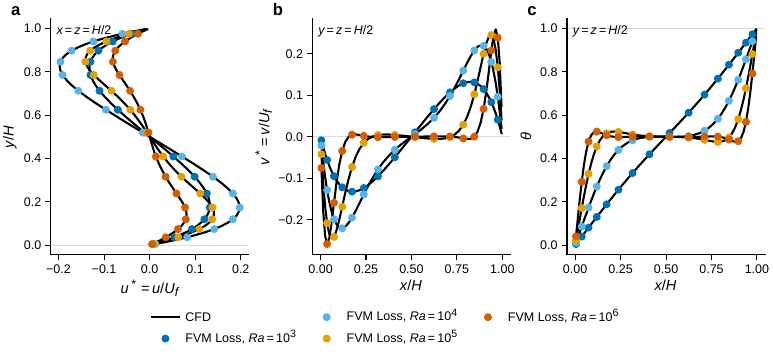}
\caption{Centerline velocity and temperature profiles of the three-dimensional natural-convection cavity.
Profiles are shown for the Rayleigh numbers $\mathrm{Ra}=10^3$--$10^6$, with the coordinates $x$ (hot-wall normal), $y$ (vertical) and $z$ (spanwise) normalized by the cube side $H$.
\textbf{a}, horizontal velocity along the vertical centerline $x=z=H/2$.
\textbf{b}, vertical velocity along the horizontal centerline $y=z=H/2$.
\textbf{c}, temperature along the horizontal centerline $y=z=H/2$.
Black lines are the computational fluid dynamics (CFD) solution and colored markers the solution of the model trained on the finite-volume-method residual loss (FVM Loss), with one color per Rayleigh number as identified in the key below the panels.
Velocity is scaled by the free-fall velocity $U_f$ ($u^{*}=u/U_f$, $v^{*}=v/U_f$), and temperature is shown as the dimensionless $\theta$, which is 0 on the cold wall and 1 on the hot wall.}
\label{fig:nc_centerlines}
\end{figure}

\begin{figure}[!htbp]
\centering
\includegraphics[width=\textwidth,height=0.72\textheight,keepaspectratio]{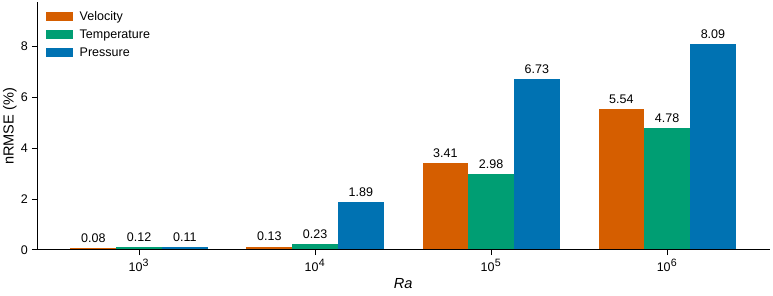}
\caption{Accuracy of the FVM-loss solution across Rayleigh numbers in the natural-convection cavity.
Bars give the velocity (vermillion), temperature (green) and modified pressure (blue) normalized root-mean-square error (nRMSE, in percent) of the solution of the model trained on the finite-volume-method (FVM) residual loss against the same-mesh computational fluid dynamics (CFD) solution, over all interior cells of the $48^3$ natural-convection mesh at each Rayleigh number $\mathrm{Ra}$,
with the value of each bar printed above it.
Velocity combines all three components,
and temperature is measured relative to the mean wall temperature.}
\label{fig:nc_error_heatmap}
\end{figure}

Figures~\ref{fig:nc_theta_planes} and~\ref{fig:nc_velocity_planes} show temperature and velocity on three spanwise planes of the cube.
The spanwise velocity component reaches 5--14\% of the maximum speed, and the flow is therefore three-dimensional,
whereas the temperature on the three planes differs by less than 3\% of the wall temperature difference.
The FVM-loss solution reproduces the structure of the CFD fields on every plane, closely at the lower Rayleigh numbers and with deviations of a few percent at the highest.
The mid-span temperature fields (Figure~\ref{fig:natural_convection_ra_comparison}) show the transition from conduction-dominated to boundary-layer-dominated heat transfer,
and the centerline velocity and temperature profiles (Figure~\ref{fig:nc_centerlines}) closely follow CFD across the sweep.
Across the four Rayleigh numbers, the velocity and temperature nRMSE over all interior cells stay below $0.25\%$ at $\mathrm{Ra}=10^3$ and $10^4$ and rise to about $5\%$ at $\mathrm{Ra}=10^6$ (Figure~\ref{fig:nc_error_heatmap}),
while the modified pressure shows larger relative errors, of up to about $8\%$, at the higher Rayleigh numbers.
Temperature errors are computed on the deviation from the mean wall temperature, $\theta-1/2$, which avoids normalization by an absolute temperature level.
These values measure the agreement between the two discrete solutions on the same mesh, and its decrease with the Rayleigh number is consistent with the thinner boundary layers at higher $\mathrm{Ra}$.

\begin{tableorg}[!htbp]
\centering
\caption{Benchmark quantities of the 3D natural-convection cavity on the $48^3$ mesh.
The peak $u^*_{\max}$ of $|u|/U_f$ is taken in the upper half of the vertical centerline $x=z=H/2$,
and the peak $v^*_{\max}$ of $|v|/U_f$ in the hot-wall half of the horizontal centerline $y=z=H/2$.
Their locations, interpolated quadratically between cell centers, are given in parentheses,
and the nearly antisymmetric profiles have peaks of the same magnitude in the opposite halves.
The Nusselt number $\overline{\mathrm{Nu}}$ is area-averaged over the hot wall.
The reference values are the spectral solution of Tric et al.~\cite{tric2000cubic} for a differentially heated cubic cavity of air at the same Rayleigh numbers and with the same Nusselt definition;
their centerline velocity peaks are rescaled from the thermal-diffusion velocity $\alpha/H$ to $U_f=\sqrt{\mathrm{Ra}\,\mathrm{Pr}}\,\alpha/H$ and mapped to the present coordinates.}
\label{tab:nc_benchmark}
\small\setlength{\tabcolsep}{4pt}
\begin{tabular}{llccc}
\toprule
 & Quantity & Tric et al.~\cite{tric2000cubic} & CFD & FVM Loss \\
\midrule
\multirow{3}{*}{$\mathrm{Ra}=10^{3}$}
  & $u^*_{\max}$ ($y/H$) & $0.133$ ($0.815$) & $0.132$ ($0.817$) & $0.132$ ($0.817$) \\
  & $v^*_{\max}$ ($x/H$) & $0.133$ ($0.815$) & $0.133$ ($0.823$) & $0.133$ ($0.823$) \\
  & $\overline{\mathrm{Nu}}$ & $1.07$ & $1.07$ & $1.07$ \\
\midrule
\multirow{3}{*}{$\mathrm{Ra}=10^{4}$}
  & $u^*_{\max}$ ($y/H$) & $0.198$ ($0.824$) & $0.198$ ($0.826$) & $0.198$ ($0.826$) \\
  & $v^*_{\max}$ ($x/H$) & $0.221$ ($0.880$) & $0.221$ ($0.884$) & $0.221$ ($0.884$) \\
  & $\overline{\mathrm{Nu}}$ & $2.05$ & $2.06$ & $2.06$ \\
\midrule
\multirow{3}{*}{$\mathrm{Ra}=10^{5}$}
  & $u^*_{\max}$ ($y/H$) & $0.141$ ($0.853$) & $0.141$ ($0.853$) & $0.141$ ($0.853$) \\
  & $v^*_{\max}$ ($x/H$) & $0.245$ ($0.933$) & $0.246$ ($0.936$) & $0.246$ ($0.936$) \\
  & $\overline{\mathrm{Nu}}$ & $4.34$ & $4.35$ & $4.35$ \\
\midrule
\multirow{3}{*}{$\mathrm{Ra}=10^{6}$}
  & $u^*_{\max}$ ($y/H$) & $0.0810$ ($0.854$) & $0.0812$ ($0.859$) & $0.0813$ ($0.859$) \\
  & $v^*_{\max}$ ($x/H$) & $0.258$ ($0.967$) & $0.259$ ($0.962$) & $0.259$ ($0.962$) \\
  & $\overline{\mathrm{Nu}}$ & $8.64$ & $8.72$ & $8.72$ \\
\bottomrule
\end{tabular}
\end{tableorg}

Table~\ref{tab:nc_benchmark} compares the mean hot-wall Nusselt number and the centerline velocity peaks of both discrete solutions with the spectral reference of Tric et al.~\cite{tric2000cubic}.
Residual training thus approximates the buoyancy-driven flow and heat transfer of the reference solution on the same mesh, a coupling that also shapes the rack plumes of the data center hall.

\subsection{3D cylinder wake flow with heat transfer}\label{subsec:cylinder_wake}

The heated cylinder wake introduces two challenges absent from the cavity and natural-convection benchmarks.
Vortex shedding makes the flow unsteady,
and the unstructured, body-fitted hexahedral mesh of about $1.47\times10^{5}$ cells has cell volumes spanning more than two orders of magnitude, as is common in practical applications.
The preceding steady benchmarks test the accuracy of a trained discrete solve.
Vortex shedding, fluctuating aerodynamic loads and convective mixing are instead inherently unsteady, and their autoregressive rollout is harder to learn.
The network advances each step from its own previous prediction, and errors can therefore accumulate~\cite{lippe2023pde_refiner}.

\begin{figure}[!htbp]
\centering
\includegraphics[width=0.75\columnwidth]{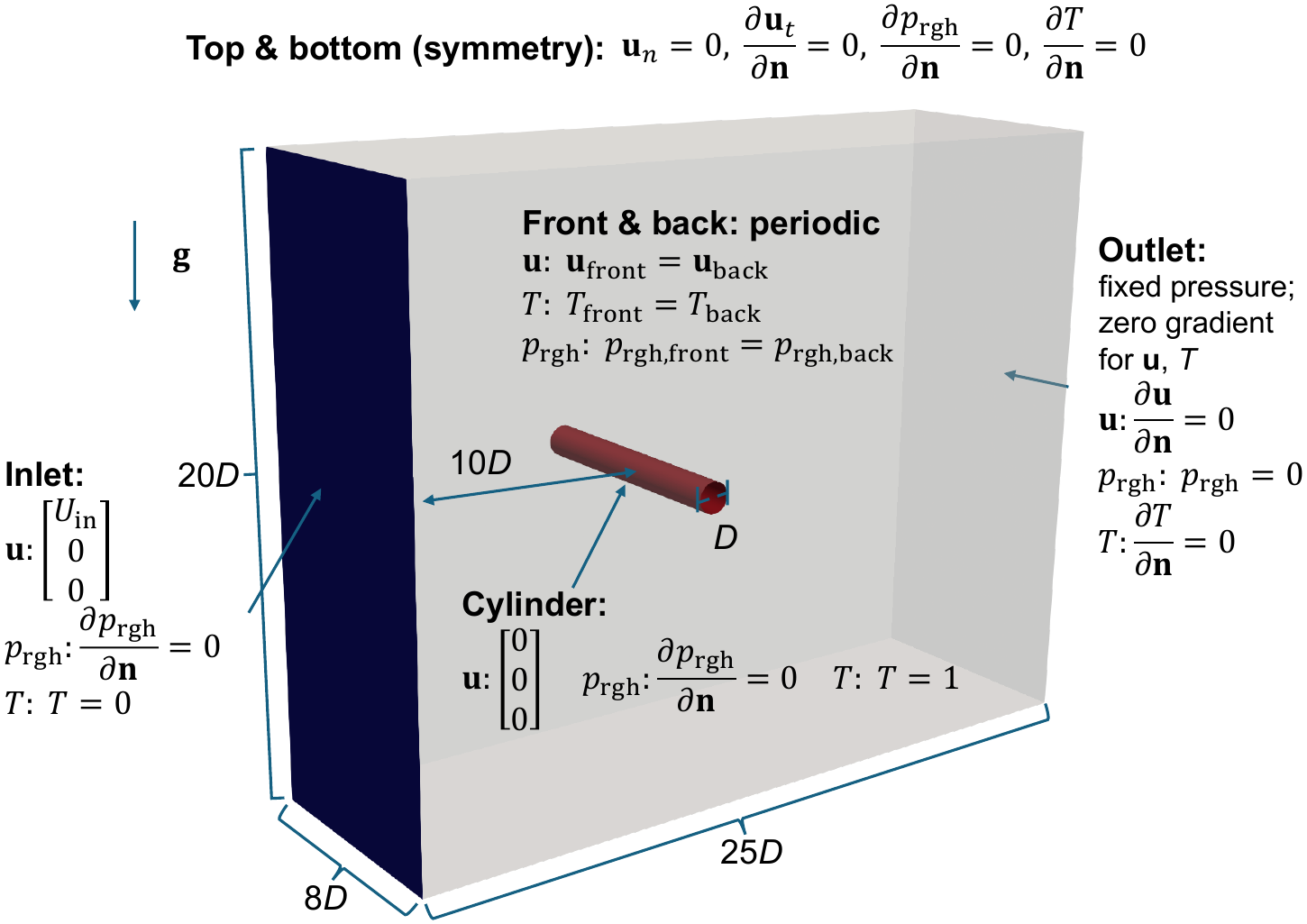}
\caption{Configuration of the three-dimensional heated cylinder wake.
Dimensions are given in units of the cylinder diameter $D=0.1$~m.
The computational domain ($25D\times20D\times8D$) contains the cylinder $10D$ downstream of the inlet, and gravity $\mathbf{g}$ acts in $-y$.
The boundary conditions are a uniform inlet velocity $U_{\mathrm{in}}$ with inlet temperature $T_{\infty}=0$, a no-slip isothermal cylinder wall at $T_{\mathrm{w}}=1$, a zero-gradient outlet at fixed pressure, periodic spanwise boundaries (front and back), and symmetry on the top and bottom boundaries.
In the annotations, $\mathbf{u}$ is the velocity, $\mathbf{u}_{\mathrm{n}}$ and $\mathbf{u}_{\mathrm{t}}$ its boundary-normal and tangential parts, $p_{\mathrm{rgh}}$ the modified pressure, $T$ the dimensionless temperature and $\partial/\partial\mathbf{n}$ the derivative along the boundary normal.}
\label{fig:cylinder_setup}
\end{figure}

As shown in Figure~\ref{fig:cylinder_setup},
a circular cylinder of diameter $D=0.1$~m is placed $10D$ downstream of the inlet, midway between the top and bottom boundaries of a $25D\times20D\times8D$ domain;
a uniform inlet velocity $U_{\mathrm{in}}$ and a dimensionless inlet temperature $T_{\infty}=0$ are prescribed upstream,
with the cylinder wall held at $T_{\mathrm{w}}=1$.
The temperature is solved directly in this dimensionless form, with $\beta=3\times10^{-3}$ per unit temperature and $T_{\mathrm{w}}-T_{\infty}=1$, and the exported fields and thermal diagnostics use the same variable.
The buoyancy acceleration is therefore $g\beta(T_{\mathrm{w}}-T_{\infty})\approx0.029$~m\,s$^{-2}$.
The corresponding Richardson number $\mathrm{Ri}=g\beta(T_{\mathrm{w}}-T_{\infty})D/U_{\mathrm{in}}^2$ ranges from about $9\times10^{-4}$ at $\mathrm{Re}=180$ to $4.6\times10^{-3}$ at $\mathrm{Re}=80$.
The wake is thus in the forced-convection regime, and the Boussinesq coupling (see ``Governing equations'' in Methods) remains active but weak.
Periodic boundary conditions are applied in the spanwise direction ($L_z=8D$),
and symmetry conditions on the top and bottom boundaries.
The Reynolds number $\mathrm{Re}=U_{\mathrm{in}}D/\nu$ spans $[80,180]$, covering the laminar vortex-shedding regime up to the onset of the Mode~A three-dimensional transition~\cite{williamson1996cylinder_wake}.
Following the protocol in Methods (``Experimental setup''),
300 equally spaced Reynolds numbers in this range are split $8\!:\!1\!:\!1$ into training, validation, and test conditions, with the 30 test values spread across the range.
All held-out test conditions lie inside the trained parameter range;
both parametric benchmarks therefore evaluate interpolation within that range, and only the elliptical-cylinder study is a zero-shot transfer to unseen geometries.
Both surrogates share architecture, optimizer, and training conditions and differ primarily in the training signal.
The FVM-loss model minimizes the label-free finite-volume residual (see ``FVM-loss training'' in Methods),
while the data-loss baseline regresses the labeled CFD trajectories;
the figures and tables label the two models FVM Loss and Data Loss.
Both are evaluated by rolling out the same 30 held-out test Reynolds numbers, testing generalization to unseen Reynolds numbers within the trained range.
Each rollout advances 500 steps of $\Delta t=0.01$~s, a 5~s horizon.
Here the velocity error uses the in-plane components $(u,v)$, because the spanwise component carries little reference energy under the spanwise-periodic conditions;
the elliptical-cylinder and data-center results use all three components.
Per-step curves, cumulative-horizon errors and errors of time-averaged fields are distinct statistics and are identified in the captions.

\begin{figure}[!htbp]
\centering
\includegraphics[width=\textwidth,height=0.72\textheight,keepaspectratio]{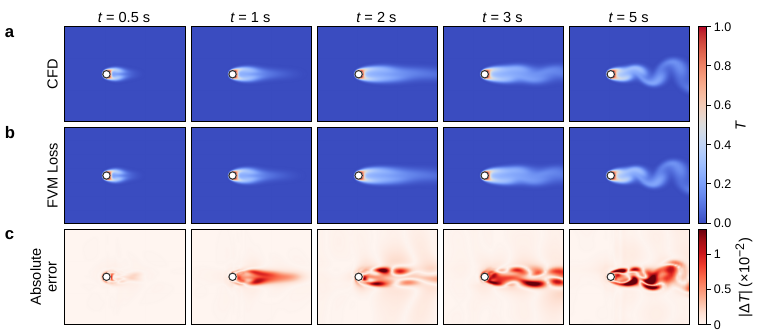}
\caption{Mid-span temperature fields of the heated cylinder wake for the canonical case at Reynolds number $\mathrm{Re}=100$.
Columns correspond to the rollout times $t=0.5$, 1, 2, 3 and 5~s.
\textbf{a}, computational fluid dynamics (CFD) reference.
\textbf{b}, prediction of TransFVGN, the attention graph neural network of this study, trained on the finite-volume-method residual loss (FVM Loss).
\textbf{c}, absolute error $|\Delta T|$.
Temperature $T$ is dimensionless, equal to 0 at the inlet and 1 on the cylinder wall.
The error row has its own scale.
The model frame is the phase-matched snapshot described in the text, and both fields are rendered from the three-dimensional mesh.
Panels show the region $-5.6\le x/D\le10.6$, $|y/D|\le6.5$ of the mid-span plane, where $D$ is the cylinder diameter and $x$ and $y$ are the streamwise and cross-stream coordinates.}
\label{fig:cylinder_temp_re100}
\end{figure}

\begin{figure}[!htbp]
\centering
\includegraphics[width=\textwidth,height=0.72\textheight,keepaspectratio]{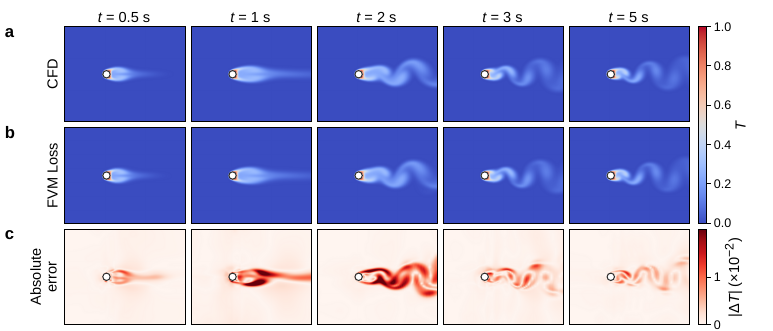}
\caption{Mid-span temperature fields of the heated cylinder wake for the canonical case at Reynolds number $\mathrm{Re}=180$.
Columns correspond to the rollout times $t=0.5$, 1, 2, 3 and 5~s.
\textbf{a}, computational fluid dynamics (CFD) reference.
\textbf{b}, prediction of the FVM-loss TransFVGN.
\textbf{c}, absolute error $|\Delta T|$ of the dimensionless temperature $T$, which is 0 at the inlet and 1 on the cylinder wall.
TransFVGN is the attention graph neural network of this study, and FVM Loss denotes training on the finite-volume-method residual.
The model frame and the displayed region follow Figure~\ref{fig:cylinder_temp_re100}, and the error row has its own scale.}
\label{fig:cylinder_temp_re180}
\end{figure}

\begin{figure}[!htbp]
\centering
\includegraphics[width=\textwidth,height=0.72\textheight,keepaspectratio]{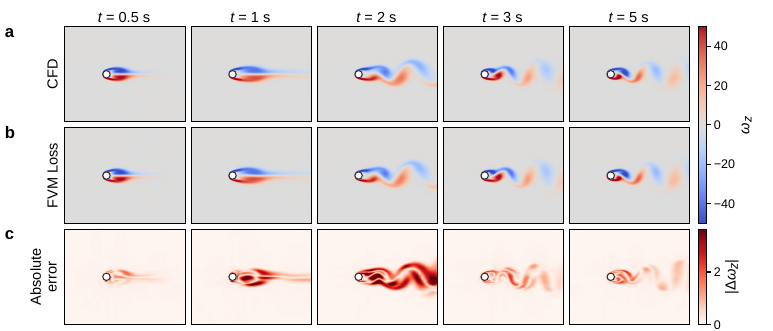}
\caption{Mid-span vorticity fields of the heated cylinder wake at Reynolds number $\mathrm{Re}=180$.
Columns correspond to the rollout times $t=0.5$, 1, 2, 3 and 5~s.
\textbf{a}, computational fluid dynamics (CFD) reference.
\textbf{b}, prediction of the FVM-loss TransFVGN.
\textbf{c}, absolute error $|\Delta\omega_z|$.
Here $\omega_z$ is the spanwise vorticity, given in s$^{-1}$.
TransFVGN is the attention graph neural network of this study, and FVM Loss denotes training on the finite-volume-method residual.
The model frame and the displayed region follow Figure~\ref{fig:cylinder_temp_re100}, and the error row has its own scale.}
\label{fig:cylinder_re180_snapshots}
\end{figure}

\begin{figure}[!htbp]
\centering
\includegraphics[width=\textwidth,height=0.72\textheight,keepaspectratio]{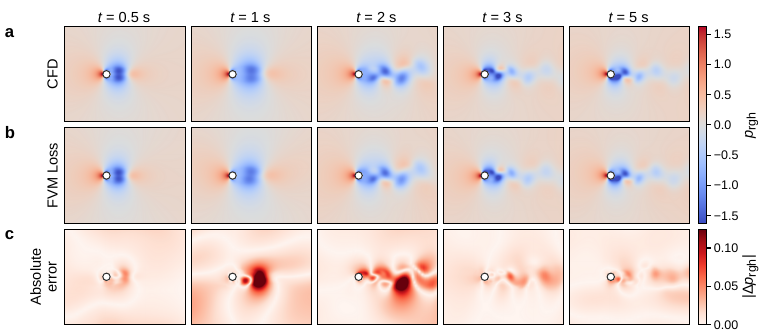}
\caption{Mid-span modified-pressure fields of the heated cylinder wake at Reynolds number $\mathrm{Re}=180$.
Columns correspond to the rollout times $t=0.5$, 1, 2, 3 and 5~s.
\textbf{a}, computational fluid dynamics (CFD) reference.
\textbf{b}, prediction of the FVM-loss TransFVGN.
\textbf{c}, absolute error $|\Delta p_{\mathrm{rgh}}|$.
Here $p_{\mathrm{rgh}}$ is the modified pressure in Pa, displayed after a plane fitted on the outer part of the slice and the remaining mean are removed separately from each field, as described in the text.
TransFVGN is the attention graph neural network of this study, and FVM Loss denotes training on the finite-volume-method residual.
The model frame and the displayed region follow Figure~\ref{fig:cylinder_temp_re100}, and the error row has its own scale.}
\label{fig:cylinder_re180_pressure}
\end{figure}

\begin{figure}[!htbp]
\centering
\includegraphics[width=\textwidth,height=0.72\textheight,keepaspectratio]{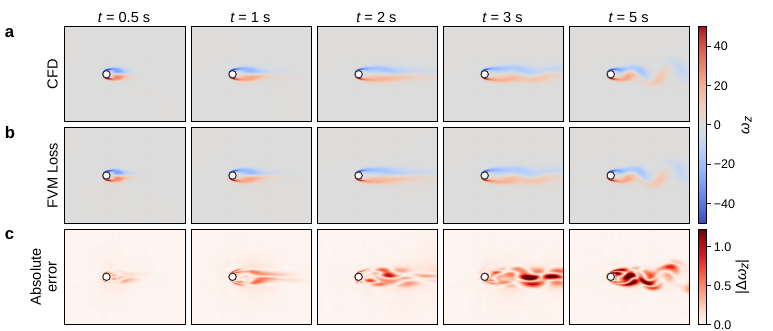}
\caption{Mid-span vorticity fields of the heated cylinder wake for the canonical case at Reynolds number $\mathrm{Re}=100$.
Columns correspond to the rollout times $t=0.5$, 1, 2, 3 and 5~s.
\textbf{a}, computational fluid dynamics (CFD) reference.
\textbf{b}, prediction of the FVM-loss TransFVGN.
\textbf{c}, absolute error $|\Delta\omega_z|$.
Here $\omega_z$ is the spanwise vorticity, given in s$^{-1}$.
TransFVGN is the attention graph neural network of this study, and FVM Loss denotes training on the finite-volume-method residual.
The model frame and the displayed region follow Figure~\ref{fig:cylinder_temp_re100}, and the error row has its own scale.}
\label{fig:cylinder_re100_vorticity}
\end{figure}

\begin{figure}[!htbp]
\centering
\includegraphics[width=\textwidth,height=0.72\textheight,keepaspectratio]{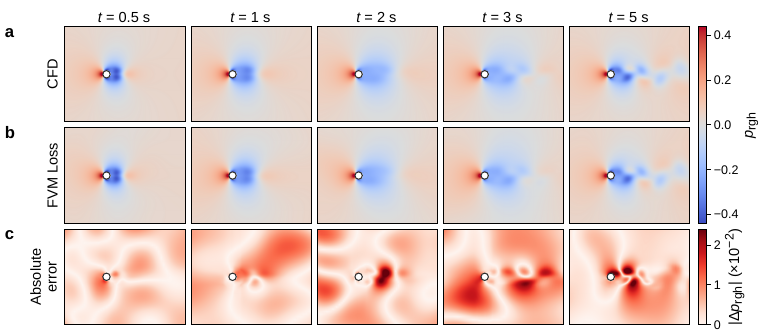}
\caption{Mid-span modified-pressure fields of the heated cylinder wake for the canonical case at Reynolds number $\mathrm{Re}=100$.
Columns correspond to the rollout times $t=0.5$, 1, 2, 3 and 5~s.
\textbf{a}, computational fluid dynamics (CFD) reference.
\textbf{b}, prediction of the FVM-loss TransFVGN.
\textbf{c}, absolute error $|\Delta p_{\mathrm{rgh}}|$.
Here $p_{\mathrm{rgh}}$ is the modified pressure in Pa, displayed after a plane fitted on the outer part of the slice and the remaining mean are removed separately from each field, as described in the text.
TransFVGN is the attention graph neural network of this study, and FVM Loss denotes training on the finite-volume-method residual.
The model frame and the displayed region follow Figure~\ref{fig:cylinder_temp_re100}, and the error row has its own scale.}
\label{fig:cylinder_re100_pressure}
\end{figure}

The two canonical cases at $\mathrm{Re}=100$ and 180 are examined in detail below.
Their displayed CFD references were computed with the $10^{-6}$ residual-control target of the PIMPLE algorithm, which merges the pressure-implicit with splitting of operators (PISO) algorithm~\cite{issa1986piso} with SIMPLE;
the same settings back the label dataset, the elliptical-cylinder references and the cost benchmark.
Figures~\ref{fig:cylinder_temp_re100}--\ref{fig:cylinder_re100_pressure} show the temperature, vorticity and pressure fields at $t=0.5,1,2,3,5$~s.
The model and reference rollouts shed at slightly different frequencies.
For display, the model frame at each figure time is accordingly the snapshot whose spanwise vorticity matches the reference most closely, searched over one shedding period either side of that time, one-sided at the last time, where the rollout ends;
the comparison therefore isolates the wake itself from the shedding-phase offset.
The FVM-loss model reproduces the thermal wake and the vortex-shedding structures.
Its largest remaining differences arise while shedding develops, near $t=2$~s at $\mathrm{Re}=180$ and toward $t=5$~s at $\mathrm{Re}=100$,
and they reflect the wake amplitude and the shape of the separated shear layers rather than a phase offset.
For visualization only, a plane fitted on the outer part of each pressure slice and its remaining mean are removed separately from the model and reference;
the displayed pressure error therefore excludes those large-scale components.
Full-field pressure statistics are computed from the stored rollout records rather than from the display fields.

Shedding frequency and phase are evaluated separately from field error, using the lift coefficient and the cross-stream velocity at a monitoring point $5D$ downstream of the cylinder over $t=3$--$5$~s,
with the Strouhal number $\mathrm{St}=f_{\mathrm{s}}D/U_{\mathrm{in}}$ based on the shedding frequency $f_{\mathrm{s}}$.
The spectral Strouhal numbers of the lift signal are 0.142 (CFD), 0.138 (FVM-loss model) and 0.124 (data-loss model) at $\mathrm{Re}=100$, and 0.182, 0.183 and 0.174 at $\mathrm{Re}=180$.
They are read from the peak of the Hann-windowed lift spectrum over $t=3$--$5$~s, whose 2~s length limits the spectral resolution to $\Delta\mathrm{St}\approx0.05$ at $\mathrm{Re}=100$ and $0.028$ at $\mathrm{Re}=180$.
At this resolution both models lie close to the CFD shedding frequency, and the FVM-loss peak is the closer one at both Reynolds numbers;
the phase statistics below separate the two models more clearly.
The lift amplitude of the FVM-loss model is also closer to CFD ($-19.0\%$ and $-10.1\%$, against $-36.5\%$ and $-18.5\%$ for the data-loss model),
and its circular-mean phase lag behind the CFD lift and wake-velocity signals stays within about $12^{\circ}$ at both Reynolds numbers, whereas the data-loss wake drifts by $69$--$166^{\circ}$.
At $\mathrm{Re}=100$ the shedding is still developing over this window (the CFD lift amplitude grows from 0.03 to 0.14), and these statistics therefore characterize the developing wake rather than saturated shedding.

\begin{figure}[!htbp]
\centering
\includegraphics[width=\textwidth,height=0.72\textheight,keepaspectratio]{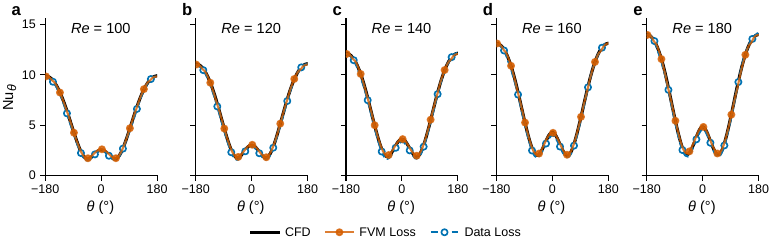}
\caption{Time-averaged circumferential Nusselt-number profiles on the heated cylinder over $t=3$--$5$~s.
\textbf{a}, Reynolds number $\mathrm{Re}=100$.
\textbf{b}, held-out test condition near $\mathrm{Re}=120$.
\textbf{c}, held-out test condition near $\mathrm{Re}=140$.
\textbf{d}, held-out test condition near $\mathrm{Re}=160$.
\textbf{e}, $\mathrm{Re}=180$.
Here $\mathrm{Nu}_\theta=-D\,(\partial T/\partial n)_{\mathrm{w}}/(T_{\mathrm{w}}-T_{\infty})$ is the local Nusselt number, with $D$ the cylinder diameter, $T$ the dimensionless temperature, $T_{\mathrm{w}}$ and $T_{\infty}$ the wall and inlet temperatures, and $n$ the wall normal pointing into the fluid;
$\theta$ is the angle measured from the rear stagnation point.
Black lines show the computational fluid dynamics (CFD) reference, vermillion lines with filled markers the model trained on the finite-volume-method residual loss (FVM Loss), and blue dashed lines with open markers the model trained on CFD labels (Data Loss).}
\label{fig:cylinder_nusselt_re}
\end{figure}

The circumferential Nusselt profiles add a wall-heat-flux test (Figure~\ref{fig:cylinder_nusselt_re}).
Averaged over $t=3$--$5$~s and around the cylinder, the FVM-loss Nusselt number agrees with CFD to within $0.36\%$ at all five Reynolds numbers,
including $+0.35\%$ and $+0.12\%$ at the canonical $\mathrm{Re}=100$ and 180 (CFD means 5.18 and 7.26),
whereas the data-loss model underpredicts it by $1.86$--$2.76\%$.
The three interior panels show held-out test conditions near $\mathrm{Re}=120$, 140 and 160, compared against their label-dataset references.

\begin{figure}[!htbp]
\centering
\includegraphics[width=\textwidth,height=0.72\textheight,keepaspectratio]{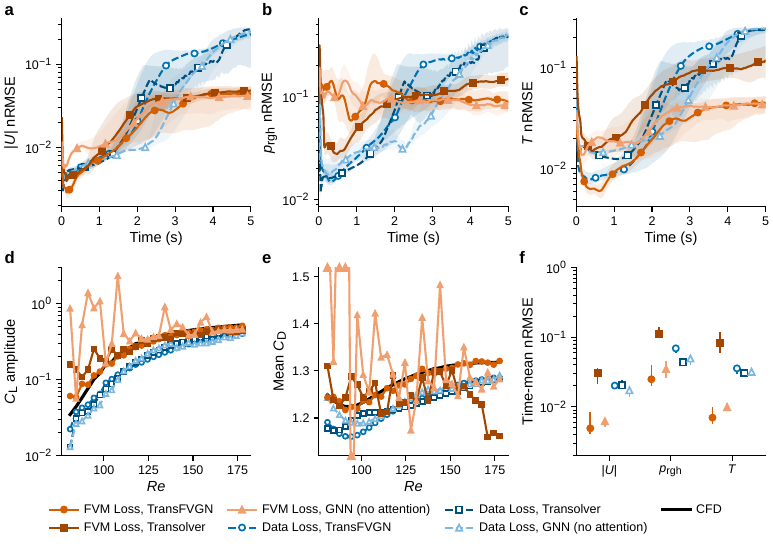}
\caption{Comparison of six cylinder-wake models on 30 held-out Reynolds numbers.
TransFVGN (the attention graph neural network of this study), Transolver and an attention-free graph neural network (GNN) are each trained on the finite-volume-method residual loss (FVM Loss) or on computational fluid dynamics (CFD) labels (Data Loss), as identified in the key.
\textbf{a}--\textbf{c}, median per-step normalized root-mean-square error (nRMSE) of the in-plane velocity $|U|$ (\textbf{a}), modified pressure $p_{\mathrm{rgh}}$ (\textbf{b}) and temperature $T$ (\textbf{c}), with interquartile bands.
\textbf{d}, lift-coefficient ($C_{\mathrm{L}}$) amplitude against the Reynolds number $\mathrm{Re}$.
\textbf{e}, mean drag coefficient $C_{\mathrm{D}}$ (off-scale values on the panel edges).
\textbf{f}, median and interquartile range of the time-mean field nRMSE.}
\label{fig:cylinder_testset_arms}
\end{figure}

Figure~\ref{fig:cylinder_testset_arms} compares TransFVGN, Transolver and an attention-free GNN, each trained with the FVM loss and with the data loss, on the 30 held-out Reynolds numbers.
Compared at coincident times, the per-step errors of the FVM-loss TransFVGN grow more slowly than those of its data-loss counterpart.
Its velocity and temperature errors are lower over most of the rollout,
and its pressure error, higher during the first 2.2~s, is lower from then on.
The early pressure error is concentrated in the first steps from the uniform initial field, where the median per-step pressure error is largest;
from $t=0.5$~s onward it stays between 0.05 and 0.16, whereas the data-loss pressure error grows to 0.39 by the final step.
Over the full 5~s horizon, the mean cumulative velocity and temperature errors of the FVM-loss model are less than half those of the data-loss model,
and its pressure error is comparable (Table~\ref{tab:per_channel_accuracy}).
At the final step the case medians of the per-step error are 0.043 versus 0.228 in velocity, 0.089 versus 0.387 in pressure and 0.045 versus 0.242 in temperature.
The mean field-averaged score is 0.047 for the FVM-loss model against 0.062 for the data-loss model (see ``Computational cost and amortization'').
Across the three architectures, the FVM loss mainly suppresses the late-time error growth.
At the final step every FVM-loss model has a lower median error than every data-loss model in all three fields.
Over the full horizon, the FVM-loss TransFVGN has the lowest velocity and temperature errors of the six models,
and its field-averaged score is the second lowest of the six, close to that of the attention-free data-loss GNN;
its pressure error is the highest of the six, although the six models differ less in pressure than in velocity and temperature.

The per-step errors of the two FVM-loss message-passing models are close (Figure~\ref{fig:cylinder_testset_arms}a--c),
with the contribution of attention appearing most clearly in the force statistics and in the early part of the rollout.
The FVM-loss TransFVGN has the lowest median time-mean errors of the six models in velocity, pressure and temperature, below those of the attention-free FVM-loss GNN in every field.
The attention-free FVM-loss GNN also shows large lift-amplitude and mean-drag outliers at several lower Reynolds numbers (Figure~\ref{fig:cylinder_testset_arms}d,e), whereas the FVM-loss TransFVGN stays within 0.036 of the CFD lift amplitude and within about 1\% of its mean drag at every held-out test Reynolds number over $t=3.5$--$5$~s.
All three data-loss models underpredict both the lift amplitude and the mean drag at every test Reynolds number, on average by 34--40\% and 3.2--4.1\%, respectively.
Transolver, which has no mesh message passing, behaves differently in this statistic and reaches lower time-mean errors with the data loss than with the FVM loss,
although its FVM-loss version ends the rollout with lower per-step errors and transfers better to unseen geometries (see ``Transfer to unseen cross-section geometries'').
These results suggest that the residual objective is more effective in combination with local message passing.

\subsection{Transfer to unseen cross-section geometries}\label{subsec:geometry_extrapolation}

\begin{figure}[!htbp]
\centering
\includegraphics[width=\textwidth,height=0.72\textheight,keepaspectratio]{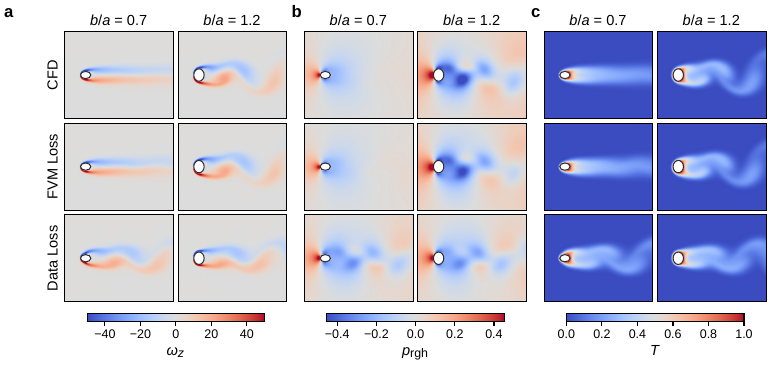}
\caption{Fields on two unseen elliptical cylinders at $t=5$~s and Reynolds number $\mathrm{Re}=100$.
\textbf{a}, spanwise vorticity $\omega_z$ (s$^{-1}$).
\textbf{b}, display-detrended modified pressure $p_{\mathrm{rgh}}$ (Pa).
\textbf{c}, dimensionless temperature $T$.
In each block the columns are the aspect ratios $b/a=0.7$ and 1.2, with $a$ and $b$ the streamwise and cross-stream semi-axes,
and the rows are the computational fluid dynamics (CFD) reference and TransFVGN trained on the finite-volume-method residual loss (FVM Loss) or on CFD labels (Data Loss).
TransFVGN, the attention graph neural network of this study, was trained on the circular cylinder only.}
\label{fig:ellipse_fields}
\end{figure}

\begin{figure}[!htbp]
\centering
\includegraphics[width=\textwidth,height=0.72\textheight,keepaspectratio]{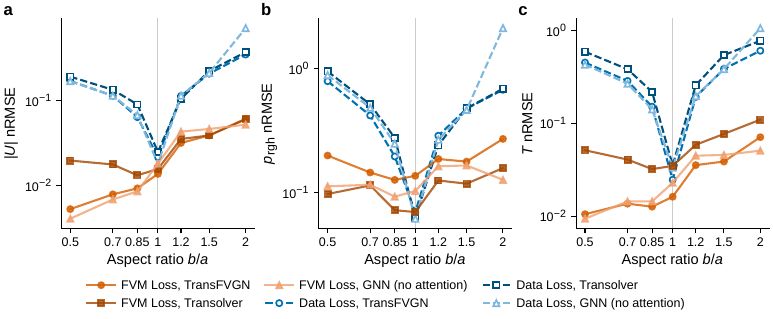}
\caption{Geometry transfer of the six models trained on the circular cylinder.
\textbf{a}, full-horizon normalized root-mean-square error (nRMSE) of the velocity vector $|\mathbf{U}|$ against the aspect ratio $b/a$ of the elliptical cross-section, with $a$ and $b$ the streamwise and cross-stream semi-axes.
\textbf{b}, the same for the modified pressure $p_{\mathrm{rgh}}$.
\textbf{c}, the same for the temperature $T$.
The vertical line marks the training geometry $b/a=1$.
Vermillion solid lines denote the finite-volume-method residual loss (FVM Loss) and blue dashed lines supervised regression on computational fluid dynamics labels (Data Loss);
circles, squares and triangles denote TransFVGN (the attention graph neural network of this study), Transolver (darker shade) and the attention-free graph neural network (GNN, lighter shade).}
\label{fig:ellipse_generalization}
\end{figure}

All six cylinder-trained models are applied without retraining to heated elliptical cylinders at $\mathrm{Re}=100$, based on the streamwise axis $2a$;
expressed with the cross-stream width $2b$ instead, the effective Reynolds number is $100\,b/a$, that is, $50$--$200$ across this family.
The streamwise semi-axis $a$ stays fixed and the cross-stream semi-axis $b$ varies, giving the slender shapes $b/a=0.5,0.7,0.85$ and the bluff shapes $b/a=1.2,1.5,2.0$, with the circle $b/a=1$ re-evaluated for context;
mesh operators are rebuilt for each geometry (Figure~\ref{fig:ellipse_fields} shows two of the shapes).
Figure~\ref{fig:ellipse_generalization} reports the full-horizon cumulative geometric-mean nRMSE of each field separately (Eq.~\eqref{eq:nrmse} with $N=500$).
The FVM-loss TransFVGN spans 0.0053--0.061 in velocity, 0.127--0.272 in pressure and 0.0105--0.0709 in temperature on the six unseen geometries.
Its data-loss counterpart spans 0.064--0.350, 0.197--0.794 and 0.151--0.605, respectively.
For this matched pair, the ratios of data-loss to FVM-loss error range from 3.6 to 32.1 in velocity, from 1.54 to 3.97 in pressure and from 5.4 to 43.0 in temperature, with the largest velocity and temperature ratios at $b/a=0.5$.

All three FVM-loss models have smaller errors than their own data-loss counterparts in every field at every unseen ratio.
Among them, the attention-free GNN gives the lowest velocity error at four of the six unseen ratios and TransFVGN the lowest temperature error at four, whereas Transolver gives the lowest pressure error at five.
From the training circle to the unseen shapes, the errors of the three data-loss models grow by factors of 2.8 to 40, whereas those of the three FVM-loss models change by factors of 0.23 to 4.4.
For this cross-section family, the residual objective thus improves transfer consistently across all three architectures.
At the training geometry itself, the data-loss TransFVGN and attention-free GNN reach lower pressure errors than their FVM-loss counterparts, an advantage that does not carry over to any unseen ratio.
This circle point is one rollout at $\mathrm{Re}=100$, close to a training Reynolds number, and is therefore a different statistic from the mean over the 30-condition cylinder test set.

Figure~\ref{fig:ellipse_fields} shows the corresponding fields at $t=5$~s for $b/a=0.7$ and 1.2, with the model frame phase-matched to the reference as for the circular cylinder.
The FVM-loss predictions reproduce the wake of both shapes closely, including the still-symmetric wake behind $b/a=0.7$ at $t=5$~s.
The data-loss model, in contrast, already sheds vortices behind that body and shifts the $b/a=1.2$ wake.

\subsection{Data center hall}\label{subsec:multi_rack}

\begin{figure}[!htbp]
\centering
\includegraphics[width=\columnwidth]{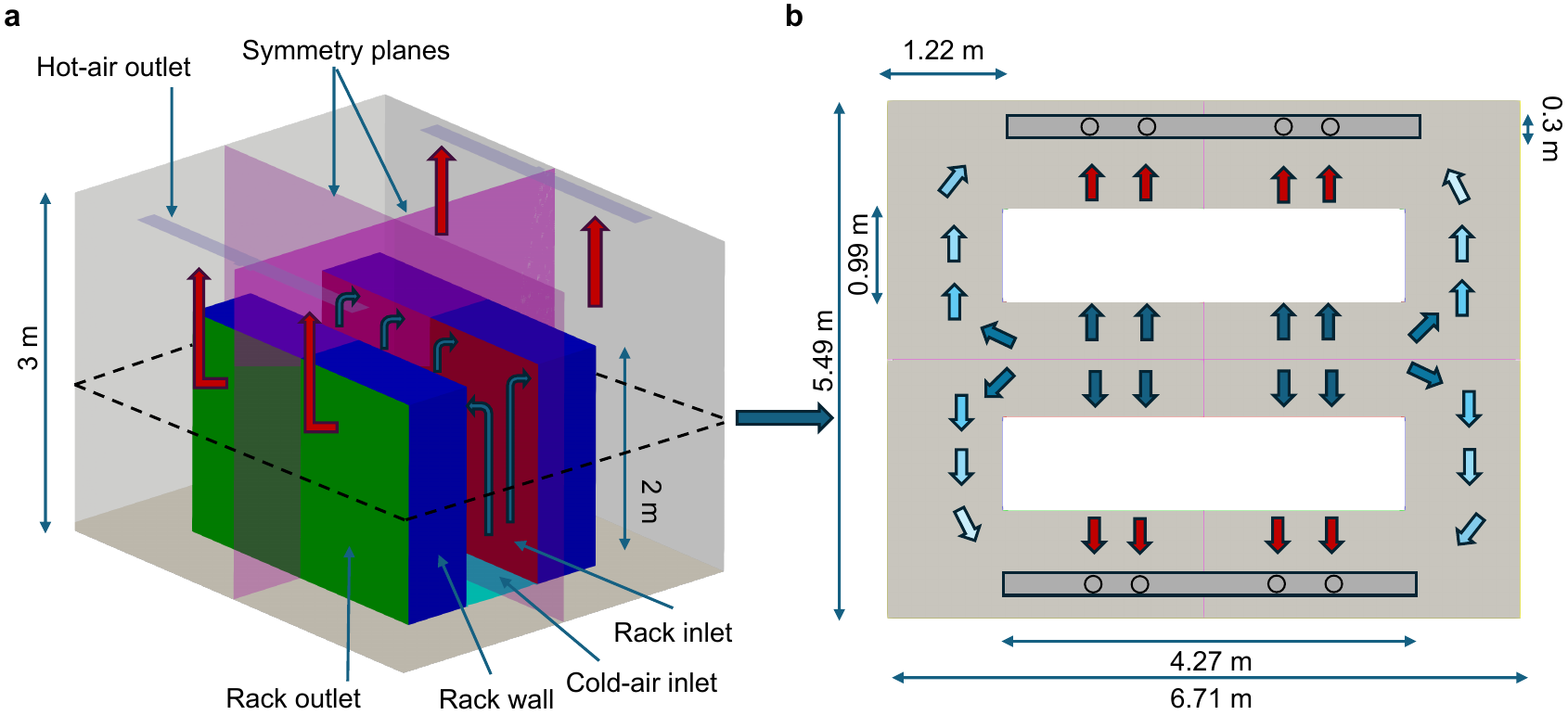}
\caption{Layout of the three-dimensional (3D) data center hall.
\textbf{a}, 3D view with the server racks, the cold-air inlet, the hot-air outlets and the two symmetry planes (translucent magenta) that reduce the computational domain to a quarter of the hall.
Rack inlet faces are red, rack outlet faces green, rack walls dark blue and the cold-air inlet cyan;
red and dark blue arrows indicate the warm exhaust and the cold supply.
\textbf{b}, horizontal section at the dashed outline in \textbf{a}, with the principal dimensions and the air paths (dark blue, cold supply; red, rack exhaust; light blue, hall circulation).}
\label{fig:dc_3d_layout}
\end{figure}

The data center hall case measures $6.71\,\text{m}\times5.49\,\text{m}\times3\,\text{m}$ and houses two server racks arranged in a hot/cold aisle configuration.
The dataset sweep varies the total rack heat load over the symmetry-reduced rack block from 5 to 25~kW,
and the volumetric flow rate through that block from $0.60$ to $1.80\,\text{m}^3\,\text{s}^{-1}$.
The conditioned cold-air supply is held fixed.
The rack-outlet temperature rise is imposed as $\Delta T_{\mathrm{rack}}=Q/(\rho_0 c_p q_v)$,
giving $\Delta T_{\mathrm{rack}}=2.38$--$35.7$~K over the sweep.
The 3D layout, a simplified version of the configuration of Bai et al.~\cite{bai2024improved}, is illustrated in Figure~\ref{fig:dc_3d_layout}.
The resulting multi-scale thermo-fluid problem on a mesh of about $1.4\times10^{5}$ cells stresses accuracy, efficiency, generalization, and label-free training at once.

\begin{figure}[!htbp]
\centering
\includegraphics[width=\textwidth,height=0.8\textheight,keepaspectratio]{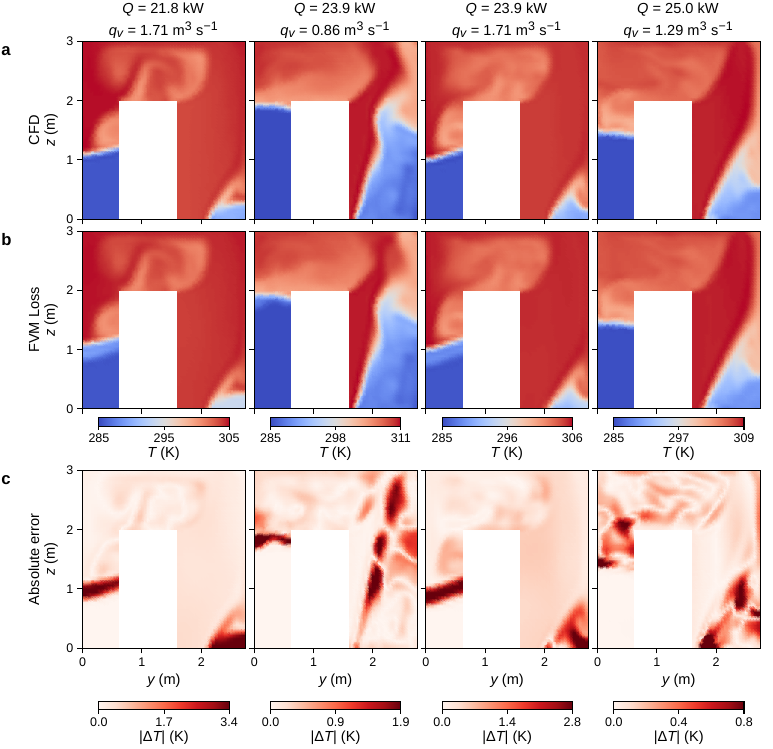}
\caption{Temperature fields of the data center hall after 500 autoregressive steps.
The fields are shown on the $x$-normal mid-plane of the computational domain at $t=50$~s for four held-out conditions (columns), labeled by their rack heat load $Q$ and rack volumetric flow rate $q_v$,
with $y$ and $z$ the horizontal and vertical coordinates.
\textbf{a}, computational fluid dynamics (CFD) reference.
\textbf{b}, prediction of TransFVGN, the attention graph neural network of this study, trained on the finite-volume-method residual loss (FVM Loss).
\textbf{c}, absolute temperature error $|\Delta T|$ in kelvin.
Rows a and b share the temperature ($T$) scale of each column, and the white rectangle is the rack block.}
\label{fig:dc_fvm_symmetry_results}
\end{figure}

The FVM-loss model and a data-loss baseline trained on the labeled trajectories of the same 240 training conditions are both TransFVGN models, and they receive the same 2400-epoch training budget (see ``Experimental setup'' in Methods).
The 300 conditions form a $20\times15$ grid in $(Q,q_v)$, split $8\!:\!1\!:\!1$ into training, validation and test conditions, and both models are evaluated on the 30 held-out conditions.
Four of these, $(Q,q_v)=(21.8~\mathrm{kW},\,1.71~\mathrm{m^3\,s^{-1}})$, $(23.9,\,0.86)$, $(23.9,\,1.71)$ and $(25.0,\,1.29)$, lie at the high-load end of the sweep and are visualized in Figure~\ref{fig:dc_fvm_symmetry_results}.

The slice comparison of Figure~\ref{fig:dc_fvm_symmetry_results} shows the cold-supply region, warm return flow and rack-adjacent plume at $t=50$~s.
After 500 autoregressive steps, the full-volume temperature root-mean-square errors (RMSE) of the four conditions are 0.801, 0.563, 0.725 and 0.257~K,
and the mean absolute errors (MAE) are 0.433, 0.310, 0.463 and 0.140~K, over a temperature range of 285--311~K.

\begin{figure}[!htbp]
\centering
\includegraphics[width=\textwidth,height=0.72\textheight,keepaspectratio]{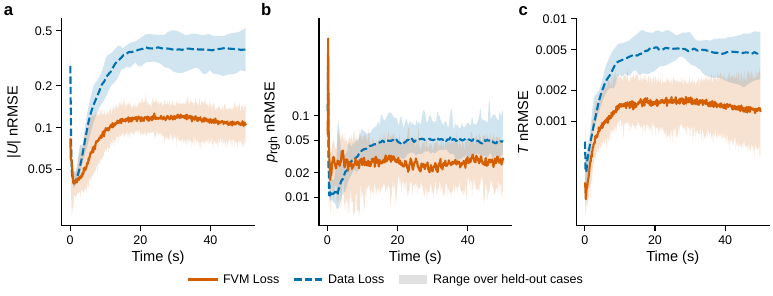}
\caption{Rollout errors of the two data-center models on 30 held-out conditions.
Lines are case medians of the per-step normalized root-mean-square error (nRMSE) and bands the full case range, for TransFVGN (the attention graph neural network of this study) trained on the finite-volume-method residual loss (FVM Loss, vermillion solid) or on computational fluid dynamics labels (Data Loss, blue dashed).
\textbf{a}, velocity vector $|\mathbf{U}|$.
\textbf{b}, modified pressure $p_{\mathrm{rgh}}$.
\textbf{c}, temperature $T$, normalized by the absolute reference field.}
\label{fig:dc_rollout_latest}
\end{figure}

Averaged over the 30 held-out conditions, the full-horizon errors of the FVM-loss model are lower than those of the data-loss model in all three fields (Table~\ref{tab:fieldwise_nrmse}),
with a velocity error lower by a factor of about 2.7 and field-averaged scores of 0.042 and 0.106.
The gap widens toward the end of the rollout;
in kelvin, the final-step median mean absolute temperature error is 0.23~K for the FVM-loss model against 0.8~K for the data-loss model.
The FVM-loss median errors are lower at every step in temperature, from about $t=2$~s onward in velocity and over most of the horizon in pressure (Figure~\ref{fig:dc_rollout_latest}).

\begin{figure}[!htbp]
\centering
\includegraphics[width=\textwidth,height=0.72\textheight,keepaspectratio]{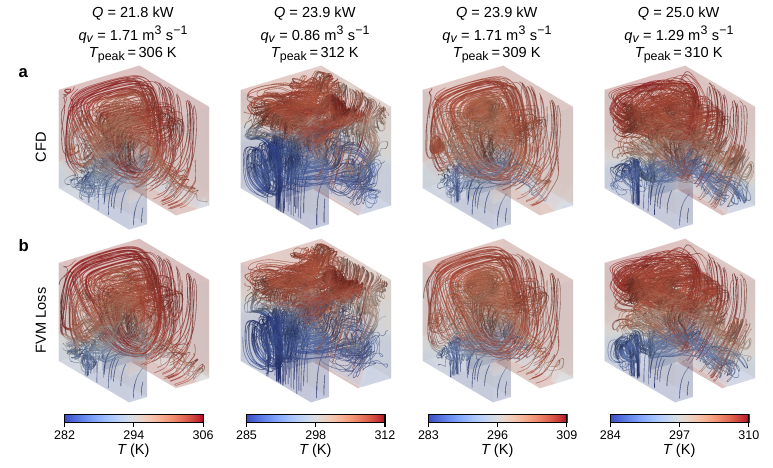}
\caption{Three-dimensional streamlines in the data center hall at $t=50$~s, after 500 autoregressive steps.
Columns are the four held-out conditions of Figure~\ref{fig:dc_fvm_symmetry_results}, labeled by the rack heat load $Q$ and the rack volumetric flow rate $q_v$, and both rows start from the same seed points in the quarter domain.
\textbf{a}, computational fluid dynamics (CFD) reference.
\textbf{b}, TransFVGN, the attention graph neural network of this study, trained on the finite-volume-method residual loss (FVM Loss).
Color denotes temperature $T$ on a common range per column taken from the CFD field, with $T_{\mathrm{peak}}$ its upper limit.}
\label{fig:datacenter_streamline_grid}
\end{figure}

The paired three-dimensional streamlines in Figure~\ref{fig:datacenter_streamline_grid} provide a complementary spatial view of the supply jet and warm recirculation.
The FVM-loss streamlines reproduce the supply jet, the recirculation around the racks and the warm return path at all four operating conditions.
One 500-step FVM-loss rollout takes about 158~s on one RTX 4090, against 954~s for a four-core CFD run of the $(25.0~\mathrm{kW},\,1.29~\mathrm{m^3\,s^{-1}})$ condition (see ``Computational cost and amortization'').
The data center hall is thus consistent with the cylinder findings in an application-oriented thermal problem, with lower full-horizon errors than the supervised baseline in all three fields and a 500-step rollout of under three minutes on one GPU.

\subsection{Pressure--velocity decoupling under residual optimization}\label{subsec:null_modes}

The accuracy results above are obtained with a design choice of the objective, namely that the momentum residual is evaluated on both cell and face control volumes.
On a collocated mesh, where velocity and pressure share the cell centers, a residual that is minimized rather than solved exactly can leave some patterns of the discrete fields weakly constrained.
This section examines, on a controlled test, which patterns each of the two residuals responds to and why the two are therefore used together.
On point-symmetric collocated stencils, the reconstructed cell pressure gradient is insensitive to grid-parity pressure fields (cell-to-cell alternating patterns such as the classical checkerboard), while averaging neighboring momentum residuals can attenuate alternating velocity contributions.
The combined cell and face objective of Eq.~\eqref{eq:l2_norms} is motivated by this complementarity.

\begin{figure}[!htbp]
\centering
\includegraphics[width=\textwidth,height=0.72\textheight,keepaspectratio]{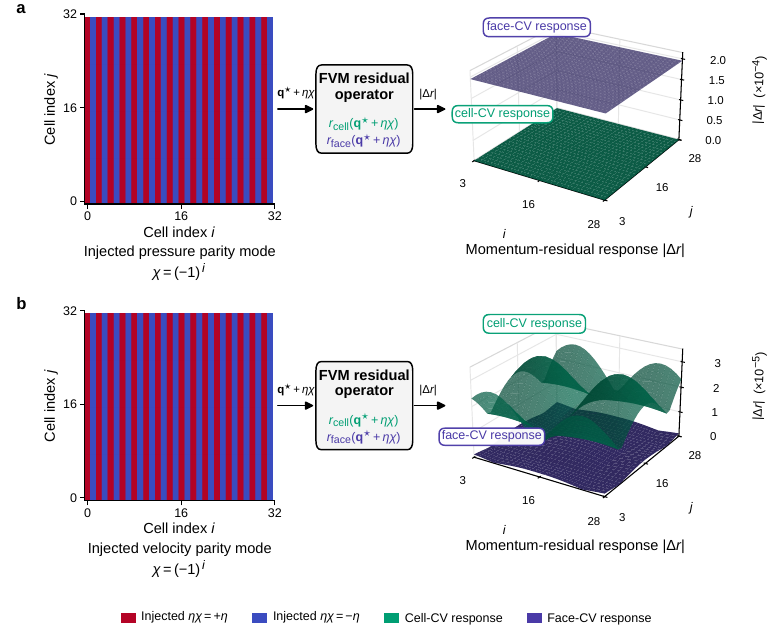}
\caption{Single-axis parity tests on a manufactured state on a uniform $32^3$ mesh.
\textbf{a}, pressure injection.
\textbf{b}, velocity injection.
Each row shows the parity pattern $\chi=(-1)^i$ on the mid-plane $k=16$ (red, injected $\eta\chi=+\eta$; blue, injected $\eta\chi=-\eta$), the finite-volume-method (FVM) residual operator, and the magnitude $|\Delta r|$ of the momentum-residual response on the deep interior of that plane (green, cell-control-volume (cell-CV) response; violet, face-control-volume (face-CV) response).
Here $i$, $j$ and $k$ are the cell indices, $\mathbf{q}^{\star}$ the manufactured state, $\eta$ the injection amplitude, and $r_{\mathrm{cell}}$ and $r_{\mathrm{face}}$ the cell and face momentum residuals.
Face responses are mapped to cells for the surface display.}
\label{fig:null_mode_visibility}
\end{figure}

The parity tests of Figure~\ref{fig:null_mode_visibility} add a single-axis parity perturbation $\chi=(-1)^i$, separately to the pressure and to the velocity, of a manufactured state on a uniform $32^3$ hexahedral mesh.
Each row shows the injected field, the two residuals it enters and the response of both momentum residuals over the deep interior, which excludes the three cell layers next to every boundary.
The pressure injection leaves the cell-control-volume (cell-CV) response at the level of numerical roundoff, whereas the face-control-volume (face-CV) response forms a flat plateau of approximately $2\times10^{-4}$, constant to $2\times10^{-5}$ relative over the slice.
The reconstructed cell gradient is therefore blind to the stripe, while the face-CV residual detects it.
For the velocity injection the face-CV response is attenuated but nonzero, reaching $0.62$ of its response to an equal-amplitude smooth control (ratio of geometric means over the deep interior), whereas the cell-CV response is not attenuated and stays well above it.
The two residuals thus respond to complementary perturbations, which the combined objective exploits.
In short, the face residual detects the grid-parity pressure stripe that the cell residual misses, and the cell residual detects the velocity stripe that the face residual weakens.

All reported FVM-loss models use the combined cell and face objective.
For a smooth pressure field, the compact face-normal pressure gradient and the face average of the reconstructed cell gradients differ only by a truncation error (see ``Designed to suppress pressure--velocity decoupling'' in Methods).
At a state that satisfies the cell balances, the face residuals therefore add only a truncation-level term.

\subsection{Computational cost and amortization}\label{subsec:performance}

Accuracy alone does not establish the practical value of a surrogate, which also depends on the cost of building it and of querying it afterwards.
We therefore compare the two training objectives, and CFD, on the same hardware.
Tables~\ref{tab:upfront_cost} and~\ref{tab:inference_cost} report the upfront training cost, the inference time and the memory of every model.
Costs were benchmarked on one machine with eight RTX 4090 GPUs and two Xeon Silver 4410Y CPUs;
neural inference uses one GPU, while CFD uses 1, 2, 4 or 8 CPU cores.
Training times are wall-clock hours of the data-parallel run on the eight GPUs, inference times are wall-clock times on one RTX 4090, and CFD times are wall-clock times on the stated number of cores.
Apart from label generation, all of these times are measured on the same machine.
The training hours are budget estimates from measured steady per-epoch throughput, for 2400 epochs, a 2400-slot pool and global batch size eight.
For TransFVGN, the cylinder upfront estimates are 232~h (FVM-loss model) and 254~h (data-loss model, including 52.2~h for labels);
the data-center estimates are 205 and 228~h (including 45.2~h for labels).
Both data-center models use the TransFVGN backbone, and each estimate therefore prices the model whose accuracy is reported.
Each label-generation cost covers the 240 training trajectories of a 300-case $8\!:\!1\!:\!1$ split,
recorded as the total wall clock of a sweep on a shared computing cluster under varying load.
The converted label datasets occupy 520~GB and 476~GB, and the data pools approximately 8.3~GB and 7.6~GB, respectively.
Peak training memory stays between 10.7 and 20.0~GiB per GPU, within the 24~GB of one RTX 4090.
The message-passing models reach this by recomputing each processor layer in the backward pass (activation checkpointing);
without it, the FVM-loss TransFVGN does not fit in 24~GB, whereas Transolver fits without it.
\begin{tableorg}[!htbp]
\centering
\caption{Full-horizon field-wise nRMSE of the six models on the 30 held-out cylinder conditions, given as the cumulative geometric mean in time followed by the arithmetic mean over the cases.
The velocity entry is the relative error of the in-plane velocity vector $(u,v)$.}
\label{tab:per_channel_accuracy}
\small\setlength{\tabcolsep}{6pt}
\begin{tabular}{@{}llccc@{}}
\toprule
Architecture & Loss & $|\mathbf{U}|$ & $p_{\mathrm{rgh}}$ & $T$ \\
\midrule
TransFVGN            & FVM Loss  & $0.0178$ & $0.102$ & $0.0213$ \\
Transolver           & FVM Loss  & $0.0233$ & $0.0828$ & $0.0555$ \\
GNN (no attention)   & FVM Loss  & $0.0227$ & $0.0931$ & $0.0287$ \\\midrule
TransFVGN            & Data Loss & $0.0393$ & $0.097$ & $0.0485$ \\
Transolver           & Data Loss & $0.0332$ & $0.0777$ & $0.0479$ \\
GNN (no attention)   & Data Loss & $0.0251$ & $0.0668$ & $0.0433$ \\
\bottomrule
\end{tabular}
\end{tableorg}
\begin{tableorg}[!htbp]
\centering
\caption{Full-horizon field-wise rollout nRMSE, the cumulative geometric mean in time followed by the arithmetic mean over the 30 held-out cases of each benchmark.
All four models are TransFVGN networks.
``All fields'' is the average of the three unrounded field entries.
The velocity row uses the in-plane vector $(u,v)$ for the cylinder and the full vector $(u,v,w)$ for the data center.
The temperature row uses the dimensionless cylinder temperature and the absolute data-center temperature.}
\label{tab:fieldwise_nrmse}
\small\setlength{\tabcolsep}{6pt}
\begin{tabular}{@{}lcccc@{}}
\toprule
\multirow{2}{*}{Field} & \multicolumn{2}{c}{Heated cylinder wake} & \multicolumn{2}{c}{Data center hall} \\
\cmidrule(lr){2-3} \cmidrule(lr){4-5}
& \shortstack{Data Loss} & \shortstack{FVM Loss} & \shortstack{Data Loss} & \shortstack{FVM Loss} \\
\midrule
$|\mathbf{U}|$ & 0.0393 & 0.0178 & 0.271 & 0.0994 \\
$p_{\mathrm{rgh}}$ & 0.0970 & 0.102 & 0.0423 & 0.0263 \\
$T$ & 0.0485 & 0.0213 & 0.00384 & 0.00130 \\
\midrule
All fields & 0.0616 & 0.0469 & 0.106 & 0.0423 \\
\bottomrule
\end{tabular}
\end{tableorg}
\begin{tableorg}[!htbp]
\centering
\caption{Upfront cost of every model.
The epochs column is the training budget in epochs, the s/epoch column the steady wall time of one epoch, and the training hours multiply the two;
the data-generation (Data gen) column is the recorded total wall clock of computing the 240 labeled training trajectories, which only Data Loss needs.
Peak memory is the allocation per GPU, and Act.\ ckpt.\ marks whether activation checkpointing is enabled.}

\label{tab:upfront_cost}
\footnotesize\setlength{\tabcolsep}{3pt}
\resizebox{\textwidth}{!}{%
\begin{tabular}{llrrrrrrl}
\toprule
 & & & & Training & Data & Upfront & Peak per & Act. \\
Architecture & Loss & Epochs & s/epoch & (h) & gen (h) & total (h) & GPU (GiB) & ckpt. \\
\midrule
\multicolumn{9}{@{}l}{\emph{Heated cylinder wake}} \\
TransFVGN & FVM Loss & 2400 & 347.86 & 231.91 & --- & 231.91 & 11.89 & on \\
GNN (no attention) & FVM Loss & 2400 & 285.78 & 190.52 & --- & 190.52 & 19.96 & on \\
Transolver & FVM Loss & 2400 & 226.50 & 151.00 & --- & 151.00 & 16.80 & off \\
TransFVGN & Data Loss & 2400 & 302.77 & 201.85 & 52.19 & 254.04 & 11.39 & on \\
GNN (no attention) & Data Loss & 2400 & 251.82 & 167.88 & 52.19 & 220.07 & 19.11 & on \\
Transolver & Data Loss & 2400 & 157.62 & 105.08 & 52.19 & 157.27 & 16.30 & off \\
\addlinespace
\multicolumn{9}{@{}l}{\emph{Data center hall}} \\
TransFVGN & FVM Loss & 2400 & 307.00 & 204.67 & --- & 204.67 & 11.64 & on \\
TransFVGN & Data Loss & 2400 & 273.61 & 182.41 & 45.20 & 227.61 & 10.66 & on \\
\bottomrule
\end{tabular}%
}
\end{tableorg}

The network-forward column measures the synchronized encoder--processor--decoder evaluation after warm-up;
it does not include graph preprocessing or host-to-device transfer.
Per-step rollout is timed from the graph fetch to the state write-back and additionally includes the boundary conditions and, for an FVM-loss model, the finite-volume operators.
The CFD times were measured separately, in dedicated OpenFOAM~\cite{weller1998openfoam} runs of a representative operating condition of each benchmark on the same machine, each covering the complete 500-step horizon.
The transient and data-center references behind these numbers share one solve setting, a $10^{-6}$ residual-control target with Crank--Nicolson time integration (off-centering 0.5);
the two steady benchmarks use the settings given in their sections.
The residual model itself advances one time step per network evaluation, and its increment is first-order in time with the $\vartheta$-weighted convective flux of Eq.~\eqref{eq:imex_convection}.

FVM-loss TransFVGN requires 0.325~s per cylinder step (162~s per 500 steps) with 1.81~GiB peak GPU allocation and 0.317~s per data-center step (158~s) with 1.69~GiB.
Across the six cylinder models, the 500-step inference time ranges from 116 to 162~s and peak allocation from 1.19 to 3.63~GiB.
Every neural model therefore fits on a single consumer GPU, using at most 3.63~GiB of its 24~GB.
The data-center data-loss model takes 139~s per 500 steps.
\begin{tableorg}[!htbp]
\centering
\caption{Inference cost of every model for one 500-step case, against CFD on 1--8 CPU cores.
Neural rows use one RTX 4090, and s/step is the complete timed rollout step, a median of the steady steps for the neural models and a mean over the complete run for CFD.
The 500-step column multiplies s/step by 500.
Peak memory is the GPU allocation for a neural model and, for CFD, the resident CPU memory summed over the ranks.
The CFD rows use the reference conditions $\mathrm{Re}=180$ and $(Q,q_v)=(25~\mathrm{kW},\,1.29~\mathrm{m^3\,s^{-1}})$.}
\label{tab:inference_cost}
\footnotesize\setlength{\tabcolsep}{4pt}
\begin{tabular}{llrrrr}
\toprule
 & & Network & & 500 steps & Peak memory \\
Solver & Loss / hardware & forward (ms) & s/step & (s) & (GiB) \\
\midrule
\multicolumn{6}{@{}l}{\emph{Heated cylinder wake}} \\
TransFVGN & FVM Loss, 1 RTX 4090 & 219.57 & 0.3248 & 162.40 & 1.81 \\
GNN (no attention) & FVM Loss, 1 RTX 4090 & 159.31 & 0.2630 & 131.50 & 3.63 \\
Transolver & FVM Loss, 1 RTX 4090 & 160.29 & 0.2549 & 127.45 & 1.26 \\
TransFVGN & Data Loss, 1 RTX 4090 & 218.62 & 0.2929 & 146.45 & 1.74 \\
GNN (no attention) & Data Loss, 1 RTX 4090 & 159.66 & 0.2318 & 115.90 & 3.57 \\
Transolver & Data Loss, 1 RTX 4090 & 159.86 & 0.2333 & 116.65 & 1.19 \\
CFD & 1 core & --- & 54.85 & 27427 & 0.33 \\
CFD & 2 cores & --- & 28.16 & 14080 & 0.49 \\
CFD & 4 cores & --- & 15.84 & 7918 & 0.72 \\
CFD & 8 cores & --- & 9.4 & 4702 & 1.21 \\
\addlinespace
\multicolumn{6}{@{}l}{\emph{Data center hall}} \\
TransFVGN & FVM Loss, 1 RTX 4090 & 201.44 & 0.3168 & 158.40 & 1.69 \\
TransFVGN & Data Loss, 1 RTX 4090 & 201.47 & 0.2782 & 139.10 & 1.63 \\
CFD & 1 core & --- & 6.8 & 3400.5 & 0.36 \\
CFD & 2 cores & --- & 4.77 & 2384 & 0.58 \\
CFD & 4 cores & --- & 1.91 & 953.5 & 0.86 \\
CFD & 8 cores & --- & 0.922 & 461 & 1.48 \\
\bottomrule
\end{tabular}
\end{tableorg}

\begin{figure}[!htbp]
\centering
\includegraphics[width=\textwidth,height=0.72\textheight,keepaspectratio]{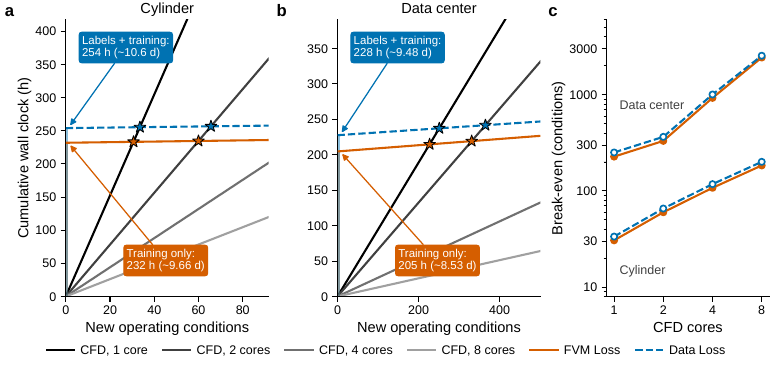}
\caption{Amortized wall-clock cost of TransFVGN surrogates against computational fluid dynamics (CFD).
TransFVGN, the attention graph neural network of this study, is trained on the finite-volume-method residual loss (FVM Loss) or on CFD labels (Data Loss),
and CFD runs on 1, 2, 4 or 8 cores, as identified in the key.
\textbf{a}, cumulative wall clock for the heated cylinder wake.
\textbf{b}, the same for the data center hall.
Boxes give upfront budgets in hours (h) and days (d), including label generation for Data Loss;
stars mark break-even points.
\textbf{c}, break-even number of new conditions against the CFD core count (filled circles, FVM Loss; open circles, Data Loss).}
\label{fig:amortization_analysis}
\end{figure}

For $N_{\mathrm{q}}$ new conditions, cumulative wall time is modeled as $C(N_{\mathrm{q}})=C_{\mathrm{upfront}}+N_{\mathrm{q}}\,C_{\mathrm{query}}$ for a surrogate and as $N_{\mathrm{q}}\,C_{\mathrm{CFD}}$ for CFD,
and the break-even count is $C_{\mathrm{upfront}}/(C_{\mathrm{CFD}}-C_{\mathrm{query}})$.
Figure~\ref{fig:amortization_analysis} uses the budgets and timings of Tables~\ref{tab:upfront_cost} and~\ref{tab:inference_cost}.
Against four-core CFD, FVM-loss TransFVGN breaks even at approximately 108 cylinder conditions and 927 data-center conditions;
the data-loss model reaches approximately 118 and 1006.
Against eight-core CFD the corresponding FVM-loss thresholds increase to approximately 184 and 2435, versus 201 and 2545 for the data-loss model.
Although an FVM-loss epoch takes somewhat longer than a data-loss epoch, the FVM-loss model needs no label generation; it therefore starts from an upfront cost about 9--10\% lower and breaks even with 4--9\% fewer queries than the data-loss model.

Table~\ref{tab:fieldwise_nrmse} places the full-horizon accuracy of the same TransFVGN models beside these costs.
Over the full horizon, the FVM-loss models have the lower error in almost every field of both benchmarks, and the two cylinder pressure errors are nearly equal.

Table~\ref{tab:per_channel_accuracy} lists the full-horizon errors of all six cylinder models field by field, on which the six-model comparison of the ``3D cylinder wake flow with heat transfer'' section is based.

%%==========================================================================%%
%%  DISCUSSION -- Nature style: brief and focused, absorbs Conclusion
%%==========================================================================%%
\section{Discussion}\label{sec:discussion}

Taken together, the experiments indicate that finite-volume residuals can serve as an effective source of supervision for thermo-fluid surrogates on unstructured three-dimensional meshes.
``Label-free'' describes the absence of solution-field targets during training, while the discretized conservation laws continue to supervise the model and CFD remains the reference for validation.
In the terminology of physics-informed learning~\cite{karniadakis2021physics_informed_ml}, the residual supplies a learning bias while mesh message passing and slice attention supply architectural inductive biases.
The steady cavity benchmarks examine the trained discrete solves, whereas the cylinder and data-center rollouts evaluate the reuse of a trained update across operating conditions.

The cylinder experiment gives a controlled comparison.
The FVM-loss TransFVGN reduces the mean field-averaged cumulative rollout error from 0.062 to 0.047, with velocity and temperature errors less than half those of its data-loss counterpart,
and it matches the CFD mean drag, wall heat transfer and shedding phase more closely.
The advantage of residual training lies mainly in velocity, temperature and the later part of the rollout; in the full-horizon pressure error it does not improve on the supervised baseline, for any of the three architectures.
Its time-mean field errors are the lowest of the six models and lower than those of the attention-free FVM-loss GNN in every field;
only the Transolver pair reverses in this statistic, where its data-loss version is the more accurate one.
Together with the force statistics, these comparisons suggest that slice attention complements local message passing, most visibly in the time-mean fields, the integral forces and the early part of the rollout,
whereas the residual objective improves transfer to new cross-sections consistently across all three architectures.
On unseen elliptical cross-sections, every FVM-loss model improves on its own data-loss counterpart in every field.
Transolver, which has no mesh message passing, gains from the residual objective in late-horizon rollout error and in geometry transfer but not in time-mean accuracy.
The objective thus appears to work best together with mesh message passing.
The data-center experiment adds an application-oriented stress test on 30 held-out conditions.
There the FVM-loss model has lower full-horizon velocity, pressure and temperature errors than its data-loss counterpart, with a velocity error lower by a factor of about 2.7.
Two properties of the objective are plausible, though untested, contributors to these gains.
The discrete residual couples each predicted cell to its stencil neighbors, whereas the data loss regresses each cell toward its own label (Supplementary Note~1).
The residual also scores each update from the state that the model actually reached, whereas the data loss pulls that state toward the next reference snapshot.

The face-flux operators are conservative by construction, and the conservation diagnostic confirms that the shared-face cancellation holds to floating-point precision.
On the cylinder test cases, residual training also lowers the mass imbalance of the predicted fields relative to the data-loss model, although the CFD reference remains better balanced (see ``Conservation property of the discrete operators'' in Methods).
Conservation of the predicted fields is thus encouraged, though not imposed exactly, by the residual norm that training minimizes.
Beyond accuracy and conservation, the residual objective lowers the cost of building the surrogate.
Because no labels have to be generated, the upfront budget remains below that of supervised training, even though an FVM-loss epoch takes somewhat longer.
Once trained, the model answers a 500-step query in a few minutes on one GPU with under 2~GiB of allocated memory,
and the training investment is repaid over repeated queries (see ``Computational cost and amortization''), a regime typical of the many-query design and control studies that motivate surrogate modeling.

A limitation of the method itself remains open.
Relative to the exact solution, the error of the surrogate combines the discretization error of the finite-volume scheme it approximates with the learning error of the network (see ``Problem definition and framework overview''), as is also the case for a data-driven model;
training on the residual changes the learning target but does not remove this superposition.
Our results suggest that the balance between the two depends on how the model is used.
For the steady benchmarks, where each condition is solved by training and no generalization is required, the deviation of the trained discrete solution from the same-mesh CFD solution is small at the lower Reynolds and Rayleigh numbers and reaches a few percent in the most demanding cases, and the remaining error appears to stem mainly from the scheme differences between the residual and the reference solver.
In the parametric transient benchmarks, where the model must roll out unseen conditions over long horizons, the learning error tends to dominate instead.
The present study does not analyze or decompose the two contributions rigorously;
quantifying them for each mode of use, and reducing the learning error accordingly, is an important direction for future work.
In addition, the validation is confined to laminar flows with constant viscosity and time-constant boundary conditions.
At these Richardson numbers the temperature acts almost as a passive scalar, and because $\mathrm{Re}\le180$ lies below the Mode A threshold ($\mathrm{Re}\approx188.5$), the reference wake is two-dimensional; two-way momentum--energy coupling in three dimensions is tested by the data-center hall.
The transient surrogates are trained on one mesh per benchmark, and geometry transfer is demonstrated on one elliptical-cylinder family.
The controlled comparisons are made against supervised training rather than against other physics-informed learning methods.
Natural extensions include turbulence closures, time-dependent boundary conditions, broader shape and mesh transfer, and deployment on other hardware.

Within this scope, the results indicate that a neural surrogate trained without labeled CFD solution fields can predict coupled three-dimensional thermo-fluid fields across operating conditions and transfer to related cross-sections without retraining.
Finite-volume residuals thereby offer a route to neural surrogates whose development does not depend on a labeled simulation campaign.

%%==========================================================================%%
%%  METHODS -- Nature style: placed after Discussion
%%==========================================================================%%
\section{Methods}\label{sec:methods}

This section describes the FVM-loss model in terms of its governing physics, graph representation, model architecture, training objective, and training procedure.
Figure~\ref{fig:framework_overview} summarizes the neural-numerical pipeline.
Following its bottom row,
the evolving flow state is streamed from the two-tier data pool (see ``Training strategy'') to the GPU, batched into the compound-cell graph, advanced by the attention GNN, and written back to the pool as the next state.
We first present the governing equations (``Governing equations''),
then define the mesh preliminaries and their graph representation (``Heterogeneous graph representation'') and the neural network architecture that operates on it (``Neural network architecture'').
The ``Finite-volume discretization residual'' section then constructs the discrete residual objective,
including the weighted least-squares (WLSQ) gradient reconstruction (``3D gradient reconstruction via weighted least squares''),
the face-value reconstruction and flux assembly (``Face-value reconstruction and flux assembly''),
the face-control-volume momentum residual that suppresses pressure--velocity decoupling (``Designed to suppress pressure--velocity decoupling''),
and the boundary-condition encoding and enforcement (``GPU-parallel residual assembly and boundary-condition enforcement'').
Finally, we describe the training strategy and the experimental setup (from ``Training strategy'' to ``Experimental setup'').
Supplementary Note~2 collects the notation and the main formulas of the method in one place (Supplementary Tables~1 and~2).

\subsection{Governing equations}\label{subsec:governing_equations}

We consider the 3D incompressible Navier--Stokes equations coupled with the energy equation under the Boussinesq approximation~\cite{gray1976boussinesq}.
The hydrostatic contribution is absorbed into a modified pressure $p_{\mathrm{rgh}}=p-\rho_k\,g_h$, where $\rho_k=\rho_0\bigl[1-\beta\,(T-T_0)\bigr]$ is the Boussinesq-linearized density,
so that the buoyancy coupling appears as a gradient term rather than a body force:
\begin{align}
\nabla \cdot \mathbf{u} &= 0, \label{eq:continuity}\\
\frac{\partial \mathbf{u}}{\partial t} + \nabla \cdot (\mathbf{u} \otimes \mathbf{u}) &= -\tfrac{1}{\rho_0}\nabla p_{\mathrm{rgh}} + \nu \nabla^2 \mathbf{u} + g_h\,\beta\,\nabla T, \label{eq:momentum}\\
\frac{\partial T}{\partial t} + \nabla \cdot (\mathbf{u} T) &= \alpha \nabla^2 T + \tfrac{\dot{q}}{\rho_0 c_p}, \label{eq:energy}
\end{align}
where $\mathbf{u} = (u, v, w)^{\top}$ is the velocity vector, $p$ the static pressure, $p_{\mathrm{rgh}}$ the modified pressure, $T$ the temperature, $g_h=\mathbf{g}\cdot(\mathbf{x}-\mathbf{x}_h)$ the negative of the gravitational potential at position $\mathbf{x}$, measured from a point $\mathbf{x}_h$ at the reference height, $\rho_0$ the constant reference density, $\nu$ the kinematic viscosity, $\alpha$ the thermal diffusivity, $\beta$ the thermal expansion coefficient, $T_0$ the reference temperature, $\mathbf{g}$ the gravitational acceleration, $\dot{q}$ a volumetric heat source, and $c_p$ the specific heat capacity.
Their finite-volume discretization and the residual-based training objective are constructed in the sections from ``Finite-volume discretization residual'' to ``GPU-parallel residual assembly and boundary-condition enforcement''.

\subsection{Heterogeneous graph representation}\label{subsec:graph_representation}

\begin{figure}[!htbp]
\centering
\includegraphics[width=\textwidth,height=0.72\textheight,keepaspectratio]{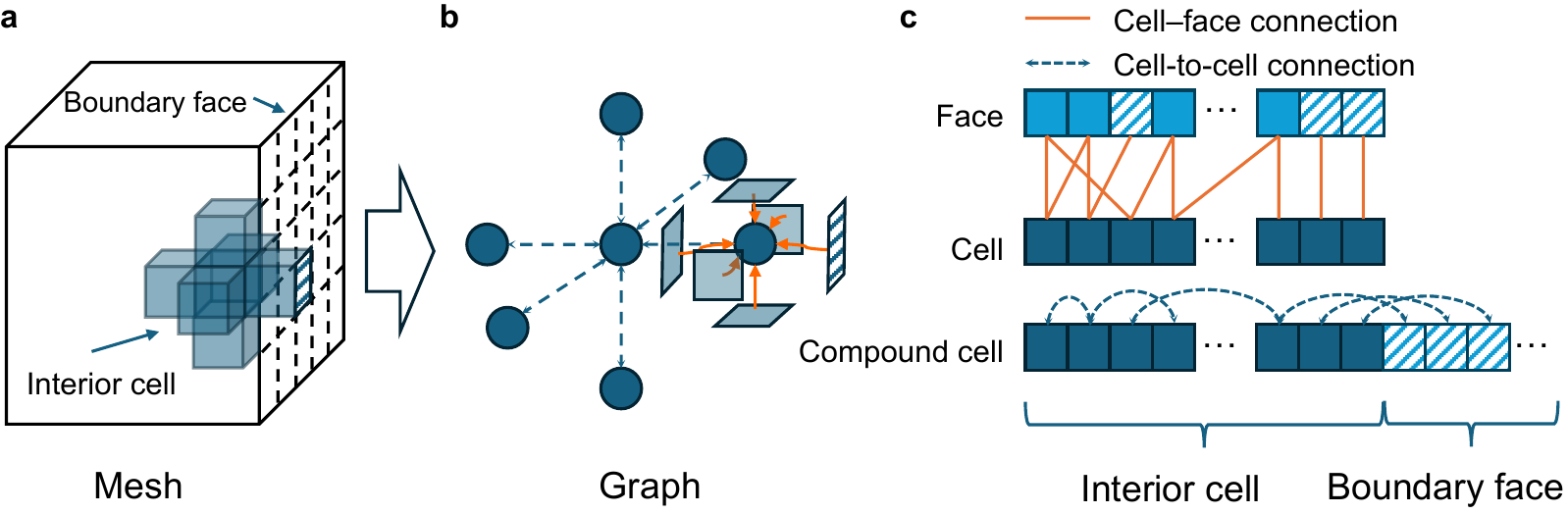}
\caption{Graph representation of a three-dimensional unstructured mesh.
\textbf{a}, mesh with interior control volumes and boundary faces.
\textbf{b}, graph in which cell centers are nodes (dark blue circles), coupled to face-adjacent neighbors by cell-to-cell connections (blue dashed lines) and to their bounding faces (gray-blue parallelograms) by cell--face connections (orange solid lines);
hatched faces are boundary faces.
\textbf{c}, attribute layout, with face attributes in light blue, cell attributes in dark blue and the compound-cell array concatenating the interior cells with the boundary-face entries (hatched).}
\label{fig:graph_representation}
\end{figure}

We represent the mesh as a typed heterogeneous graph~\cite{wu2020gnn_survey} tailored to finite-volume discretization, so that geometry, topology, and the finite-volume operators are explicit and suited to GPU execution.
The representation has three parts, described in turn below, namely the entity sets (vertices, faces, and control volumes) with their attributes,
the typed connectivity that encodes the mesh topology,
and the face-to-cell scatter reduction that carries the conservative assembly.
A key concept is the \emph{compound cell},
which unifies the interior control volumes with the boundary-face entries so that the finite-volume operators apply uniformly across the mesh.
Figure~\ref{fig:graph_representation} sketches the mapping from the mesh to the graph and the resulting compound-cell data layout.

\emph{Entity sets.}
Let $\mathcal{V}_{\mathrm{n}}$, $\mathcal{V}_{\mathrm{f}}$, and $\mathcal{V}_{\mathrm{c}}$ denote the sets of mesh vertices, faces, and control volumes, with cardinalities $N_{\mathrm{n}}$, $N_{\mathrm{f}}$, and $N_{\mathrm{c}}$.
Each vertex carries its position in $\mathbb{R}^{3}$;
each face $f\in\mathcal{V}_{\mathrm{f}}$ carries a centroid $\mathbf{x}_f$, an area $A_f$, a unit normal $\mathbf{n}_f$ pointing out of its owner cell, and a boundary-type label;
and each control volume $i\in\mathcal{V}_{\mathrm{c}}$ carries a centroid $\mathbf{x}_i$, a volume $V_i$, and the unit normals of its bounding faces.
The faces bounding cell $i$ form the set $\mathcal{F}_i$.
A graph-level attribute set stores global quantities such as the governing-PDE coefficients and a case identifier.

\emph{Compound cell.}
The finite-volume operators must treat interior cells and domain boundaries uniformly,
yet the two are structurally different.
In particular, an interior face is shared by two control volumes,
whereas a boundary face is adjacent to only one control volume.
It therefore lacks the second value that face interpolation and flux assembly require.
We remove this asymmetry by augmenting the $N_{\mathrm{c}}$ interior control volumes with one entry per boundary face, located at the face center,
and indexing the two together as a single set of \emph{compound cells} $\mathcal{V}_{\bar{\mathrm{c}}}$.
The cardinality is $N_{\bar{\mathrm{c}}}=N_{\mathrm{c}}+N_{\mathrm{bf}}$, with $N_{\mathrm{bf}}$ the number of boundary faces.
The interior control volumes occupy the first $N_{\mathrm{c}}$ entries and the boundary-face entries the remaining $N_{\mathrm{bf}}$,
and a single contiguous index accordingly addresses both (Figure~\ref{fig:graph_representation}c).
Each boundary-face entry holds the state of its boundary face,
and the boundary condition is enforced there (see ``GPU-parallel residual assembly and boundary-condition enforcement'').
These entries differ from the ghost cells of conventional finite-volume solvers, which are fictitious cells outside the domain, mirrored across the boundary;
here the boundary faces themselves are appended to the cell array.
Dirichlet values are written directly into the boundary face attribute,
and Neumann conditions are imposed by setting the boundary gradient flux.
This construction gives every face, interior or boundary, an owner and a neighbor in $\mathcal{V}_{\bar{\mathrm{c}}}$.
Interior and boundary faces are thus handled by the same interpolation and flux-assembly operators.
Each compound cell $i\in\mathcal{V}_{\bar{\mathrm{c}}}$ stores a centroid $\mathbf{x}_i$,
a state vector $\mathbf{q}_i$ (velocity, pressure, and temperature), the governing-PDE coefficients as separate node features, type and global-index labels, and the precomputed WLSQ gradient operator that maps neighboring states to the reconstructed cell gradient (see ``3D gradient reconstruction via weighted least squares'').
The network state, the gradient reconstruction, and the residual assembly all act on this same index;
we therefore call the resulting graph the \emph{compound-cell graph},
and the predicted field on the interior cells is the quantity advanced in time.

\emph{Typed connectivity.}
Connectivity is encoded as typed edge sets,
each stored in compressed sparse row (CSR) form to accommodate the variable number of neighbors per entity.
These sets comprise the face--vertex and cell--vertex incidences $\mathcal{E}_{\mathrm{fn}}$ and $\mathcal{E}_{\mathrm{cn}}$;
the cell--face incidence $\mathcal{E}_{\mathrm{cf}}$ together with its transpose $\mathcal{E}_{\mathrm{cf}}^{\top}$ (the face-to-owner/neighbor map that carries the conservative flux assembly);
the first-ring (face-adjacent) compound-cell adjacency $\mathcal{E}^{(1)}\subseteq\mathcal{V}_{\bar{\mathrm{c}}}\times\mathcal{V}_{\bar{\mathrm{c}}}$ used for face interpolation and flux assembly, with the neighbors $\mathcal{N}(i)=\{j:(i,j)\in\mathcal{E}^{(1)}\}$ of cell $i$;
and the extended neighborhood $\mathcal{E}^{\mathrm{w}}\supseteq\mathcal{E}^{(1)}$ used by the WLSQ gradient reconstruction,
which augments the face-adjacent neighbors of each cell with their own face neighbors and gives the WLSQ stencil $\mathcal{N}^{\mathrm{w}}(i)=\{j:(i,j)\in\mathcal{E}^{\mathrm{w}}\}$ (see ``3D gradient reconstruction via weighted least squares'').
Periodic boundaries are represented by an additional pairing that identifies matching periodic faces.

\begin{figure}[!htbp]
\centering
\includegraphics[width=\textwidth,height=0.72\textheight,keepaspectratio]{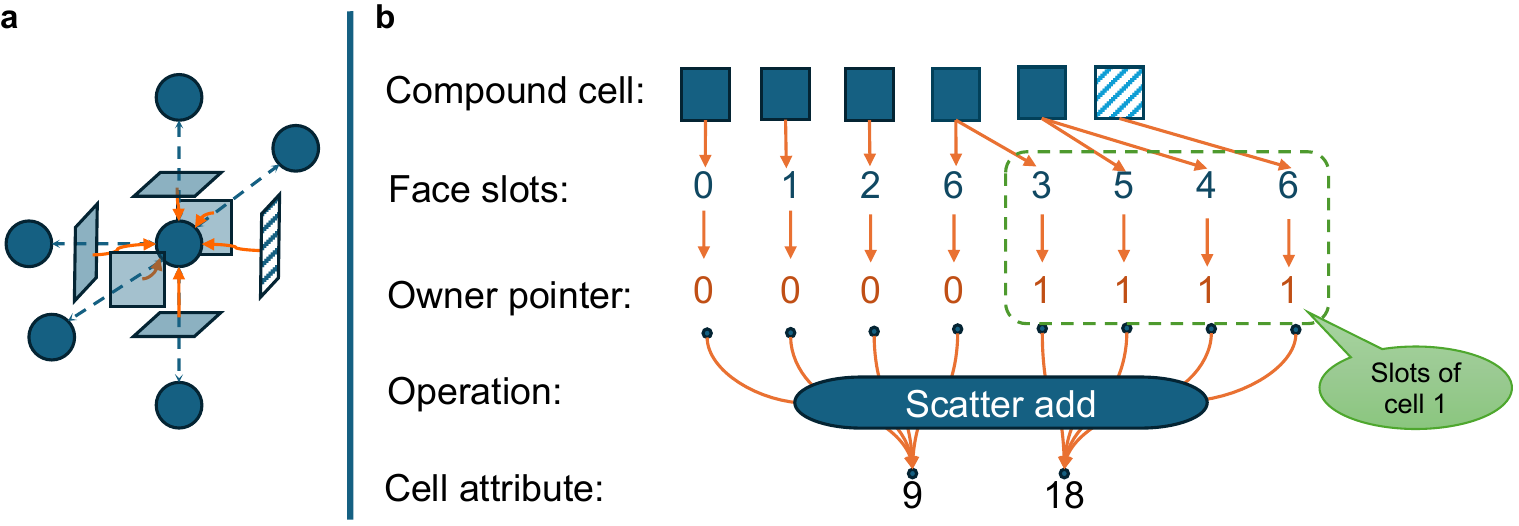}
\caption{Local stencil of a compound cell and the face-to-cell scatter reduction.
\textbf{a}, a compound cell (center) linked to its face-adjacent neighbors (dark blue circles) by cell-to-cell connections (blue dashed lines) and to its six faces (gray-blue squares) by cell--face connections (orange arrows);
hatching marks a boundary face.
\textbf{b}, scatter-add of face slots into cell attributes.
Compound cells (dark blue squares; hatched, a boundary-face entry) write illustrative values into slots tagged with owner pointers,
and the slots of each owner sum to its attribute (9 and 18).}
\label{fig:graph_scatter}
\end{figure}

\emph{Face-to-cell scatter reduction.}
The cell--face incidence $\mathcal{E}_{\mathrm{cf}}$, stored in CSR form, turns the conservative flux assembly into a segment reduction over the graph (Figure~\ref{fig:graph_scatter}).
Each compound cell owns a contiguous block of face slots,
and a per-slot cell pointer records the cell to which every face contribution belongs, reusing the CSR offsets defined above.
Given the face-centered fluxes $\{\Phi_f\}$,
the cell residual is formed by a scatter-add that groups all slots sharing a cell pointer and sums them into that cell,
\begin{equation}
\mathcal{R}_i = \sum_{f\in\mathcal{F}_i} \Phi_f,
\qquad
\mathcal{F}_i=\{f:(i,f)\in\mathcal{E}_{\mathrm{cf}}\}.
\label{eq:scatter_reduction}
\end{equation}
Each face flux enters with its area vector oriented outward from cell $i$, and an interior face therefore contributes with opposite signs to its two cells.
Once grouped by cell,
the reduction becomes one vectorized scatter-add over all face slots, and a single operator assembles interior and boundary residuals without case-by-case branching.
The same indexing serves both the WLSQ gradient reconstruction and the conservative residual assembly detailed in the ``Face-value reconstruction and flux assembly'' section.

The compound-cell representation is a data-layout choice that lets one set of operators serve interior and boundary faces alike;
separate boundary-face kernels would also be possible.
The boundary-face entries serve two purposes.
They supply the missing second state at every boundary face,
and the interpolation and flux-assembly operators therefore apply uniformly, with the boundary-specific treatment confined to the boundary-condition operators that set the boundary-face states and the boundary gradient fluxes.
They also inject the boundary data into the \emph{face fluxes} (through the convective face states and the diffusive boundary fluxes described in ``Face-value reconstruction and flux assembly''),
in addition to entering the network as node features of the boundary-face entries.
Dirichlet and Neumann conditions thus enter the discrete conservation balances at the boundary faces, where a finite-volume solver applies them.
The boundary-face entries are not additional unknowns,
since the network output is restricted to the interior control volumes and their states are set by the boundary-condition operators.
Imposing the conditions on the first layer of interior cells instead would remove the conservation equations of those cells and would place the boundary value half a cell away from the boundary.
An analysis of neural operators trained on labeled data~\cite{shikhman2026boundary_indexed} indicates that the learned map remains conditioned on the boundary conditions seen in training;
such feature-side conditioning of the network is complementary to this flux-side mechanism.

\subsection{Neural network architecture}\label{subsec:nn_architecture}

The learnable model operates on the same compound-cell graph used by the finite-volume residual operator;
the neural state and the residual operator therefore share the same indexing.

\emph{Model inputs.}
For each compound cell $i$,
the feature vector concatenates the current field $\mathbf{q}_i^{n}$, initialization features, cell-type indicators, PDE parameters, and Fourier positional encodings of the cell centroid $\mathbf{x}_i$ (four frequency bands)~\cite{tancik2020fourier}.
Edge attributes encode the relative position and the current relative state of neighboring compound cells.
This choice lets the network act directly on the cell-centered variables required by flux assembly.

\emph{Processor.}
Between the encoder and decoder (Figure~\ref{fig:framework_overview}),
the processor stacks blocks that each pair a local message-passing step~\cite{gilmer2017message_passing,battaglia2018relational_inductive} with a geometry-aware slice-attention layer ($\mathcal{A}_{\ell}$)~\cite{wu2024transolver}.
In block $\ell$, a node-to-edge update first computes the edge increments $\Delta\mathbf{e}_{ij}^{(\ell)}=\psi_{\ell}^{\mathrm{e}}\big(\mathbf{h}_i^{(\ell)},\mathbf{h}_j^{(\ell)},\mathbf{e}_{ij}^{(\ell)}\big)$.
An edge-to-node step then sums, for every cell, the increments of its incident edges, $\mathbf{a}_i=\sum_{j\in\mathcal{N}(i)}\Delta\mathbf{e}_{ji}^{(\ell)}$,
and a node-to-node step averages these sums over the face neighbors, $\bar{\mathbf{a}}_i=|\mathcal{N}(i)|^{-1}\sum_{j\in\mathcal{N}(i)}\mathbf{a}_j$,
so that each block also draws on the edges of the neighboring cells.
The cell and edge latents are updated as
$\mathbf{h}_i^{(\ell+1/2)}=\mathbf{h}_i^{(\ell)}+\big[\mathbf{h}_i^{(\ell)}+\psi_{\ell}^{\mathrm{c}}\big(\bar{\mathbf{a}}_i,\mathbf{h}_i^{(\ell)}\big)\big]$ and $\mathbf{e}_{ij}^{(\ell+1)}=\mathbf{e}_{ij}^{(\ell)}+\Delta\mathbf{e}_{ij}^{(\ell)}$,
where the bracket is the residual connection of the message-passing step and the outer sum a skip connection around it,
and the slice-attention layer returns $\mathbf{h}_i^{(\ell+1)}=\mathcal{A}_{\ell}\big(\mathbf{h}^{(\ell+1/2)}\big)_i$.
Here $\mathbf{h}_i^{(\ell)}$ is the latent state of cell $i$, $\mathbf{e}_{ij}^{(\ell)}$ the latent attribute of the edge between cells $i$ and $j$ (initialized from the edge attributes above), and $\mathcal{N}(i)$ the face-adjacent neighborhood of cell $i$ (the adjacency $\mathcal{E}^{(1)}$ of the ``Heterogeneous graph representation'' section).
Each edge latent holds one half-width vector for each direction, and $\Delta\mathbf{e}_{ji}^{(\ell)}$ denotes the half oriented toward cell $i$.
The maps $\psi_{\ell}^{\mathrm{e}}$ and $\psi_{\ell}^{\mathrm{c}}$ are multilayer perceptrons with two hidden layers, GELU activations and layer normalization of their outputs,
and $\mathcal{A}_{\ell}$ applies slice attention and a feed-forward layer, each after layer normalization and with a residual connection.
The local message-passing layer preserves stencil locality,
while the slice-attention layer $\mathcal{A}_{\ell}$ acts on the full latent field $\mathbf{h}^{(\ell+1/2)}$ rather than on a local neighborhood and can thus exchange information between distant regions, such as the two ends of an elongated domain or separated recirculation zones, within one layer (top row of Figure~\ref{fig:framework_overview}).
This layer assigns the cell-centered latent states to a small set of learnable \emph{slices} through a softmax weighting computed from each cell's features.
It then aggregates the cells of each slice into a single slice token,
and applies multi-head self-attention~\cite{vaswani2017attention} among these slice tokens.
The result is broadcast back to the cells using the same assignment weights.
Attention therefore operates on the small number of slice tokens rather than on all cells.
Its token-to-token cost depends on the number of slices, whereas the assignment and broadcasting steps scale with the number of cells.
Because a slice can gather cells from anywhere in the domain, a single layer can exchange information between distant regions of the mesh.
The pairing is a heuristic design based on the complementary reach of the two operators.
Graph message passing acts locally by construction, much like the stencils of a finite-volume discretization, and resolves the local, high-frequency content of the flow.
Slice attention, adopted from Transolver~\cite{wu2024transolver}, instead provides a global communication space for the long-range coupling.
Figure~\ref{fig:cylinder_testset_arms} evaluates the contribution of attention on the cylinder test set through an attention-free GNN with a matched parameter budget.

\emph{Model output.}
The model output, produced by the decoder at the top right of Figure~\ref{fig:framework_overview}, is a correction $\Delta\mathbf{q}_{\theta,i}$ for each interior cell $i\in\mathcal{V}_{\mathrm{c}}$,
and the next iterate is updated as
\begin{equation}
\mathbf{q}_i^{n+1} = \mathbf{q}_i^{n} + \Delta\mathbf{q}_{\theta,i}.
\label{eq:residual_update}
\end{equation}
This increment parameterization~\cite{he2016deep_residual,pfaff2021meshgraphnets} tends to stabilize training.
Each network evaluation thereby corrects the current state over one time step (a pseudo-time step in the two steady benchmarks) rather than generating the field in one shot~\cite{kochkov2021ml_accelerated_cfd}.

\emph{Architecture hyperparameters.}
The TransFVGN backbone is an encoder--processor--decoder graph network with latent width $128$.
The processor stacks 8 layers of message-passing and slice-attention blocks with residual connections and layer normalization,
and each slice-attention layer has 8 heads with 32 slices per head.
Every TransFVGN model reported with the FVM loss uses this eight-layer processor (2,169,349 parameters).
The optimizer, learning rate, and distributed data-parallel (DDP) configuration shared by both training modes are given in the ``Experimental setup'' section.

\emph{Baseline architectures.}
The controlled comparison on the cylinder wake (see ``3D cylinder wake flow with heat transfer'' in Results) adds two architectures of matched size.
The first is a Transolver network~\cite{wu2024transolver}, which uses slice-attention layers without message passing on the mesh graph (latent width $192$).
The second is an attention-free GNN, which keeps the message-passing blocks and removes the slice-attention layers (latent width $170$).
With these widths and 8 processor layers each, the three architectures have $2.16$--$2.20$~M trainable parameters.
Each architecture is trained once with the FVM loss and once with the data loss,
on the same cylinder training conditions and with the same optimizer.
Both model families are trained for 2400 epochs; the FVM-loss training performs more than one inner pass per epoch, the data-loss training one (Table~\ref{tab:upfront_cost}).
The resulting six models are compared in Figure~\ref{fig:cylinder_testset_arms}.

\subsection{Finite-volume discretization residual}\label{subsec:discrete_residual}

The FVM-loss objective measures whether the predicted state satisfies the cell-wise algebraic balances of a finite-volume discretization on the mesh.
Minimizing $\mathcal{L}_{\mathrm{FVM}}$ therefore drives the predicted state toward a weighted least-squares compromise of the cell balances in Eqs.~\eqref{eq:discrete_continuity}--\eqref{eq:discrete_energy} and the face balance of the ``Designed to suppress pressure--velocity decoupling'' section, which agree up to truncation error.
In practice the loss converges to a small finite value,
and the cell-wise residual magnitude provides a direct, per-cell diagnostic of the remaining conservation deviation.

For an interior control volume $i\in\mathcal{V}_{\mathrm{c}}$ with volume $V_i$, face set $\mathcal{F}_i$, and face-area vectors $\mathbf{S}_f=A_f\mathbf{n}_f$ (see ``Heterogeneous graph representation''; in every sum over $f\in\mathcal{F}_i$, $\mathbf{S}_f$ is oriented outward from cell $i$),
the discrete finite-volume balances~\cite{versteeg2007fvm} of the governing equations (see ``Governing equations'') at time level $n+1$ can be written as
\begin{align}
\mathcal{R}_{\mathrm{cont},i} &=
\sum_{f\in\mathcal{F}_i}\mathbf{u}^{n+1}_f\cdot \mathbf{S}_f, \label{eq:discrete_continuity}\\
\mathcal{R}_{\mathrm{mom},i} &=
\frac{V_i}{\Delta t}\left(\mathbf{u}^{n+1}_i-\mathbf{u}^{n}_i\right) \notag\\
&\quad + \sum_{f\in\mathcal{F}_i}\left(\mathbf{u}^{n+1}_f\cdot \mathbf{S}_f\right)\mathbf{u}^{n+1}_f \notag\\
&\quad + \frac{1}{\rho_0}(\nabla_h p_{\mathrm{rgh}})_i^{n+1}\,V_i \notag\\
&\quad - g_{h,i}\,\beta\,(\nabla_h T)_i^{n+1}\,V_i
- \nu\sum_{f\in\mathcal{F}_i}\left(\nabla_h \mathbf{u}\right)_f^{n+1}\cdot \mathbf{S}_f, \label{eq:discrete_momentum}\\
\mathcal{R}_{\mathrm{energy},i} &=
\frac{V_i}{\Delta t}\left(T_i^{n+1}-T_i^n\right) \notag\\
&\quad + \sum_{f\in\mathcal{F}_i}\left(\mathbf{u}^{n+1}_f\cdot \mathbf{S}_f\right)T_f^{n+1}
- \alpha\sum_{f\in\mathcal{F}_i}\left(\nabla_h T\right)_f^{n+1}\cdot \mathbf{S}_f
- \frac{\dot{q}_iV_i}{\rho_0 c_p}. \label{eq:discrete_energy}
\end{align}
Here $\mathbf{u}_f$ and $T_f$ are face-centered values obtained from the graph state, interpolation operators, and boundary-condition enforcement, and $\nabla_h$ denotes the discrete gradient operator realized by the WLSQ reconstruction (see ``3D gradient reconstruction via weighted least squares'').
The pressure gradient and buoyancy source are evaluated at cell centers through the same WLSQ gradient reconstruction used for the diffusive fluxes,
a mesh-only operator that is fitted over each cell's WLSQ stencil and precomputed once per cell (see ``3D gradient reconstruction via weighted least squares'', Eq.~\eqref{eq:wlsq_3d}).
The residual vector is assembled on the mesh of each case.

We train the network by minimizing a log-scaled discrete residual objective.
For each case $s$ of a batch, the per-equation $\ell_2$ norms of the cell-wise residuals are first computed,
\begin{equation}
\begin{aligned}
\mathcal{L}_{\mathrm{cont}}^{(s)} &= \sqrt{\textstyle\sum_{i\in\mathcal{V}_{\mathrm{c}}} \mathcal{R}_{\mathrm{cont},i}^2+\varepsilon},\\
\mathcal{L}_{\mathrm{mom}}^{(s)} &= \sum_{k=1}^{d}\sqrt{\textstyle\sum_{i\in\mathcal{V}_{\mathrm{c}}} \mathcal{R}_{\mathrm{mom},i,k}^2+\varepsilon}\;+\;\sqrt{\textstyle\sum_{f\in\mathcal{V}_{\mathrm{f}}^{\ast}} \mathcal{R}_{\mathrm{mom},f}^2+\varepsilon},\\
\mathcal{L}_{\mathrm{energy}}^{(s)} &= \sqrt{\textstyle\sum_{i\in\mathcal{V}_{\mathrm{c}}} \mathcal{R}_{\mathrm{energy},i}^2+\varepsilon},
\end{aligned}
\label{eq:l2_norms}
\end{equation}
and then combined into a weighted sum, passed through a natural logarithm, and averaged over the batch:
\begin{equation}
\mathcal{L}_{\mathrm{FVM}} =
\frac{1}{B}\sum_{s=1}^{B}
\ln\!\Big(
\lambda_{\mathrm{cont}}\,\mathcal{L}_{\mathrm{cont}}^{(s)}
+ \lambda_{\mathrm{mom}}\,\mathcal{L}_{\mathrm{mom}}^{(s)}
+ \lambda_{\mathrm{energy}}\,\mathcal{L}_{\mathrm{energy}}^{(s)}
+ \lambda_{\mathrm{gauge}}\,\mathcal{L}_{\mathrm{gauge}}^{(s)} + \varepsilon
\Big),
\label{eq:discrete_loss}
\end{equation}
where $B$ is the batch size;
the sums in Eq.~\eqref{eq:l2_norms} run over the interior cells $\mathcal{V}_{\mathrm{c}}$ and over the set $\mathcal{V}_{\mathrm{f}}^{\ast}\subseteq\mathcal{V}_{\mathrm{f}}$ of faces that carry a face residual (faces that only serve to make a mesh two-dimensional are excluded, and the four benchmarks here have none), with all residuals evaluated on the state of case $s$;
$\mathcal{R}_{\mathrm{mom},i,k}$ denotes the $k$-th component of the momentum residual, $\mathcal{R}_{\mathrm{mom},f}$ is the face-normal momentum residual constructed in the ``Designed to suppress pressure--velocity decoupling'' section (Eq.~\eqref{eq:face_cv_residual}), and $d{=}3$ is the number of velocity components.
The fixed weights $\lambda_{\mathrm{cont}}{=}10^{5}$, $\lambda_{\mathrm{mom}}{=}10^{4}$, and $\lambda_{\mathrm{energy}}{=}10^{4}$ are empirical scale factors that bring the three conservation residuals to comparable numerical magnitude so that no single equation dominates the gradient~\cite{wang2021gradient_pathology};
the auxiliary gauge term $\mathcal{L}_{\mathrm{gauge}}$ (defined below) carries unit weight, $\lambda_{\mathrm{gauge}}{=}1$, and enters only when no Dirichlet pressure boundary is present.
The three conservation weights were set once from the measured residual magnitudes at initialization and then kept fixed for every benchmark and every case, rather than tuned per benchmark.
The natural logarithm, floored by a small constant $\varepsilon>0$ for numerical safety, keeps cases with very different residual magnitudes on a comparable scale within a batch.
The weighted residual sum can span several orders of magnitude across cases and training stages,
and the logarithm prevents the largest cases from dominating the gradient.
When no Dirichlet pressure boundary is present,
the pressure field is determined only up to an additive constant,
and one auxiliary term fixes this constant:
\begin{equation}
\mathcal{L}_{\mathrm{gauge}}^{(s)} = \left|\frac{\sum_{i\in\mathcal{V}_{\mathrm{c}}} V_i\,p_{\mathrm{rgh},i}}{\sum_{i\in\mathcal{V}_{\mathrm{c}}} V_i} - p_0\right|,
\label{eq:auxiliary_losses}
\end{equation}
where $\mathcal{L}_{\mathrm{gauge}}^{(s)}$ penalizes deviations of the volume-averaged modified pressure of case $s$ from a prescribed gauge level $p_0$.
Constraining the volume-weighted mean, rather than pinning a single reference cell as FVM solvers usually do, matters under gradient-based optimization.
A single-cell pin has only local gradient support,
whereas the volume-weighted mean couples to every interior cell in one backward pass and removes the undetermined constant-pressure mode without biasing the pressure distribution.
Boundary conditions are not represented by loss terms;
they enter as hard constraints during residual assembly (detailed in ``GPU-parallel residual assembly and boundary-condition enforcement'').

This training objective differs from an AD-based PINN loss.
Instead of penalizing the pointwise residual of a continuous ansatz,
the FVM loss drives the network toward the discrete state that satisfies the algebraic balances of an implicit finite-volume update on the mesh itself (Supplementary Note~1).
The two objectives also differ in computational cost.
An AD-based loss obtains spatial derivatives by differentiating the network output with respect to the input coordinates,
and each higher derivative order adds a further reverse-mode sweep through the network graph,
at a cost that grows with the depth and width of the network.
The FVM loss instead draws its spatial derivatives from the precomputed sparse mesh operators detailed below,
so that one residual evaluation reduces to sparse gather--scatter operations whose cost scales with the mesh rather than with the network size.
Training thus requires a single standard backward pass through this fixed operator pipeline rather than nested differentiation of the network.
The sections that follow detail, in turn, the WLSQ gradient reconstruction, the face-value reconstruction and flux assembly, the conservation property of these operators, the complementary face-control-volume residual, the time integration, and the boundary-condition enforcement.

\subsection{3D gradient reconstruction via weighted least squares}\label{subsec:wlsq_3d}

The pressure-gradient and buoyancy terms of the discrete momentum residual (Eq.~\eqref{eq:discrete_momentum}), the diffusive fluxes, and the face-value reconstruction (see ``Face-value reconstruction and flux assembly'') all rely on cell-centered gradients.
Accurate gradient reconstruction is hence central to the quality of the assembled residuals.
We therefore use WLSQ reconstruction,
a standard gradient operator on unstructured meshes~\cite{mavriplis2003revisiting},
as a deterministic finite-volume operator on the compound-cell graph rather than as another learnable message-passing layer.

This distinction is important because the two operators use different neighborhoods for different purposes.
The GNN processor aggregates learnable features over the face-adjacent graph $\mathcal{E}^{(1)}$ (see ``Neural network architecture''),
preserving the local stencil used by face interpolation and conservative flux assembly.
The WLSQ gradient, by contrast, is built on the wider stencil $\mathcal{E}^{\mathrm{w}}$ defined in the ``Heterogeneous graph representation'' section.
This stencil contains the face-adjacent compound cells and their two-hop extension on the compound-cell graph.
On a hexahedral mesh it thereby augments the face-sharing stencil with the cells that share an edge with the target cell and with the second cell along each face direction,
giving the least-squares fit enough spatial support on non-orthogonal 3D meshes.
For a scalar field $\phi$ at cell center $i$,
the gradient $(\nabla_h\phi)_i$ is reconstructed from the values on the WLSQ stencil $\mathcal{N}^{\mathrm{w}}(i)$ by solving the weighted least-squares normal equations
\begin{equation}
\mathbf{M}_i\,(\nabla_h\phi)_i=\sum_{j\in\mathcal{N}^{\mathrm{w}}(i)} w_{ij}\,\Delta\mathbf{x}_{ij}\,(\phi_j-\phi_i),
\qquad
\mathbf{M}_i=\sum_{j\in\mathcal{N}^{\mathrm{w}}(i)} w_{ij}\,\Delta\mathbf{x}_{ij}\,\Delta\mathbf{x}_{ij}^{\top},
\label{eq:wlsq_3d}
\end{equation}
where $\Delta\mathbf{x}_{ij}=\mathbf{x}_j-\mathbf{x}_i=(\Delta x_{ij},\Delta y_{ij},\Delta z_{ij})^{\top}$ is the offset between the two cell centroids and $w_{ij}=1/|\Delta\mathbf{x}_{ij}|$ is the inverse-distance weight.
Because the $3\times3$ moment matrix $\mathbf{M}_i$ depends only on mesh geometry,
it is precomputed and stored for each cell during preprocessing.
During training and inference on a fixed mesh,
each gradient evaluation then reduces to applying the stored inverse $\mathbf{M}_i^{-1}$ to the right-hand side assembled from the current field values,
rather than rebuilding the local least-squares system at every update.
The resulting per-cell gradient operator, the linear map $\mathbf{G}_{h,i}$ from the stencil values to $(\nabla_h\phi)_i$, is stored with each compound cell (see ``Heterogeneous graph representation'') and is reused for all scalar channels;
vector fields are reconstructed component-wise by the same operator.

\subsection{Face-value reconstruction and flux assembly}\label{subsec:fvm_loss}
All face-centered quantities required by Eqs.~\eqref{eq:discrete_continuity}--\eqref{eq:discrete_energy} are obtained from cell-center values via a two-step procedure.
For a face $f$ shared by owner cell $C$ and neighbor cell $F$ with interpolation weight $\gamma_f = |\mathbf{x}_F - \mathbf{x}_f|/|\mathbf{x}_F - \mathbf{x}_C|$,
the face value of a field $\phi$ is
\begin{equation}
  \phi_f = \gamma_f\,\phi_C + (1{-}\gamma_f)\,\phi_F
           + (\nabla_h\phi)_f\cdot\bigl[\mathbf{x}_f - \gamma_f\,\mathbf{x}_C - (1{-}\gamma_f)\,\mathbf{x}_F\bigr],
  \label{eq:face_interp}
\end{equation}
where the last term corrects for the offset of the face centroid from the interpolation point on the line joining the two cell centroids (a skewness correction) and vanishes on Cartesian grids~\cite{moukalled2016fvm}.
Diffusive fluxes use the reconstructed surface-normal gradient $(\nabla_h\phi)_f\cdot\mathbf{n}_f$.
Here the face gradient $(\nabla_h\phi)_f$ is interpolated from the cell-centered WLSQ gradients of the two adjacent cells,
and the diffusive flux is evaluated from it in full.
Implicit solvers instead split this flux into a two-point normal difference, which enters the implicit matrix,
and a non-orthogonal correction lagged from the previous iterate (deferred correction)~\cite{moukalled2016fvm}.
Because the residual is evaluated on the predicted state rather than solved as a linear system,
this split is not required,
and the complete diffusive flux remains differentiable with respect to the network output.
On non-periodic boundary faces, the boundary-condition operators supply the diffusive flux instead (see ``GPU-parallel residual assembly and boundary-condition enforcement''), as a two-point normal difference to a prescribed boundary value or as a prescribed normal gradient;
periodic faces are treated as interior faces through the cyclic pairing.
For the convective fluxes, the benchmarks replace the central face value of Eq.~\eqref{eq:face_interp} by a second-order linear-upwind reconstruction,
which extrapolates the upwind cell value with its WLSQ gradient,
and the data center hall uses a bounded variant of this scheme.
After boundary-condition enforcement on the boundary-face entries (see ``GPU-parallel residual assembly and boundary-condition enforcement''), face contributions are scatter-summed over the face--cell connectivity to form the residuals $\mathcal{R}_{\mathrm{cont},i}$, $\mathcal{R}_{\mathrm{mom},i}$, and $\mathcal{R}_{\mathrm{energy},i}$.

\subsection{Conservation property of the discrete operators}\label{subsec:conservation_audit}

Because the training signal \emph{is} a conservation balance, we state how the discrete operators treat the conserved quantities.
Each internal face flux is computed once and enters the balances of its two adjacent control volumes with opposite signs.
Summed over all interior cells, the assembled face-flux divergences therefore reduce to the flux through the boundary faces.
The convective and diffusive surface transfers are consequently conservative by construction,
independently of mesh quality and of the WLSQ reconstruction that supplies the face values.
This telescoping property belongs to the discrete face-flux construction;
it does not imply that predicted fields conserve physical quantities in the continuum sense.
The volume terms (pressure gradient and buoyancy) are not surface transfers and are balanced only to the accuracy of the WLSQ gradient of Eq.~\eqref{eq:wlsq_3d}.
For a predicted field, the global imbalance of an equation is the sum of its cell residuals,
and by the Cauchy--Schwarz inequality its magnitude is bounded by $\sqrt{N_{\mathrm{c}}}$ times the residual norm of Eq.~\eqref{eq:l2_norms} that the loss minimizes.
Conservation of the network outputs is thus promoted, but not imposed exactly, through the residual norm that training minimizes,
while the face-flux identity underlying each residual holds by construction.

\begin{figure}[!htbp]
\centering
\includegraphics[width=\textwidth,height=0.72\textheight,keepaspectratio]{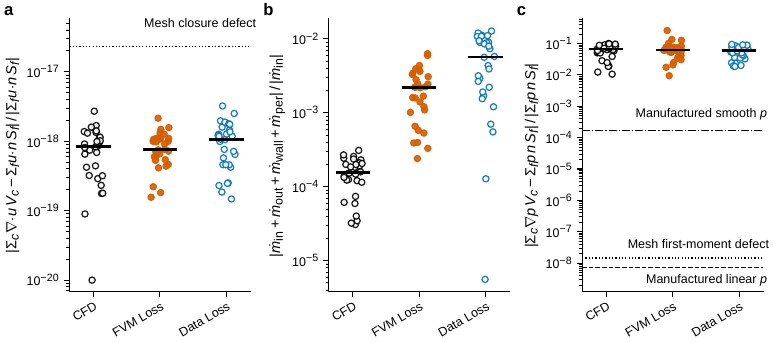}
\caption{Conservation diagnostic on final cylinder-wake states of 30 held-out conditions.
Columns show the computational fluid dynamics (CFD) reference and TransFVGN, the attention graph neural network of this study, trained with the finite-volume-method residual (FVM Loss) or CFD labels (Data Loss);
bars mark medians.
\textbf{a}, face-flux telescoping defect.
\textbf{b}, net mass imbalance, with $\dot m$ the mass flow rate through the inlet (in), outlet (out), wall and periodic (per) faces.
\textbf{c}, pressure--volume defect.
Here $c$ indexes cells of volume $V_c$ and $f$ boundary faces of area $S_f$ and outward unit normal $\mathbf{n}$;
$\mathbf{u}$ is the velocity, $p$ the pressure, and $\nabla$ and $\nabla\cdot$ the discrete gradient and divergence.}
\label{fig:conservation_balance}
\end{figure}

The conservation diagnostic evaluates the final step of all 30 cylinder test cases for the two TransFVGN models and the CFD reference (Figure~\ref{fig:conservation_balance}).
The stored fields, whatever precision the model produced them in, are promoted to double precision before the diagnostic operators are applied,
so that the roundoff of the diagnostic itself stays far below the imbalances it measures.
All 30 cases of each model pass the acceptance threshold of a final-step velocity error below one, and none is excluded.
The symmetry planes, where the boundary condition prescribes zero mass flux, are left out of the balance,
because the face reconstruction of the diagnostic evaluates a small residual outflow there in all three series, including the CFD reference.
On the remaining boundary faces, the FVM-loss model has a clearly smaller net mass imbalance than the data-loss model, both in the median and in the paired per-case comparison.
Part of the data-loss imbalance arises at the inlet and the cylinder wall, where the data-loss states depart slightly from the prescribed boundary fluxes;
the ordering of the two models is unchanged when those fluxes are replaced by their prescribed values.
The CFD reference, solved to a tight residual tolerance at every step, remains the best balanced.
The algebraic telescoping defect lies at the level of double-precision roundoff for all three series;
this tests the discrete identity rather than prediction quality.

The same diagnostic compares the volume-integrated WLSQ pressure gradient with the boundary pressure force.
For manufactured linear and smooth pressure fields the relative defect is small, which confirms the implementation.
On final rollout fields the median defect amounts to a few percent of the net boundary force and to well below one percent of the summed boundary-force magnitudes, nearly the same for all three series.
The gap between the two normalizations indicates that the net boundary force is small because its contributions largely cancel.
The remaining defect reflects the reconstruction accuracy of the WLSQ volume gradients, which, unlike the face fluxes, are not balanced by pairwise telescoping.

\subsection{Designed to suppress pressure--velocity decoupling}\label{subsec:dual_test_space}

In the collocated arrangement used here, velocity and pressure share the cell centers, and Eq.~\eqref{eq:discrete_momentum} evaluates the momentum balance on the cell control volumes.
The pressure therefore enters the residual only through the cell-centered reconstructed gradient $(\nabla_h p_{\mathrm{rgh}})_i$.
On interior stencils whose neighbors are placed point-symmetrically about the cell,
this gradient annihilates pressure fields that alternate in sign between adjacent cells (grid-parity fields),
leaving the classical checkerboard pressure mode of collocated grids essentially unpenalized~\cite{harlow1965mac,moukalled2016fvm}.
Conventional solvers suppress this pressure--velocity decoupling by staggering the variable arrangement~\cite{harlow1965mac} or through a Rhie--Chow-type flux correction~\cite{rhie1983numerical}.
Under gradient-based residual minimization, a pattern to which the residual barely responds (see ``Pressure--velocity decoupling under residual optimization'' in Results) spans a nearly flat direction of the loss landscape,
and the optimizer can accumulate error along such a direction when doing so lowers the remaining residual components.

We suppress this decoupling by evaluating the same momentum balance a second time, on staggered face control volumes.
For each interior face $f$ with owner $C$ and neighbor $F$ we form a face control volume of measure $V_f=\tfrac12(V_C+V_F)$,
and we split the cell residual into the pressure term and the pressure-free remainder.
With $\langle\cdot\rangle_f$ denoting the mean of the values in cells $C$ and $F$,
the face-normal momentum residual reads
\begin{equation}
\mathcal{R}_{\mathrm{mom},f}=
\Bigl[\mathbf{n}_f\cdot\Bigl\langle\frac{\mathcal{R}_{\mathrm{mom}}}{V}-\frac{1}{\rho_0}\nabla_h p_{\mathrm{rgh}}\Bigr\rangle_f
+\frac{1}{\rho_0}\,(\delta_n p_{\mathrm{rgh}})_f\Bigr]\,V_f,
\label{eq:face_cv_residual}
\end{equation}
where the averaged quantity is the pressure-free momentum residual per unit volume, $\mathcal{R}_{\mathrm{mom},i}/V_i-\rho_0^{-1}(\nabla_h p_{\mathrm{rgh}})_i$, of the two cells,
and $(\delta_n p_{\mathrm{rgh}})_f$ is the compact surface-normal pressure gradient between the two adjacent cells,
evaluated with a two-point normal derivative and a non-orthogonal correction.
The face residual is evaluated on every face of the mesh.
Periodic faces use the same two-sided form, with the periodic partner cell as the neighbor.
On the other boundary faces the average is replaced by the owner-cell value with $V_f=V_C$,
and the normal pressure gradient is formed with the boundary value where a pressure value is prescribed,
whereas zero-gradient and symmetry faces carry no pressure term.
The pressure enters Eq.~\eqref{eq:face_cv_residual} only through this two-point gradient,
which responds most strongly to the cell-to-cell alternation that the reconstructed gradient misses;
the unsteady, convective, diffusive, and buoyancy contributions enter through the averaged pressure-free residual per unit volume.
The half average $\langle\cdot\rangle_f$ attenuates velocity patterns that alternate in sign between the two adjacent cells,
and the cell and face residuals consequently respond most strongly to complementary mode families.

The momentum norm in Eq.~\eqref{eq:l2_norms} evaluates the union of the two.
The face-normal component is appended to the $d$ Cartesian cell components at equal weight,
adding one scalar residual per face at no additional gradient-reconstruction cost.
All benchmarks in this paper are trained with this combined form.
The ``Pressure--velocity decoupling under residual optimization'' section of Results measures the complementarity that motivates it.

The two residuals are consistent with each other.
At a state that satisfies the cell balances,
Eq.~\eqref{eq:face_cv_residual} reduces to the mismatch
$V_f\bigl[(\delta_n p_{\mathrm{rgh}})_f-\mathbf{n}_f\cdot\langle\nabla_h p_{\mathrm{rgh}}\rangle_f\bigr]/\rho_0$
between the compact normal pressure gradient and the averaged reconstructed gradient,
which corresponds to the core term of the classical Rhie--Chow flux correction~\cite{rhie1983numerical}.
On smooth fields this mismatch is a truncation-level quantity.
Unlike a Rhie--Chow-stabilized flux,
the construction modifies no discrete equation and needs no interpolation coefficient from the momentum matrix.
Both members are standard physical residuals on different control volumes,
in the spirit of staggered and covolume discretizations~\cite{harlow1965mac,perot2000conservation}.

\subsection{Time integration}
Equations~\eqref{eq:discrete_continuity}--\eqref{eq:discrete_energy} evaluate every flux and source term at the new time level~$n{+}1$.
In practice, the convective flux is evaluated as a convex combination of the old and new time steps,
which blends the implicit new-time contribution with the explicitly known old-time flux.
We parameterize this choice through a weighting factor $\vartheta\in[0,1]$ (distinct from the network parameters $\theta$).
For a generic transported quantity $\phi$,
the convective flux across face $f$ is evaluated as
\begin{equation}
  \Phi^{\mathrm{conv}}_f
  = \vartheta\,\dot{m}^{n+1}_f\,\phi^{n+1}_f
    + (1{-}\vartheta)\,\dot{m}^{n}_f\,\phi^{n}_f,
  \label{eq:imex_convection}
\end{equation}
where $\dot{m}_f=\mathbf{u}_f\cdot\mathbf{S}_f$ is the volumetric face flux.
Setting $\vartheta{=}1$ recovers the fully implicit residual in Eqs.~\eqref{eq:discrete_momentum}--\eqref{eq:discrete_energy};
$\vartheta{=}\tfrac{1}{2}$ yields a time-centered (Crank--Nicolson-type) weighting in which the convective flux is averaged over the old and new time steps while diffusion, pressure, and buoyancy are evaluated implicitly.
The cylinder-wake and data-center benchmarks use this weighting.
The new-time term $\vartheta\,\dot{m}^{n+1}_f\,\phi^{n+1}_f$ is a product of two updated-state quantities,
the face flux and the transported face value.
The convective term therefore remains nonlinear in the updated state for every $\vartheta>0$.
We retain the unsteady term for all benchmarks,
including the nominally steady cavity and natural-convection cases,
which are advanced in pseudo-transient form.
As discussed in Supplementary Note~1,
the time-derivative term adds the diagonal weight $V_i/\Delta t$ to the residual Jacobian while leaving its neighbor coupling unchanged;
this shift moves the transport blocks toward diagonal dominance and is expected to support the convergence of residual-based training.

\subsection{GPU-parallel residual assembly and boundary-condition enforcement}\label{subsec:bc_enforcement}
Once the face fluxes are computed,
the final step aggregates them into the cell-centered residuals of Eqs.~\eqref{eq:discrete_continuity}--\eqref{eq:discrete_energy}.
For computational efficiency on large 3D meshes,
this aggregation uses the GPU-parallel scatter reduction introduced in the ``Heterogeneous graph representation'' section (Figure~\ref{fig:graph_scatter}, Eq.~\eqref{eq:scatter_reduction}).
Cell-wise residuals are assembled in parallel on the GPU from the owner-pointer grouping of the cell--face incidence, combined with vectorized face-flux and volume-integral operations, memory-coalesced attribute arrays, and a single scatter-add for aggregation.
The same indexing assembles the continuity, momentum, and energy residuals in a single vectorized pass over the mesh.

Boundary types (Dirichlet, Neumann, symmetry, periodic) and their values are encoded as face-level attributes of the compound-cell graph (see ``Heterogeneous graph representation''),
and a single trained model can therefore be evaluated on new boundary values within its training range without retraining (see ``3D cylinder wake flow with heat transfer'' and ``Data center hall'' in Results).
During residual assembly, the conditions are enforced as hard constraints on the graph-resident boundary-face entries,
before and after gradient reconstruction.
Dirichlet values are written directly into the boundary attributes.
Neumann (zero- and fixed-gradient) conditions are imposed by setting the boundary-normal gradient flux directly on the boundary faces.
Periodic boundaries are handled through the cyclic face pairing defined in the ``Heterogeneous graph representation'' section.
The only auxiliary loss term is the pressure-gauge constraint of Eq.~\eqref{eq:auxiliary_losses}.
Data-loss training uses the same boundary-condition operators to set the boundary-face entries (see ``Data-loss training'').

\subsection{Training strategy}\label{subsec:training_strategy}

Both training modes share the same data pool, the same way of advancing its cases, and the same reset rule.
They differ in their loss objective and in the wall clock of a training epoch.
An \emph{epoch} consists of $K$ inner passes over the pool, each of which sweeps all mini-batches with one network evaluation and one parameter update per mini-batch;
the updated states are written back after the last pass, and every epoch therefore advances each case in the pool by one time step (Algorithm~\ref{alg:fvm_training}).
The FVM-loss models perform more than one inner pass per epoch, and the data-loss models one.

\emph{Data pool.}
Training advances many parametric cases in parallel, and each case carries a large snapshot of the flow state on the 3D mesh.
Keeping every state in GPU memory scales poorly, because the aggregate size grows with the number of cases and the mesh size and competes with the memory needed for training.
We therefore organize training around a customized \emph{data pool}.
Pools of evolving states that are written back after each update and periodically re-initialized have been used for label-free training on regular grids~\cite{wandel2021from_scratch} and in a finite-volume graph network~\cite{li2025learning}; here the pool follows the data-pool mechanism of Min et al.~\cite{min2025efficient}, is kept on disk and memory-mapped, and is reset on a first-in-first-out schedule tied to the target rollout length.
The pool is a data structure with a fixed number of indexed entries, which we call \emph{slots}.
A \emph{case} is one instance with a specific boundary condition, together with its snapshot of flow state;
a slot is the pool position that holds one case.
Concretely, a slot stores the state of one case over the compound cells and the physical channels ($u$, $v$, $w$, $p_{\mathrm{rgh}}$, $T$), advanced to that case's current time step.
In data-loss training it additionally reads the labeled reference trajectory of its condition from the converted label dataset.
Only the case currently being advanced must be materialized on the GPU;
the pool is in this sense two-tier, with persistent slots on disk and the active mini-batch in GPU memory.
For the pool sizes used in this work, slots are kept on disk and are exposed to the training process through memory mapping.
This implementation detail lets on-disk states be addressed as ordinary in-memory arrays.
The mesh information (geometry, connectivity, and the precomputed WLSQ operators of Eq.~\eqref{eq:wlsq_3d}) is identical for all cases that share a mesh.
It is therefore packed once into a static, read-only file shared by all slots and all distributed processes.
Adding a new condition consequently allocates one slot without duplicating any mesh data,
and the pool size is bounded by disk capacity rather than by GPU or host memory.

Each case's updated state is written back to its slot and becomes the input of the next forward pass.
The network is therefore trained on its own previous, error-carrying output rather than on a clean reference state.
This aligns the training input distribution with the autoregressive-rollout distribution encountered at inference,
and is intended to mitigate the error accumulation that affects surrogates trained on one-step ground-truth transitions.
The same cycle drives both training modes;
the data-loss baseline is thus likewise trained under rollout-consistent conditions.

\subsection{FVM-loss training}\label{subsec:fvm_training}

The FVM-loss objective is label-free,
since the network minimizes the discrete conservation residuals (Eq.~\eqref{eq:discrete_loss}) directly on the mesh.
For the parametric benchmarks, an FVM-loss epoch assembles the discrete residual on the mesh,
whereas data-loss training regresses one labeled case per pass.
At inference both models evaluate the network once per physical time step.
In each pass the network predicts an increment $\Delta\mathbf{q}_{\theta}$ and the boundary conditions are enforced.
WLSQ gradients and face fluxes are then assembled,
and the back-propagated residual loss updates the parameters through the SOAP optimizer~\cite{vyas2024soap}.
The state $\mathbf{q}^{n+1}$ is written back at the end of the epoch.
After write-back the time-step counter is incremented,
and each epoch therefore advances every case by one time step (Algorithm~\ref{alg:fvm_training}).

\emph{Periodic reset.}
Every epoch advances each case in the pool by one time step.
Without intervention, the pool would therefore fill with states that lie ever farther from their initial conditions.
The training distribution would then concentrate on a narrow portion of the rollout horizon,
and the network may overfit to those late-time states.
To avoid this, the pool is managed as a first-in-first-out queue.
At the end of every epoch, the oldest cases (those that have advanced for the most epochs, and hence the most time steps, since their last initialization) are removed from the pool.
Each vacated slot is immediately re-populated with a fresh case.
A new boundary condition is drawn from the fixed training partition (validation and test conditions are held out).
The flow state is then re-initialized to a uniform field given by the inflow values of that condition, without any CFD solve,
while the precomputed mesh operators are reused.
Data-loss training follows the same rule, draws its conditions from the same training partition, and starts each case from the same uniform initial field.
The number of cases replaced per epoch is the pool size divided by a target rollout length, rounded up.
With a $2400$-slot pool and a $500$-step target equal to the inference horizon, this gives $\lceil 2400/500\rceil=5$ cases per epoch for both benchmarks.
Because every epoch advances each case by one time step,
a case resides in the pool for $2400/5=480$ epochs and is rolled out for up to $480$ time steps before being replaced,
close to the $500$-step horizon evaluated at inference.
Because every reset draws a new condition from the training partition, the same training condition can occupy a slot more than once.
The resident states consequently spread over the whole training horizon instead of accumulating at its far end.

\begin{algorithm}[t]
\caption{FVM-loss training with the data pool}\label{alg:fvm_training}
\begin{algorithmic}[1]
\Require $\mathcal{M}$, $\mathcal{B}$, $\boldsymbol{\mu}$ (Eq.~\eqref{eq:problem_definition})
\Ensure trained parameters $\theta$
\State Initialize $\theta$ and the data pool from $(\mathcal{M},\mathcal{B},\boldsymbol{\mu})$
\For{epoch $= 1$ to $N_{\mathrm{epochs}}$}
    \For{$m = 1$ to $K$}
        \For{each mini-batch of $B$ case states $\mathbf{q}^{n}$ in the pool}
            \State $\mathbf{q}^{n+1}\gets\mathbf{q}^{n}+\Delta\mathbf{q}_{\theta}$ \hfill (Eq.~\eqref{eq:residual_update})
            \State Enforce $\mathcal{B}$ on the boundary-face entries
            \State $\nabla_{h}\mathbf{q}^{n+1}\gets\mathbf{G}_{h}\,\mathbf{q}^{n+1}$ \hfill (Eq.~\eqref{eq:wlsq_3d})
            \State $\mathcal{R}_{\mathrm{cont}},\,\mathcal{R}_{\mathrm{mom}},\,\mathcal{R}_{\mathrm{energy}}\gets$ flux assembly \hfill (Eqs.~\eqref{eq:discrete_continuity}--\eqref{eq:discrete_energy})
            \State $\mathcal{L}_{\mathrm{FVM}}\gets$ Eq.~\eqref{eq:discrete_loss}
            \State $\theta\gets\operatorname{SOAP}(\theta,\,\nabla_{\theta}\mathcal{L}_{\mathrm{FVM}})$
            \If{$m = K$}
                \State $\mathbf{q}^{n}\gets\mathbf{q}^{n+1}$ (excluded from back-propagation)
                \State $n\gets n+1$
            \EndIf
        \EndFor
    \EndFor
    \If{reset due}
        \State reset oldest cases with $(\mathcal{B}',\boldsymbol{\mu}')$ drawn from the training set
    \EndIf
\EndFor
\end{algorithmic}
\end{algorithm}

\begin{figure}[!t]
\centering
\includegraphics[width=\textwidth,height=0.72\textheight,keepaspectratio]{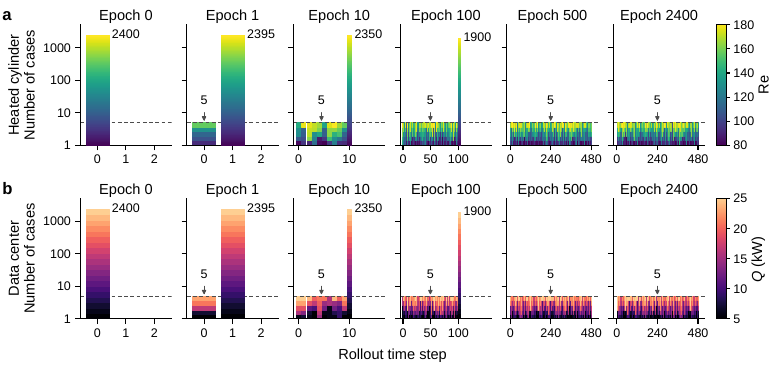}
\caption{Sampling schedule induced by the periodic case reset of the data pool.
\textbf{a}, heated-cylinder pool, colored by the Reynolds number $\mathrm{Re}$.
\textbf{b}, data-center pool, colored by the rack heat load $Q$ (kW).
Bars count the active cases at each rollout time step at epochs 0, 1, 10, 100, 500 and 2400, on a logarithmic count axis.
The reset re-initializes five cases per epoch, and occupancy builds toward a uniform level of five cases per time step (dashed line);
the tall bar is the initial cohort that has not yet been reset, labeled with its size.
Colors of reset cases follow a shuffled list of the training conditions and illustrate the mixing of conditions across the parameter sweep.}
\label{fig:curriculum_pool}
\end{figure}

The reset determines which rollout positions and which flow conditions the network sees during training.
Because the reset rule prescribes at every epoch which slots are re-initialized,
we refer to it as a sampling \emph{schedule}.
Figure~\ref{fig:curriculum_pool} shows the pool occupancy that this schedule induces at a sequence of training epochs.
At epoch $0$ every case sits at time step $0$.
The unreset cases then advance together as one cohort,
and until this cohort is recycled the tall bar in each panel sits at the time step equal to the training epoch.
After one epoch five cases at the end of the data pool have been reset to time step $0$,
while the remaining cases have advanced to time step $1$.
As training proceeds the reset repeats, and the re-initialized cases advance again from time step $0$.
Each reset hence leaves a small group of cases at earlier time steps than the main cohort,
visible as the small trailing bars in the histograms.
The pool thus builds toward a uniform occupancy of cases per time step ($2400/480=5$ for both pools).
After $480$ epochs the initial cohort has been recycled,
and the pool covers the training horizon at close to this uniform occupancy.
The bar colors show conditions drawn from across the parameter sweep at every populated time step.
From epoch 480 onward the pool therefore holds a nearly equal number of cases at every time step of the horizon,
and a case's position along the rollout is essentially independent of the flow condition it carries.
The network is thus trained over the joint range of rollout position and flow condition rather than at a single fixed time step.
The overall procedure is similar to the no-gradient unrolled-training variant analyzed by List et al.~\cite{list2025differentiability}.

\subsection{Data-loss training}\label{subsec:dd_training}

\begin{figure}[!t]
\centering
\includegraphics[width=\textwidth,height=0.72\textheight,keepaspectratio]{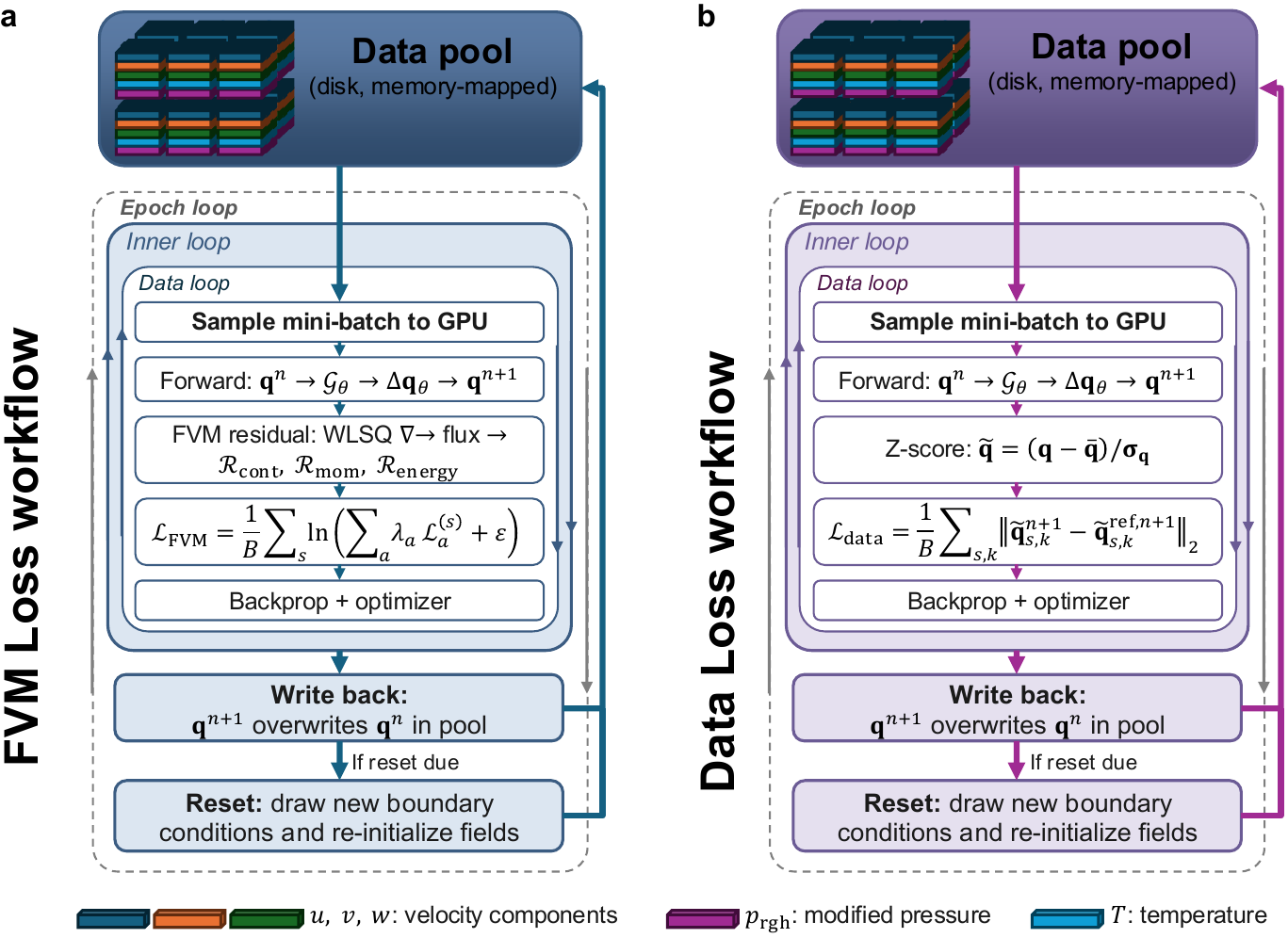}
\caption{Comparison of the FVM-loss and data-loss training workflows.
\textbf{a}, training on the finite-volume-method (FVM) residual loss.
\textbf{b}, training on computational fluid dynamics (CFD) labels (Data Loss).
The two workflows share the data-pool and reset design,
whereas their objectives and their per-epoch costs differ, the FVM-loss workflow taking more than one inner pass per epoch and the data-loss workflow one.
Here $\mathbf{q}^{n}$ is the state at time level $n$, $\mathcal{G}_{\theta}$ the network and $\Delta\mathbf{q}_{\theta}$ its increment, WLSQ the weighted least-squares reconstruction of the gradient $\nabla$, and $\mathcal{R}_{\mathrm{cont}}$, $\mathcal{R}_{\mathrm{mom}}$ and $\mathcal{R}_{\mathrm{energy}}$ the continuity, momentum and energy residuals.
In the losses $\mathcal{L}_{\mathrm{FVM}}$ and $\mathcal{L}_{\mathrm{data}}$, $B$ is the batch size, $s$ the case index, $\lambda_a$ the weight of term $\mathcal{L}_a$, $\varepsilon$ a small floor, $k$ the state channel, $\tilde{\mathbf{q}}$ the z-score of $\mathbf{q}$ with channel mean $\bar{\mathbf{q}}$ and standard deviation $\boldsymbol{\sigma}_{\mathbf{q}}$, ref the CFD label and $\|\cdot\|_2$ the $\ell_2$ norm.
GPU denotes the graphics processing unit and Backprop backpropagation.}
\label{fig:training_strategy}
\end{figure}

The data-loss baseline shares the network architectures, the two-tier data pool, and the reset rule of Algorithm~\ref{alg:fvm_training},
and each of its cases likewise starts from the uniform initial field defined by its boundary conditions.
Its objective is supervised regression against precomputed CFD reference fields.
Figure~\ref{fig:training_strategy} contrasts the two workflows,
which share the same sampling and write-back cycle and differ principally in the loss.

\emph{Z-score normalization and $\ell_2$ loss.}
Because the channels can span widely different scales (in the data center hall, velocity $\mathcal{O}(1)$~m\,s$^{-1}$ versus absolute temperature $\mathcal{O}(10^{2})$~K),
a channel-wise z-score normalization~\cite{lecun1998efficient} is applied before the loss,
using global statistics $\bar{\mathbf{q}}$ and $\boldsymbol{\sigma}_{\mathbf{q}}$ precomputed over all training labels:
\begin{equation}
\tilde{\mathbf{q}} = (\mathbf{q} - \bar{\mathbf{q}})\,/\,\boldsymbol{\sigma}_{\mathbf{q}},
\label{eq:zscore}
\end{equation}
and the loss sums the $\ell_2$ norms of the five normalized channels over interior cells and averages the result over the batch:
\begin{equation}
\mathcal{L}_{\mathrm{data}} = \frac{1}{B}\sum_{s=1}^{B}\sum_{k=1}^{5}\sqrt{\sum_{i\in\mathcal{V}_{\mathrm{c}}} \big(\tilde{q}_{i,k}^{\,n+1} - \tilde{q}_{i,k}^{\,\mathrm{ref},n+1}\big)^2},
\label{eq:dd_loss}
\end{equation}
where $\tilde{q}_{i,k}^{\,n+1}$ is channel $k$ of the normalized prediction in cell $i$ and $\tilde{q}_{i,k}^{\,\mathrm{ref},n+1}$ the same channel of the normalized CFD label of case $s$ at the same time level.
In training and inference alike, the boundary-condition operators (see ``GPU-parallel residual assembly and boundary-condition enforcement'') set the boundary-face entries,
and the loss compares normalized interior values.
The prediction is denormalized via $\mathbf{q} = \tilde{\mathbf{q}}\,\boldsymbol{\sigma}_{\mathbf{q}} + \bar{\mathbf{q}}$ before write-back,
and the model thus receives original-scale inputs at every step.

\subsection{Experimental setup}\label{subsec:experimental_setup}

The CFD references use OpenFOAM~\cite{weller1998openfoam}.
Steady LDC cases use \texttt{simpleFoam};
as an exception, the LDC reference at $\mathrm{Re}=3200$ is a time average of a \texttt{pimpleFoam} computation, because the steady solver does not converge at this Reynolds number.
The natural-convection reference uses \texttt{buoyantBoussinesqSimpleFoam},
and the transient thermo-fluid cases use the buoyant PIMPLE formulation.
The transient references, both label datasets and the timing benchmark share one solve setting, stated below.
All cases are run as laminar flows.
These solvers generate the reference data and define the finite-volume schemes with which the differentiable residual of the FVM-loss model is designed to be consistent.
The differentiable residual follows the conservative control-volume structure, using linear-upwind convection and reconstructed diffusive fluxes.
Its WLSQ gradient reconstruction, face interpolation, and convective time weighting are implemented as differentiable operations.
They do not reproduce every numerical choice of the reference solver (e.g.\ the specific limiter, linear-solver tolerance, under-relaxation, or PIMPLE/PISO outer-loop settings).

Table~\ref{tab:experiment_setup} summarizes the dataset sample counts, on-disk label and data-pool sizes, train/validation/test splits, training horizon, rollout protocol, and parametric variable ranges for the two parametric 3D benchmarks (the heated cylinder wake and the data center hall) across the CFD reference,
the data-loss baseline,
and the FVM-loss model.

\begin{table}[!htbp]
\centering
\caption{Experimental setup for the two parametric 3D benchmarks.
For each benchmark we list three configurations, namely the CFD reference,
the data-loss baseline,
and the FVM-loss model.
``Dataset samples'' is the number of labeled CFD trajectories (the FVM-loss model uses none);
``Train:Valid:Test'' is the partition of the operating conditions, which the FVM-loss model shares without using labels;
``Label size'' is the on-disk size of the converted label dataset used by the data-loss baseline;
``Data pool'' is the on-disk size of the data pool that caches the model's own rollout states during training;
``Trained time steps'' is the target rollout length of the reset rule, the static bound given in the ``FVM-loss training'' section (each case is advanced for at most 480 steps, whereas the scored rollouts realize all 500);
and ``Rollout samples''/``Rollout time steps'' are the held-out test conditions and autoregressive horizon used for evaluation.
In the parametric range column, $\mathrm{Re}$ is dimensionless, $Q$ is in kW, $q_v$ is in m$^3$\,s$^{-1}$, and $\Delta T_{\mathrm{rack}}$ is in K.
Dashes denote not-applicable entries.
Both data-center models are evaluated on the 30-condition test split.}\label{tab:experiment_setup}
\footnotesize
\setlength{\tabcolsep}{2.0pt}
\begin{tabular}{@{}>{\centering\arraybackslash}p{1.6cm}lccccccc>{\raggedright\arraybackslash}p{2.4cm}@{}}
\hline
Benchmark                                                                       & \multicolumn{1}{c}{Method} & \makecell{Dataset\\samples} & \makecell{Label\\size\\(GB)} & \makecell{Data\\pool\\(GB)} & \makecell{Train:\\Valid:\\Test} & \makecell{Trained\\time\\steps} & \makecell{Rollout\\samples} & \makecell{Rollout\\time\\steps} & \multicolumn{1}{c}{\makecell{Parametric\\range}}                                                                              \\ \hline
\multirow{3}{*}{\begin{tabular}[c]{@{}c@{}}3D Heated \\ Cylinder\end{tabular}}  & CFD                        & ---                                                                    & ---                                                               & --- & ---                                                                                                      & ---                                                                                                      & 30                                                                     & 500                                                                                                      & \multirow{3}{*}{$\mathrm{Re}\in[80,180]$}                                                                                                                                \\
                                                                                & Data Loss                  & 300                                                                    & 520                                                          & 8.3~GB & 8:1:1                                                                                                    & 500                                                                                                      & 30                                                                     & 500                                                                                                      &                                                                                                                                                                          \\
                                                                                & FVM Loss                   & 0                                                                      & 0                                                                 & 8.3~GB & 8:1:1 & 500                                                                                                      & 30                                                                     & 500                                                                                                      &                                                                                                                                                                          \\ \hline
\multirow{3}{*}{\begin{tabular}[c]{@{}c@{}}Data \\ Center \\ Hall\end{tabular}} & CFD                        & ---                                                                    & ---                                                               & --- & ---                                                                                                      & ---                                                                                                      & 30                                                                     & 500                                                                                                      & \multirow{3}{*}{\makecell[l]{$Q$: 5--25\\ $q_v$: 0.60--1.80\\ $\Delta T_{\mathrm{rack}}$: 2.38--35.7}} \\
                                                                                & Data Loss                  & 300                                                                    & 476                                                          & 7.6~GB & 8:1:1                                                                                                    & 500                                                                                                      & 30                                                                     & 500                                                                                                      &                                                                                                                                                                          \\
                                                                                & FVM Loss                   & 0                                                                      & 0                                                                 & 7.6~GB & 8:1:1 & 500                                                                                                      & 30                                                                     & 500                                                                                                      &                                                                                                                                                                          \\
\hline
\end{tabular}
\end{table}

\emph{3D lid-driven cavity.}
A unit cube ($L=1$~m) with a moving top lid (velocity $U_{\mathrm{lid}}$ along $+x$) and stationary no-slip walls elsewhere;
the flow is incompressible with no thermal coupling (gravity and the energy equation are inactive).
The lid speed is $U_{\mathrm{lid}}=1$~m\,s$^{-1}$ and the kinematic viscosity is set per case as $\nu=U_{\mathrm{lid}}L/\mathrm{Re}$, spanning $\nu\in[3.13\times10^{-4},\,10^{-2}]$~m$^2$\,s$^{-1}$ over $\mathrm{Re}=U_{\mathrm{lid}}L/\nu\in[100,3200]$;
mesh $48^3$ (110{,}592 cells);
each case is trained until its residual loss stops decreasing.
The model is used to solve five representative Reynolds numbers ($\mathrm{Re}=100,400,1000,2000,3200$);
this benchmark tests solver accuracy on the incompressible velocity and pressure fields, not parametric generalization.

\emph{Natural convection.}
The unit cube has a 1~K temperature difference between the cold wall at $x=0$ and the hot wall at $x=1$~m;
all other walls are adiabatic and every wall is no-slip.
Gravity acts along $-y$, with $g=9.81$~m\,s$^{-2}$ and $\beta=3.33\times10^{-3}$~K$^{-1}$.
At $\mathrm{Pr}=0.71$, viscosity is chosen from $\mathrm{Ra}=g\beta(T_{\mathrm{hot}}-T_{\mathrm{cold}})H^3/(\nu\alpha)$ with $\alpha=\nu/\mathrm{Pr}$ for $\mathrm{Ra}=10^3$--$10^6$.
The wall-refined mesh has $48^3$ cells (grading parameter four).
The exported temperature is the deviation from the mean wall temperature, with wall values $-0.5$ and $+0.5$~K,
and the reported dimensionless temperature is $\theta=(T-T_{\mathrm{cold}})/(T_{\mathrm{hot}}-T_{\mathrm{cold}})$.
Velocity and heat transfer are compared with steady buoyant CFD on the same mesh.
Each condition is solved separately as a steady residual solve.

\emph{3D heated cylinder wake.}
Circular cylinder ($D=0.1$~m) centered $10D$ downstream of the inlet of a $25D\times20D\times8D$ domain;
dimensionless inlet temperature $T_\infty=0$ and isothermal wall $T_{\mathrm{w}}=1$ (the solved and reported temperature variable);
periodic spanwise boundaries ($L_z=8D$), symmetry on the top and bottom boundaries.
The kinematic viscosity is fixed at $\nu=10^{-3}$~m$^2$\,s$^{-1}$ and the Reynolds number $\mathrm{Re}=U_{\mathrm{in}}D/\nu\in[80,180]$ is varied through the inlet velocity $U_{\mathrm{in}}=\mathrm{Re}\,\nu/D$;
the buoyant Boussinesq formulation of the ``Governing equations'' section remains active, and the fluid has $\mathrm{Pr}=0.71$, $\beta=3\times10^{-3}$ per unit dimensionless temperature, $g=9.81$~m\,s$^{-2}$, and $\rho_0=1$~kg\,m$^{-3}$.
With $g\beta(T_{\mathrm{w}}-T_{\infty})\approx0.029$~m\,s$^{-2}$, the Richardson number lies between about $9\times10^{-4}$ and $4.6\times10^{-3}$ over the Reynolds range, which places the cases in the forced-convection regime.
The mesh is an unstructured hexahedral mesh generated block by block (147{,}200 cells);
each case is advanced for 500 time steps of $\Delta t=0.01$~s, covering a $5$~s horizon.
The label dataset comprises 300 cases split $8\!:\!1\!:\!1$ into training, validation and test sets;
the FVM-loss model is trained label-free on the same training partition of conditions (the $80\%$ training split of $\mathrm{Re}\in[80,180]$ used by the data-loss baseline, but without labels),
and all configurations are evaluated by rolling out the 30 held-out test Reynolds numbers (Table~\ref{tab:experiment_setup}).

\emph{Elliptical-cylinder transfer.}
The study reported in ``Transfer to unseen cross-section geometries'' replaces the circular cylinder by an ellipse with streamwise semi-axis $a=D/2=0.05$~m and cross-stream semi-axis $b$,
at the aspect ratios $b/a=0.5$, $0.7$, $0.85$, $1.2$, $1.5$ and $2$ together with the training circle ($b/a=1$).
Ratios above one therefore give bluffer bodies than the circle, and ratios below one more slender ones.
The Reynolds number is held at $100$ based on the streamwise axis $2a$; expressed with the cross-stream width $2b$ instead, the effective Reynolds number is $100\,b/a$, that is, $50$--$200$ across this family.
Each mesh keeps the block topology of the cylinder mesh and has $147{,}200$ cells.
All cases are run at $\mathrm{Re}=U_{\mathrm{in}}D/\nu=100$ ($D=2a$, $U_{\mathrm{in}}=1$~m\,s$^{-1}$) with the fluid properties, boundary conditions, time step, and $500$-step horizon of the cylinder benchmark,
and the OpenFOAM reference uses the same solver settings.
The six models trained on the circular cylinder are applied to these meshes directly.
Figure~\ref{fig:ellipse_generalization} reports a separate geometric-mean error for velocity, modified pressure and temperature;
the velocity error is the relative error of the velocity vector over all three components.

\emph{Cylinder references and rollout diagnostics.}
The per-step curves of the six-model comparison compare coincident physical times.
Errors of time-mean fields use $t=3.5$--$5$~s, which reduces phase sensitivity.
The lift and drag coefficients follow the standard definitions of the cylinder benchmark of Sch{\"a}fer et al.~\cite{schafer1996benchmark}.
Lift amplitude (half the peak-to-peak lift coefficient) and mean drag use the same window in the six-model comparison.
The canonical force and frequency study instead uses $t=3$--$5$~s and reports both the lift and the cross-stream velocity at a monitoring point $5D$ downstream of the cylinder.
All OpenFOAM references (the canonical $\mathrm{Re}=100,180$ fields, the test-set label dataset, the elliptical-cylinder references and the inference-cost benchmark alike) share one solve setting, a $10^{-6}$ residual-control target.
Pressure visualization removes a fitted outer-slice plane and spatial mean;
numerical full-field errors retain the stored pressure convention.

\emph{Data center hall.}
Full-hall dimensions $6.71\times5.49\times3$~m;
two racks in a hot/cold aisle configuration;
underfloor plenum heating, ventilation and air-conditioning (HVAC) distribution;
a mesh of about $1.4\times10^{5}$ cells on the symmetry-reduced computational domain (a quarter of the full hall, obtained using the two symmetry planes of Figure~\ref{fig:dc_3d_layout}).
The working fluid has the properties of air, with $\mathrm{Pr}=0.71$, $\beta=3.33\times10^{-3}$~K$^{-1}$ and $g=9.81$~m\,s$^{-2}$, except for an increased kinematic viscosity, $\nu=5.60\times10^{-5}$~m$^2$\,s$^{-1}$ (about 3.5 times that of air), chosen to lower the Reynolds number of this synthetic configuration.
The rack-face Reynolds number is then of order $10^{4}$ and the Rayleigh number based on the rack height of order $10^{9}$;
a physical hall would be turbulent under these conditions, and the laminar hall therefore serves as a numerical stress test of the surrogate rather than as a validated data-center model.
The reference state has temperature $T_0=300.15$~K, density $\rho_0=1.1614$~kg\,m$^{-3}$, and specific heat $c_p=1005$~J\,kg$^{-1}$\,K$^{-1}$.
The dataset comprises 300 CFD trajectories on a $20\times15$ grid, varying the total rack heat load over the symmetry-reduced rack block, $Q\in[5,25]$~kW, and the rack volumetric flow rate, $q_v\in[0.60,1.80]\,\text{m}^3\,\text{s}^{-1}$;
the cold-air supply flow rate is fixed at $q_{v,\mathrm{cold}}=0.9983\,\text{m}^3\,\text{s}^{-1}$.
The rack heat load is imposed as a prescribed temperature rise across the rack-outlet patch, $\Delta T_{\mathrm{rack}}=Q/(\rho_0 c_p\,q_v)$, giving $\Delta T_{\mathrm{rack}}\in[2.38,35.7]$~K.
Each trajectory contains 500 snapshots with $\Delta t=0.1$~s on a mesh with 138{,}150 interior cells (158{,}600 compound cells including the boundary-face entries).
FVM-loss training uses the same $80\%$ training split of the $(Q,q_v)$ conditions as the data-loss baseline (Table~\ref{tab:experiment_setup}), but without labeled CFD data,
and the reset of both training modes draws only from this split.
Both evaluated data-center models have eight processor layers and are scored on the same 30 held-out conditions.

\emph{Training configuration.}
All FVM-loss and data-loss models are trained using DDP and the SOAP second-order optimizer~\cite{vyas2024soap} with zero weight decay.
The two steady benchmarks use the same architecture, objective and optimizer, and each case is advanced in pseudo-time until its residual loss stops decreasing.
Training and inference are carried out on eight NVIDIA GeForce RTX 4090 GPUs with 24~GB of memory each,
and each inference rollout runs on a single GPU.
All models of the two parametric benchmarks are trained with batch one per rank (global batch eight) and a 2400-slot pool.

Both model families receive a 2400-epoch budget over the data pool, the FVM-loss training performing more than one inner pass per epoch and the data-loss training one;
the learning rate is $5\times10^{-5}$.
The data pool holds 2400 cases, ten times the 240 training conditions of each parametric benchmark.
The target rollout length is 500 steps for both benchmarks, equal to the inference horizon,
which gives $\lceil 2400/500\rceil=5$ resets per epoch and a residence of $480$ epochs per case (see ``FVM-loss training'').
The TransFVGN models use the backbone described in the ``Neural network architecture'' section, with a five-channel output for the state increment $\Delta\mathbf{q}_{\theta}$;
the next state follows as $\mathbf{q}^{n+1} = \mathbf{q}^{n} + \Delta\mathbf{q}_{\theta}$ (Eq.~\eqref{eq:residual_update}).

%%===== Back Matter =====%%
\backmatter

\bmhead{Data availability}

\ifshowauthors
The data used for the accuracy comparison of the first two steady-state cases and the data used to train the data-loss models of the remaining unsteady cases can both be found in the Zenodo repository \url{https://doi.org/10.5281/zenodo.23085680}.
\else
The data repository is withheld to preserve anonymity during review.
\fi

\bmhead{Code availability}

\ifshowauthors
The code for this research can be found in the Git repository \texttt{https://github.com/Litianyu141/Gen-FVGN-3D-unsteady}.
\else
The code repository is withheld to preserve anonymity during review.
\fi

\bmhead{Acknowledgements}

\ifshowauthors
This research is supported by the Ministry of Education (MOE), Singapore, under its Academic Research Fund Tier 1 project ``Towards Resilient Data Centre Operations via Physics-Informed Predictive Maintenance'' (Award No. RG104/23); this research is also supported by the RIE2025 Industry Alignment Fund -- Industry Collaboration Projects (IAF-ICP) (Award I2301E0026), administered by A*STAR, as well as supported by Alibaba Group and NTU Singapore through Alibaba-NTU Global e-Sustainability CorpLab (ANGEL).
The National Supercomputing Centre (NSCC), Singapore, provided support.
\else
Acknowledgements are temporarily withheld to preserve anonymity during review.
\fi

\bmhead{Funding}

\ifshowauthors
This research is supported by the Ministry of Education (MOE), Singapore, under its Academic Research Fund Tier 1 project ``Towards Resilient Data Centre Operations via Physics-Informed Predictive Maintenance'' (Award No. RG104/23); this research is also supported by the RIE2025 Industry Alignment Fund -- Industry Collaboration Projects (IAF-ICP) (Award I2301E0026), administered by A*STAR, as well as supported by Alibaba Group and NTU Singapore through Alibaba-NTU Global e-Sustainability CorpLab (ANGEL).
The first author thanks the Interdisciplinary Graduate Programme (IGP), Graduate College, Nanyang Technological University, for the NTU Research Scholarship.
The funders had no role in study design, data collection and analysis, decision to publish or preparation of the manuscript.
\else
Temporarily withheld to preserve anonymity during review.
\fi

\bmhead{Author contributions}

\ifshowauthors
T.L.: Conceptualization, Methodology, Software, Validation, Visualization, Writing -- original draft, Writing -- review \& editing.
Z.C.: Data curation, Formal analysis, Writing -- review \& editing.
Q.Z.: Writing -- review \& editing.
R.W.: Data curation, Formal analysis, Writing -- review \& editing.
B.S.: Methodology, Validation, Supervision, Writing -- review \& editing.
Y.W.: Supervision, Funding acquisition, Project administration, Writing -- review \& editing.
\else
Temporarily withheld to preserve anonymity during review.
\fi

\bmhead{Competing interests}

The authors declare no competing financial or non-financial interests.

\bibliography{references}

%%===== Figures and legends (after the references, in citation order) =====%%
\clearpage

\end{document}